\documentclass[12pt]{article}
\pdfoutput=1
\DeclareMathAlphabet{\scr}{U}{rsfs}{m}{n}
\usepackage{latexsym,url}
\usepackage[mathscr]{eucal}
\usepackage{amsfonts}
\usepackage{amscd}
\usepackage{cite}         
\usepackage{amssymb}
\usepackage{color}
\usepackage{colordvi}
\usepackage[centertags]{amsmath}
\usepackage{enumerate}
\usepackage{graphicx}
\usepackage{booktabs}
\usepackage{soul}
\usepackage{rotating}
\usepackage{pdflscape}

\newcommand{\cleqn}{\setcounter{equation}{0}}

\newcommand{\newc}{\newcommand}
\newc{\be}{\begin{equation}}
\newc{\ee}{\end{equation}}
\newc{\bea}{\begin{eqnarray}}
\newc{\eea}{\end{eqnarray}}
\newc{\ben}{\begin{equation*}}
\newc{\een}{\end{equation*}}
\newc{\bean}{\begin{eqnarray*}}
\newc{\eean}{\end{eqnarray*}}
\newc{\ol}{\overline}
\newc{\wt}{\widetilde}
\newc{\bs}{\boldsymbol}
\newc{\m}{\mathcal}
\newc{\la}{\lambda}
\newc{\lra}{\longrightarrow}
\newc{\vp}{\varphi}
\newc{\ti}{\tilde}
\newc{\vev}[1]{\langle#1 \rangle}

\begin{document}

\title{\hfill ~\\[-30mm]
          \hfill\mbox{\small  QFET-2016-03}\\[-3.5mm]
          \hfill\mbox{\small  SI-HEP-2016-08}\\[13mm]
       \textbf{
CP-odd invariants for multi-Higgs models:
applications with discrete symmetry
}}

\author{
Ivo de Medeiros Varzielas$\,^a\,$\footnote{E-mail: {\tt ivo.de@soton.ac.uk}}
,~~
Stephen F. King$\,^a\,$\footnote{E-mail: {\tt king@soton.ac.uk}}
,\\[2mm]
Christoph Luhn$\,^b\,$\footnote{E-mail: {\tt christoph.luhn@uni-siegen.de}}
,~~
Thomas Neder$\,^a\,$\footnote{E-mail: {\tt t.neder@soton.ac.uk}}\\[8mm]
$^a$\,\it{\small School of Physics and Astronomy, University of Southampton,}\\
\it{\small SO17 1BJ Southampton, United Kingdom}\\[2mm]
$^b$\,\it{\small Theoretische Physik 1, Naturwissenschaftlich-Technische
    Fakult\"at, Universit\"at Siegen,}\\
\it{\small Walter-Flex-Stra{\ss}e 3, 57068 Siegen, Germany}
}

\date{}

\maketitle

\begin{abstract}
\noindent  
CP-odd invariants provide a basis independent way of studying the CP properties of Lagrangians. 
We propose powerful methods for constructing basis invariants and determining whether they are CP-odd or CP-even, then systematically construct all of the simplest CP-odd invariants up to a given order,
finding many new ones. The CP-odd invariants are valid for general potentials
when expressed in a standard form. 
We then apply our results to scalar
potentials involving three (or six) Higgs fields which form irreducible triplets under a discrete symmetry,
including invariants for both explicit as well as spontaneous CP violation. 
The considered cases include one triplet of Standard Model (SM) gauge
singlet scalars, one triplet of SM Higgs doublets, two triplets of SM singlets, and
two triplets of SM Higgs doublets. For each case we study the potential
symmetric under one of the simplest discrete symmetries with irreducible
triplet representations, namely $A_4$, $S_4$, $\Delta(27)$ or $\Delta(54)$, 
as well as the infinite classes of discrete symmetries $\Delta(3n^2)$ or $\Delta(6n^2)$.
\end{abstract}
\thispagestyle{empty}
\vfill

\newpage 
\tableofcontents
\setcounter{page}{1}

\section{Introduction}
\cleqn

CP symmetry, which is just the combination of particle-antiparticle exchange and space inversion, is presently known to be violated only by the weak interactions involving quarks in the Standard Model (SM)
\cite{Kobayashi:1973fv}. The origin of the observed SM quark CP violation (CPV) is a natural consequence of three generations of quarks whose mixing 
is described by a complex CKM matrix. Although the CKM matrix can be parameterised in different ways, 
it was realised that the amount of CPV in physical processes always depends on a particular weak basis invariant which can be expressed in terms of the quark mass matrices \cite{Jarlskog:1985ht}.
In the SM the electroweak symmetry $SU(2)_L\times U(1)_Y$ is broken to the electromagnetic
gauge group $U(1)_{Q}$ by a single Higgs doublet, resulting in a single physical Higgs boson which 
has been observed with a mass near 125 GeV \cite{Aad:2012tfa, Chatrchyan:2012ufa}.
Although CP is automatically conserved by the Higgs potential of the SM, with more than one Higgs doublet
it is possible that the Higgs potential violates CP, providing a new source of CPV \cite{Lee:1973iz}.
This is welcome since Sakharov discovered that CPV is a necessary condition for  
baryon asymmetry generation \cite{Sakharov:1967dj} and 
CPV arising from the quark sector of the SM is insufficient \cite{Kuzmin:1985mm}. 

It is also possible, indeed likely, that CP could be violated in the lepton sector, as is hinted at by global fits \cite{Forero:2014bxa,Gonzalez-Garcia:2014bfa}, and such a source of CPV could also contribute to the baryon asymmetry via leptogenesis \cite{Fukugita:1986hr}. In this case one would like to construct models that explain the structure of the lepton mass matrices, through which CPV enters the processes for creating the baryon asymmetry. Typical examples of such models that use discrete symmetries to constrain the structure of mass matrices need several multiplets of scalar fields that also transform under the same symmetry (for reviews, cf.\ \cite{Altarelli:2010gt,Ishimori:2010au,Grimus:2011fk,King:2013eh,King:2014nza,King:2015aea}). Such models provide a motivation to study multiple SM Higgs singlets (sometimes
called ``flavons'' in this context) as well as electroweak doublets. In the context of flavour models it is natural to consider Higgs doublets or singlets which play the role of ``flavons'' and 
form irreducible triplets under some spontaneously broken discrete family symmetry.
Motivated by the above considerations, we shall also study CPV Higgs potentials of this type.

As already mentioned in the context of the CKM matrix, the study of CP is a subtle topic because of the basis dependent nature of the phases which control CPV. Similar considerations also apply to the phases which
appear in the parameters of the potentials of multiple scalars.

An important tool to assist in determining whether CP is violated or not are basis independent CP-odd invariants (CPIs), whose usefulness has been shown in the SM in addressing CP violation arising from the CKM matrix,
sourced from the Yukawa couplings. The first example of the use of such invariants was the Jarlskog invariant~\cite{Jarlskog:1985ht},
which was reformulated in~\cite{Bernabeu:1986fc} in a form which is generally valid for an arbitrary number of generations. Generalising the invariant approach ~\cite{Bernabeu:1986fc} and applying it to fermion sectors of theories with Majorana neutrinos~\cite{Branco:1986gr} or with discrete symmetries~\cite{Branco:2015hea, Branco:2015gna} leads to other relevant CPIs. 

In extensions of the Higgs sector of the SM, the CP violation arising from the 
parameters of the scalar potential can be studied in a similar basis invariant way as
for the quark sector. For example, in the general two Higgs Doublet Model (HDM)~\cite{Lee:1973iz} (see \cite{Keus:2015hva} for a recent analysis)
a CPI was identified in~\cite{Mendez:1991gp}. More generally, 
applying the invariant approach to scalar potentials has revealed relevant
CPIs~\cite{Lavoura:1994fv, Botella:1994cs, Branco:2005em}, including for the
2HDM~\cite{Davidson:2005cw, Gunion:2005ja}. 
Building on these results, the goal of this work is to consider
    yet more general Higgs potentials and adopt the powerful method of
    so-called contraction matrices in order to identify and construct new
    non-trivial CPIs,  which we subsequently apply to potentials involving
three or six Higgs fields (which can be either electroweak doublets or
singlets) which form irreducible triplets under a discrete symmetry. 

We begin by reviewing the
systematic approach to CPIs for arbitrary scalar potentials, focusing on
renormalisable potentials with quadratic and quartic couplings. The reader may
be primarily interested in cases where the Higgs fields are electroweak
$SU(2)_L$ doublets, but the formalism can also be applied to
more general scalar potentials including cases where the Higgs fields are SM
singlets. 
We develop powerful tools where basis invariants~\cite{Branco:2005em, Davidson:2005cw, Gunion:2005ja} 
can be represented pictorially by diagrams (introduced for the 2HDM in~\cite{Davidson:2005cw}) and introduce matrices later designated as contraction matrices, that identify how the parameters in the
potential are combined to form a basis invariant. The diagrams and matrices
are extremely helpful in distinguishing CPIs from basis invariants that are
CP-even, as well as cataloguing each CPI uniquely in association with an
element of a group of permutations. We emphasise that CPIs as defined  via
such matrices are valid for any potential, and then take specific expressions
when specialising to a potential (often vanishing for cases where the
potential is very symmetric, even if the potential features explicit CP
violation as shown by other non-vanishing CPIs).

Having translated the well-known technique for constructing CPIs 
to diagrams and contraction matrices, we apply this  formalism
to some physically interesting cases. We begin with the familiar example of
the general 2HDM.
We then move on to examples of potentials which involve three or six Higgs
fields which fall into irreducible triplet representations of discrete symmetries belonging to the $\Delta(3n^2)$ and $\Delta(6n^2)$ series studied extensively in the context of flavour and CP models in 
\cite{
Feruglio:2012cw,
Ding:2013hpa, Feruglio:2013hia, King:2013vna, Luhn:2013vna, Ding:2013bpa, Li:2013jya, Ding:2013nsa,
King:2014rwa, Ding:2014ssa, Ding:2014hva, Neder:2014mxa, Li:2014eia, Hagedorn:2014wha, Ding:2014ora,
Bjorkeroth:2015ora, Li:2015jxa, DiIura:2015kfa, Ballett:2015wia, Neder:2015wsa, Turner:2015uta, Ding:2015rwa, Bjorkeroth:2015uou,
Li:2016ppt}.
We consider specific cases of the 3HDM \cite{Branco:1983tn, Toorop:2010ex, Toorop:2010kt, deMedeirosVarzielas:2011zw, Varzielas:2012nn, Bhattacharyya:2012pi, Ivanov:2012fp,Ivanov:2012ry, Varzielas:2013sla, Varzielas:2013eta, Keus:2013hya, Keus:2014jha, Ivanov:2014doa, Keus:2015xya, Fallbacher:2015rea, Ivanov:2015mwl, Emmanuel-Costa:2016vej} and of the 6HDM \cite{Ivanov:2013nla,Ivanov:2010ww,Ivanov:2010wz,Hernandez:2013dta, Keus:2014isa, Nishi:2014zla, Varzielas:2015joa} where the three or six Higgses are related by the discrete symmetry as one or two (flavour) triplets.
Although many of these cases have already been studied in the literature, our systematic formalism yields
many new results. For example, although $\Delta(27)$ with a single triplet of Higgs doublets has been extensively studied in the literature~\cite{Branco:1983tn, deMedeirosVarzielas:2011zw, Varzielas:2012nn, Bhattacharyya:2012pi, Varzielas:2013sla, Varzielas:2013eta, Fallbacher:2015rea}, using the invariant approach and the CPIs we are able to identify several new results of interest.

Using the invariant approach, the considered cases include one triplet of SM gauge
singlets, one triplet of SM Higgs doublets, two triplets of SM singlets, and
two triplets of SM Higgs doublets, where for each case we study the potential
symmetric under one of the simplest discrete symmetries with irreducible
triplet representations, namely $A_4$, $S_4$, $\Delta(27)$ or $\Delta(54)$, 
as well as the infinite classes of discrete symmetries $\Delta(3n^2)$ or $\Delta(6n^2)$.
In each case, we show which potentials are in general CP conserving (all the CPIs vanish, and we provide a CP symmetry that leaves the potential invariant) or in general CP violating (in which case it is sufficient to show a single non-vanishing CPI). For the CP violating potentials we further demonstrate the effects of imposing specific CP symmetries, that in constraining the parameters of the potential in one way or the other, lead all CPIs to vanish.
Furthermore, we extend our formalism by including also Vacuum Expectation Values (VEVs), obtaining Spontaneous CPIs (SCPIs) that are non-vanishing if CP is spontaneously violated (as considered earlier in~\cite{Lavoura:1994fv, Botella:1994cs}). We illustrate one of these SCPIs by applying it to the better studied $\Delta(27)$ potential, exploring different CP symmetries and VEVs that either conserve or spontaneously violate the imposed CP symmetry.

Since this subject has been studied extensively, it is relevant to clarify what is new in this paper:
\begin{itemize}
\item We improve on the diagram-based approach of \cite{Davidson:2005cw}.
\item With respect to the existing literature \cite{Lavoura:1994fv,
  Botella:1994cs, Branco:2005em,Davidson:2005cw, Gunion:2005ja}, we build CPIs
  up to orders in the bi-linear and quartic scalar couplings which are higher than
  those of previous work.
\item To the best of our knowledge, we discuss for the first time the use of
  CPIs in the context of discrete symmetries and give extensive examples. 
\item We introduce a very efficient new method for identifying higher
  order CPIs based on contraction matrices (in Appendix~\ref{sym_invs}).
\end{itemize}

The layout of the paper is as follows. In Section~\ref{CPoddI} we cover our general formalism. In Section~\ref{sec:2HDM} we revisit the 2HDM potential showing how our formalism applies. In Section~\ref{sec:A4}, \ref{sec:D27} and \ref{sec:D3n2}  we apply our techniques respectively to $\Delta(3n^2)$ and $\Delta(6n^2)$ groups with $n=2$ ($A_4$, $S_4$), $n=3$ ($\Delta(27)$, $\Delta(54)$) and $n>3$. A summary of the results obtained for the potentials invariant under discrete symmetries is contained in Section~\ref{sec:table} (including Table~\ref{ta:summary}). Section~\ref{spontCPI} is dedicated to SCPIs. We conclude in Section~\ref{sec:conc}. Further material is included in Appendix~\ref{sym_invs} discussing symmetries of basis invariants (including CPIs), Appendix~\ref{invariant_lists} features a complete list of the CPIs and SCPIs we found and used throughout the paper, and Appendix~\ref{6n2_from_3n2} discusses how to obtain results for $\Delta(6n^2)$ from the results of the $\Delta(3n^2)$ potentials.

\section{CP-odd invariants for scalar potentials}\label{CPoddI}
\cleqn

\subsection{General formalism}
One important aim of this paper is to explore the CP properties of the Higgs
sector of models with several copies of SM Higgs doublets. Often, scalar
potentials can be confusingly complex and it can be unclear which parameters
can contribute to CP violation. This situation is made even more difficult by
the possibility of choosing different bases which modify the explicit form of the
potential but should not change the physics described by it. Both of these
difficulties can be overcome by CPIs, in this case CP-odd (Higgs-) basis
invariants, that, when non-zero, indicate CP violation. A similar CPI for the
Yukawa sector of the SM is the well-known Jarlskog
invariant~\cite{Jarlskog:1985ht}.

Before defining and discussing CPIs of the scalar sector in detail, we first show how
to write any possible Higgs potential in a standard form which is suitable to construct 
general basis invariants. This procedure has the advantage
that basis invariants only have to be derived once in the standard
form; their explicit form for any particular Higgs potential follows almost
trivially by translating the latter into the standard parametrisation. 
Furthermore, invariants that are CP-odd (CPIs) for the standard
form of potentials are so by construction and, if non-zero, indicate CP
violation for all possible example potentials.

The relation between non-zero CPIs and CP violation can be formulated more
precisely as follows. If a potential conserves CP, then all CPIs vanish
automatically. Reversely, if one or several CPIs are non-zero, the potential
violates CP. This statement holds for both explicit and spontaneous CP
violation, and the corresponding CPIs are introduced in Sections~\ref{CPIs}
and~\ref{spontCPI}. 
Note that CPIs only guarantee CP conservation if all of them vanish. This is
equivalent to demanding a finite set of CPIs, the so-called basis out of
which all other CPIs can be produced, to vanish. Such a basis of CPIs is known
for the 2HDM~\cite{Gunion:2005ja}, but, so far, not for any other more complicated
scalar potentials.

In the following, we first introduce the standard form for scalar
potentials as discussed in~\cite{Branco:2005em,Davidson:2005cw}. 
Having established out notation, we analyse the effects of 
symmetry transformations, general basis transformations, complex conjugation
and CP transformations on the variables and parameters of the standard form.  
Adopting the procedure and notation of~\cite{Branco:2005em,Davidson:2005cw}, any even potential of $N$ scalar fields $\varphi_i$ can, with $\phi=(\varphi_1,\ldots , \varphi_N)$ and $\phi^\ast=(\varphi_1^\ast,\ldots , \varphi_N^\ast)$, be written as
\be
V_{\mathrm{scalar}}~=~ {\phi^\ast}^a  \,Y_a^b\, \phi_b 
+  {\phi^\ast}^a{\phi^\ast}^c  \,Z_{ac}^{bd}\, \phi_b \phi_d \ ,
\label{eq:potCPgeneral} 
\ee
where the notation is such that lower indices on $Y$ and $Z$ are always
contracted with $\phi^\ast$ and upper indices with  $\phi$. 
$Y$ and $Z$ are tensors that contain all possible couplings and are subject to
possible symmetries acting on $\phi$, as will be explained below.\footnote{One
  could also add a term such as e.g.\ $T^{ab}_c\phi_a \phi_b \phi^{\ast c}
  +\text{h.c.}$ to the potential to account for trilinear couplings and the
  discussion could be extended in this way.}

Any potential of several Higgs doublets can be brought into this standard form
by $\phi$ not containing doublets as such, but instead directly containing the
components of the doublets: for $n$ Higgs doublets $H_{i
  \alpha}=(h_{i,1},h_{i,2})$, where $\alpha=1,2$ denotes the $SU(2)_L$ index
and $i$ goes from $1$ to $n$, 
\begin{equation}
\phi=(\varphi_1,\varphi_2, \ldots, \varphi_{2n-1},
\varphi_{2n})=(h_{1,1},h_{1,2},\ldots,h_{n,1},h_{n,2}) \ ,
\label{eq:phi_to_h}
\end{equation}
and the invariance of the potential under $SU(2)_L\times U(1)_Y$ will be
directly reflected in the structure of $Y$ and $Z$ in a component-wise way.
This convention, which differs from the notation
of~\cite{Branco:2005em,Davidson:2005cw}, will be very useful later
on.\footnote{In~\cite{Davidson:2005cw}, for example, the $SU(2)_{L}$ indices
are summed over outside of $Z$. Our definition of $Z$ tensors can be
related to~\cite{Davidson:2005cw} by explicitly highlighting the $SU(2)_L$ subindices, 
$\{1,2,\ldots,2n-1,2n\}=\{(1,1),(1,2),\ldots,(n,1),(n,2)\}$. With this, the
$Z$ tensors in our paper become
$Z^{ab}_{cd}=Z^{(\tilde a,\alpha)(\tilde b,\beta)}_{(\tilde c,\gamma),(\tilde d,\delta)}
=\tilde{Z}^{\tilde a\tilde b}_{\tilde c\tilde d}\delta^\alpha_\gamma \delta^\beta_\delta$, 
where $\tilde{Z}$ denotes the coupling tensors of~\cite{Davidson:2005cw}.} 

More explicitly, if the theory is invariant under {\bf symmetry
transformations} of a group $G$ such that $\phi$ transforms in some (maybe
reducible) representation $\rho(g)$ of that group, where $\rho(g)$ is the matrix that corresponds to the group element $g\in G$,
\bea
 \phi_a&\mapsto& [\rho(g)]_{a}^{a'} \phi_{a'}\ ,\\[2mm]
 \phi^{\ast a}&\mapsto &\phi^{\ast a'} [\rho^\dagger(g)]^{a}_{a'}\ ,
\eea
then the invariance of the potential imposes the following constraints on the coupling tensors:
\begin{equation}
Y_a^b =  \rho_a^{a'} \, Y_{a'}^{b'} \, {\rho^\dagger}_{b'}^b \ ,
\end{equation}
\begin{equation}
Z_{ac}^{bd} =  \rho_a^{a'}\, \rho_c^{c'} \, Z_{a'c'}^{b'd'} \, 
{\rho^\dagger}_{b'}^b \,{\rho^\dagger}_{d'}^d \ ,
\label{symm_of_Z}
\end{equation}
where we have written $ \rho_a^{a'}= [\rho (g)]_a^{a'}$ and so on.
In addition to that, the quartic coupling tensor $Z^{bd}_{ac}$ is by
construction invariant under exchanging $a\leftrightarrow c$ as well as
$b\leftrightarrow d$. The reason for this is that $\phi_b$ and $\phi_d$
commute so that the indices $b$ and $d$ can be renamed into each other to restore the original ordering of the $\phi$'s, and equivalently for $\phi^\ast$ with $a$~and~$c$. 

While the theory is invariant under symmetry transformations, one also has the
possibility of applying basis transformations under which the Lagrangian is
not invariant. Of course, such a basis transformation should not change
physics. A simple example is the transformation
that diagonalises the bilinear mass terms ${\phi^\ast}^a  \,Y_a^b\, \phi_b$. 
As $Z$ is generally only invariant under a smaller group than that of all
basis changes, diagonalising $Y$ would change $Z$.\footnote{Except in the case
  where the components of $Y$ conspire in such a way that the required basis
  transformation coincides with a symmetry transformation. Furthermore, a general
  basis transformation changes the form of the potential, while only
  transformations in the automorphism group $Aut(G)$ leave the potential
  form-invariant.}   
Adopting our notation for the standard form of Higgs potentials, a unitary {\bf
  basis transformation} in the space of the $N$ dimensional vector $\phi$,
i.e.\ with $V\in U(N)$ a 
unitary $N\times N$ matrix, maps 
\bea
\phi_a & \mapsto & V_a^{a'} \phi_{a'}  \ , \\
{\phi^\ast}^a &\mapsto &  {\phi^\ast}^{a'} {V^\dagger}_{a'}^a \ .
\eea
With this definition, the kinetic terms remain unchanged while
$Y$ and $Z$ transform to
\bea
Y_a^b &\mapsto & V_a^{a'} \, Y_{a'}^{b'} \, {V^\dagger}_{b'}^b \ ,\\
Z_{ac}^{bd} &\mapsto & V_a^{a'}\, V_c^{c'} \, Z_{a'c'}^{b'd'} \, 
{V^\dagger}_{b'}^b \,{V^\dagger}_{d'}^d \ .
\eea

{\bf Complex conjugation} is an essential part of CP transformations and in the notation used here, changes the vertical position of the index of a field
so that
\bea
\phi_a & \mapsto & (\phi_a)^\ast \equiv  {\phi^\ast}^a  \ ,\\
{\phi^\ast}^a  & \mapsto &  ({\phi^\ast}^a)^\ast \equiv {\phi}_a \ .
\eea 
Complex conjugating the $Y$ term of the potential then results in
\bea
{\phi^\ast}^a  \,Y_a^b \, \phi_b ~\mapsto~
 {\phi}_a  \,(Y_a^b)^\ast\, {\phi^\ast}^b =
{\phi^\ast}^b  \,(Y_a^b)^\ast\, {\phi}_a  ={\phi^\ast}^a  \,(Y_b^a)^\ast\, {\phi}_b \ .   
\eea
Comparing this to the original term in the potential and demanding
$V_{\mathrm{scalar}}^\ast =V_{\mathrm{scalar}}$ shows that
\be
(Y_b^a)^\ast ~=~ Y_a^b \ .
\label{Y_real}
\ee 
A similar result is obtained for the quartic coupling, i.e.
\be
(Z^{ac}_{bd})^\ast ~=~ Z_{ac}^{bd} \ .
\label{Z_real}
\ee
Note that because both indices of a contracted pair interchange position under
complex conjugation, no situation can arise where one would need to sum over
two upper or two lower indices. However, expressions as
e.g.\ Eqs.~(\ref{Y_real}) and (\ref{Z_real}) where indices appear with
exchanged vertical positions without being summed over need to be understood
as conditions on the components of the tensors and not the tensors
themselves. 

Finally, all components are in place to define a {\bf (general) CP
  transformation}\footnote{This is often referred to as a generalised CP
  transformation.} with a unitary matrix $U$ on the fields as
\bea
\phi_a &\mapsto & {\phi^\ast}^{a'} U_{a'}^a \ , \\
{\phi^\ast}^a &\mapsto & {U^\dagger}_a^{a'} \phi_{a'} \ .
\eea
Again, this leaves the kinetic terms invariant, while applying the CP
transformation to the fields in the potential results for the $Y$ term in
\bea
{\phi^\ast}^a  \,Y_a^b \, \phi_b ~\mapsto~
{U^\dagger}_a^{a'} \phi_{a'}   \,Y_a^b\,  
{\phi^\ast}^{b'} U_{b'}^b   &=&
{U^\dagger}_a^{a'} \phi_{a'}   \,(Y_b^a)^\ast\,  
{\phi^\ast}^{b'} U_{b'}^b \notag \\
&=& 
{\phi^\ast}^{b'} U_{b'}^b 
\,(Y_b^a)^\ast\,  
{U^\dagger}_a^{a'} \phi_{a'} \notag\\
&=& 
{\phi^\ast}^{a} U_{a}^{a'} 
\,(Y_{a'}^{b'})^\ast\,  
{U^\dagger}_{b'}^{b} \phi_{b} \ .
\eea
Comparing this to the original term in the potential shows that a CP
transformation acting on the fields can be equally understood as the following
change of $Y$ (likewise for~$Z$),
\bea
Y_a^b &\mapsto & U_{a}^{a'} 
\,(Y_{a'}^{b'})^\ast\,  
{U^\dagger}_{b'}^{b} \ , \label{eq:CPcond1}\\
Z_{ac}^{bd} &\mapsto & U_{a}^{a'}  U_{c}^{c'} 
\,(Z_{a'c'}^{b'd'})^\ast\,  
{U^\dagger}_{b'}^{b} {U^\dagger}_{d'}^{d}\ .\label{eq:CPcond2}
\eea
The condition for CP invariance of the standard form of the potential $V_{\mathrm{scalar}}$ in
Eq.~\eqref{eq:potCPgeneral}, and thus any example potential that can be
brought into this standard form, can then be phrased as follows: \textit{CP is
  conserved if there is a $U$ such that the left- and right-hand sides  of
  Eqs.~(\ref{eq:CPcond1}) and (\ref{eq:CPcond2}) are identical.}  
As a special case of that, if the tensors $Y$ and $Z$ are real, the potential is
invariant under a CP transformation, which we refer to as  $CP_{0}$,
\be
U^{a}_{a'} =  \delta^{a}_{a'}\ .
\label{CP0}
\ee
$CP_{0}$ is often referred to as trivial or canonical CP.

In doing so, we note that a physical CP transformation will have to treat the
two components of an $SU(2)_L$ doublet consistently with that symmetry,
i.e.\ both must transform identically under the CP symmetry~\cite{Grimus:1995zi}.

In preparation for Section~\ref{spontCPI}, where CPIs for spontaneous CP
violation will be constructed, we define how VEVs behave
under basis transformations:
\bea
 v_a&\mapsto &V_a^{a'}v_{a'} \ ,\label{eq:basi}\\
 {v^\ast}^a&\mapsto &v^{\ast a'}\,{V^\dagger}^{a}_{a'}\ ,\label{eq:basi*}
\eea
where $v\equiv(v_1,\ldots)$ with $v_i=\langle\varphi_i\rangle$, and $V$ denotes the
transformation matrix of the fields~$\phi$. Similarly, under CP transformations, they become
\bea
 v_a&\mapsto& v^{\ast a'}U^{a}_{ a'}\ , \\
 {v^\ast}^a&\mapsto&   {U^\dagger}_a^{a'}v_{a'} \ .
\eea

\subsection{CP-odd invariants for explicit CP violation \label{CPIs}}

In the previous subsection, the standard form for even scalar potentials was introduced and the effects of symmetry transformations, basis transformations and CP transformations has been analysed.
This subsection starts with a discussion of simple basis invariants constructed from $Y$ and $Z$ tensors. After that, the general definition of CP-odd basis invariants (CPIs) that contain $Y$ and $Z$ is given. 

Finally, the CP properties of such invariants will be analysed. CPIs of this
type, that only consist of parameters of the potential and in particular do
not contain VEVs, indicate explicit violation of CP. The
exact statement is that if all possible CPIs are zero, then the theory is CP conserving. Vice-versa, if at least one CPI is non-zero, the theory violates CP explicitly.
Invariants including VEVs, such that they indicate spontaneous violation of
CP, will be introduced in Section~\ref{spontCPI}.

Although the material of this subsection is not
new as such, we present it here in order to make our discussion self contained.

Any product of $Y$ and $Z$ tensors where all indices are correctly contracted forms a basis invariant. Starting with $Y$ and considering $Z$ a little later, the simplest invariant (that is however not CP-odd) is 
\begin{equation}
 Y^a_a.
 \label{1Y_invariant}
\end{equation}
For products of two $Y$ tensors, the only possible contractions are
\begin{equation}
Y^a_a Y^b_b \text{ and }Y^a_bY^b_a.
\end{equation}
The above contractions correspond to the two different permutations of the two upper indices, namely firstly the identity:
\begin{equation}
Y^a_a Y^b_b \Leftrightarrow a\mapsto a \text{ and } b\mapsto b \ ,
\end{equation}
and secondly the transposition:
\begin{equation}
Y^a_b Y^b_a \Leftrightarrow a\mapsto b \text{ and } b\mapsto a\ .
\end{equation}
More formally, one can thus also express all invariants that consist of two $Y$ tensors by
\begin{equation}
 Y^a_{\sigma(a)}Y^b_{\sigma(b)} \text{ with } \sigma \in S_2 \ ,
\end{equation}
where $\sigma$ is now one of the two elements of the permutation group $S_2$. The invariant built from two $Y$ tensors that corresponds to the identity of $S_2$ is the square of the simplest invariant. Thus, only the second invariant is irreducible, which for our purposes will be defined as not being a product or power of smaller invariants. 
 
It is generally true that all possible invariants can be obtained through
permutations of indices: all conceivable invariants built from 3 $Y$ tensors are given by
\begin{equation}
  Y^a_{\sigma(a)}Y^b_{\sigma(b)}Y^c_{\sigma(c)} \text{ with } \sigma \in S_3\ ,
\end{equation}
or explicitly
\begin{equation}
Y^a_aY^b_bY^c_c,\text{ }Y^a_aY^b_cY^c_b,\text{ }Y^a_cY^b_bY^c_a,\text{ }Y^a_bY^b_aY^c_c,\text{ }Y^a_cY^b_aY^c_b,\text{ }Y^a_bY^b_cY^c_a.
\end{equation}
Here, only the last two invariants are new and irreducible, i.e.\ not products of smaller invariants. Additionally, they turn out to be equivalent as can be seen by renaming the indices $b\leftrightarrow c$ into each other. 

The identification of invariants with elements of permutation groups will be used later to systematically identify all irreducible invariants of a given order. Beyond that, it is this formalism that is going to make it possible to determine which invariants are CP-even and which are not. 

But before that, some more examples are in order, as the situation is more
complicated for invariants containing $Z$ tensors. There are already two invariants that could be built from a single $Z$ tensor that again correspond to the two possible permutations of positions of the two upper indices:
\begin{equation}
 Z^{ab}_{\sigma(a)\sigma(b)} \text{ with } \sigma \in S_2\ ,
\end{equation}
or explicitly:
\begin{equation}
Z^{ab}_{ab}\text{ and }Z^{ab}_{ba}\ .
\end{equation}
Because the $Z$ tensor of potentials considered here is symmetric under
exchanging both upper or both lower indices, cf.\ below Eq.~(\ref{symm_of_Z}),
both invariants built from one $Z$ tensor are equivalent. For larger numbers
of tensors, the number of permutations grows quickly, however, luckily, many
invariants do not need to be considered either because they are products of
smaller invariants, or because  they are equivalent due to the symmetry of
single tensors themselves or the symmetries of the invariant. For example, for two $Z$ tensors, generally all invariants would be given by
\begin{equation}
Z^{ab}_{\sigma(a)\sigma(b)}Z^{cd}_{\sigma(c)\sigma(d)} \text{ with } \sigma
\in S_4\ ,
\end{equation}
but the only new invariants can be chosen to be
\begin{equation}
Z^{ab}_{bd}Z^{cd}_{ac}\text{ and }Z^{ab}_{cd}Z^{cd}_{ab}.
\label{2Z_invariants}
\end{equation}
All other 22 invariants that correspond to the remaining elements of $S_4$ are products of smaller invariants or equivalent to the invariants in Eq.~(\ref{2Z_invariants}).

Generally, a basis invariant $I_\sigma^{(n_Z,m_Y)}$ built from $m_Y$ Y tensors and $n_Z$ $Z$ tensors can be written as\footnote{Often, not the full permutation $\sigma$ will be indicated when referring to invariants, but e.g.\ $I_2^{(3,1)}$ would be the second invariant that was found with $n_Z=3$ and $m_Y=1$.}
\begin{equation}
I_\sigma^{(n_Z,m_Y)}\equiv Y^{a_1}_{\sigma(a_1)}\ldots Y^{a_{m_Y}}_{\sigma(a_{m_Y})}Z^{b_1 b_2}_{\sigma(b_1)\sigma(b_2)}\ldots Z^{b_{2n_Z-1} b_{2n_Z}}_{\sigma(b_{2n_Z-1})\sigma(b_{2n_Z})} \text{ with } \sigma \in S_{m_Y+2n_Z}.
\label{invariant_definition}
\end{equation}
Again, $\sigma$ is a permutation of $m_Y+2n_Z$ objects, i.e.\ $\sigma\in
S_{m_Y+2n_Z}$. However, not all basis invariants are CP-odd, and in fact, all
of the examples in Eqs.~(\ref{1Y_invariant})-(\ref{2Z_invariants}) turn out to
be CP-even. To be able to make such statements, one needs to know how basis
invariants behave under CP. Under a general CP transformation, a coupling
tensor is replaced by its complex conjugate multiplied by unitary basis
transformations, earlier denoted by $U$. But, as a basis invariant is, by
definition, invariant under basis transformations, the $U$ matrices cancel, leaving only the original product of coupling tensors with tensors replaced by their complex conjugates. The complex conjugate of a coupling tensor on the other hand can be obtained by interchanging upper with lower indices, cf.\ Eqs.~(\ref{Y_real}) and (\ref{Z_real}). 
For the simplest example, $Y^a_a$, this works out in the following way:
\begin{equation}
 Y^a_a \xrightarrow{CP} (Y^{a'}_{a''})^\ast {U^\dagger}_{a'}^aU^{a''}_a=(Y^{a'}_{a''})^\ast \delta_{a'}^{a''}=(Y^a_a)^\ast=Y^a_a,
\end{equation}
where in the last step Eq.~(\ref{Y_real}) was used. As the right-hand side is the
CP conjugate of the left-hand side and is identical to the latter, this shows that this
invariant is even under CP transformations.
Similarly, and using Eqs.~(\ref{Y_real}) and (\ref{Z_real}), one can show that
the CP conjugate of a general basis invariant can be obtained by interchanging upper and lower indices:
\begin{align}
I_\sigma^{(n_Z,m_Y)} &\equiv Y^{a_1}_{\sigma(a_1)}\ldots Y^{a_{m_Y}}_{\sigma(a_{m_Y})}Z^{b_1 b_2}_{\sigma(b_1)\sigma(b_2)}\ldots Z^{b_{2n_Z-1} b_{2n_Z}}_{\sigma(b_{2n_Z-1})\sigma(b_{2n_Z})} \nonumber\\&\xrightarrow{CP}
Y^{\sigma(a_1)}_{a_1}\ldots
Y^{\sigma(a_{m_Y})}_{a_{m_Y}}Z^{\sigma(b_1)\sigma(b_2)}_{b_1 b_2}\ldots
Z^{\sigma(b_{2n_Z-1})\sigma(b_{2n_Z})}_{b_{2n_Z-1} b_{2n_Z}}=
[I_\sigma^{(n_Z,m_Y)}]^{\ast}\ .
\label{CP_of_invariant}
\end{align}
If one has found an invariant $I$ that is not CP-even, i.e.\ that does not equal
its CP conjugate $I^*$, one can 
(as is well-known~\cite{Botella:1994cs})
extract the CP-odd part by subtracting the CP-conjugated from the original invariant: 
\begin{equation}
 \mathcal{I}=I-I^{\ast}.
 \label{eq:CPI_I_c}
\end{equation}
The above procedure, of tracing over all internal spaces and identifying the remaining imaginary part, is actually more general than discussed above, and it applies also to cases with scalars and fermions \cite{Botella:1994cs}.
As a CPI is already completely defined by stating half of it, $I$, in the
following often $I^\ast$ will be omitted or abbreviated. When $\mathcal{I}$ is
given, it is implied that the quantity to follow is the difference between a
basis invariant $I$ and its CP conjugate.

For the example invariants in Eqs.~(\ref{1Y_invariant})-(\ref{2Z_invariants}),
interchanging upper and lower indices and possibly renaming indices shows that
all of them are equal to their CP conjugate and thus CP-even. 
For larger invariants, this process can become quite cumbersome. 
Even worse, in order to show that an invariant is not CP-even, one would have
to test all possible renamings of the indices, which at some point becomes too
difficult. 
Luckily, the symmetry properties of invariants can be analysed and visualised
using diagrams that encode which tensors are used and how their indices are
contracted with each other. These diagrams are the topic of the next
subsection. As the diagrams become more complicated, a more powerful technique
relies on analysing so-called contraction matrices that also encode
information about the basis invariant and reveal whether it is a CPI. We
heavily rely on the contraction matrices for our systematic searches that
revealed many new CPIs. As the discussion is somewhat technical, the 
details concerning contraction matrices are relegated to Appendix \ref{sym_invs}.

\subsection{Diagrams for invariants}
\label{subsection:invariants}

Any basis invariant consisting of contractions of $Y$ and $Z$ can be expressed by a diagram~\cite{Davidson:2005cw}. We use a slightly different notation from the one present in~\cite{Davidson:2005cw}. For each $Y$ or $Z$ draw a vertex and for any contraction of an upper index on a tensor with a lower index of a tensor draw an arrow connecting the vertices corresponding to the tensors. With $X=Y,Z$, the only rule for drawing diagrams is
\begin{equation}
X^{a.}_{..}X^{..}_{a.}=\vcenter{\hbox{\includegraphics[scale=0.2]{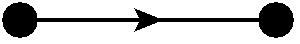}}}
\end{equation}
Additionally, as $Z^{ab}_{cd}$ is symmetric under exchange of $a\leftrightarrow b$ and/or $c\leftrightarrow d$, two lines can be attached to a vertex corresponding to a $Z$ tensor without having to distinguish them in the diagram:
\begin{equation}
Z^{ab}_{..}Z^{..}_{ab}=\includegraphics[scale=0.2]{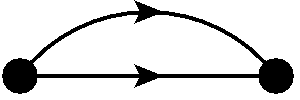}
\label{FR_2_definition}
\end{equation}
Contracting two indices on the same tensor with each other produces a loop:
\begin{equation}
X^{a.}_{a.}=\vcenter{\hbox{\includegraphics[scale=0.2]{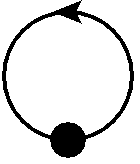}}}
\label{loop_diagram}
\end{equation}
Diagrams drawn following these rules make it possible to check if an invariant
is CP-even: from Eq.~(\ref{CP_of_invariant}) follows that \emph{the CP
  conjugate of an invariant produces exactly the same diagram but with
  inverted directions of arrows} as all upper indices have been turned into
lower indices and vice versa. An invariant is identical to its CP conjugate,
i.e.\ CP-even, if the diagrams of the invariant and its CP conjugate are identical up to the positions of the vertices. The reason for this is that in a product of $Y$ and $Z$ tensors, their position in the product is arbitrary and thus also the position of vertices (except for the type of tensor).\footnote{The internal symmetry of $Z^{ab}_{cd}$ under $a\leftrightarrow b$ and/or $c\leftrightarrow d$ is taken into account by Eq.~(\ref{FR_2_definition}).} 

\begin{figure}[t]
\begin{equation*}
 Y^a_b Y^b_a=\vcenter{\hbox{\includegraphics[scale=0.2]{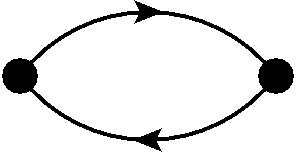}}}
\end{equation*}
\begin{equation*}
 Z^{ab}_{ab}=\vcenter{\hbox{\includegraphics[scale=0.2]{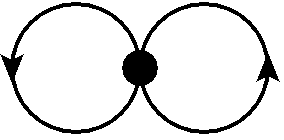}}}
\end{equation*}
\begin{equation*}
 Z^{ac}_{bc}Y^b_a=\vcenter{\hbox{\includegraphics[scale=0.2]{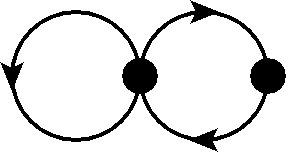}}}
\end{equation*}
\begin{equation*}
 Z^{ab}_{cd}Z^{cd}_{ab}=\vcenter{\hbox{\includegraphics[scale=0.2]{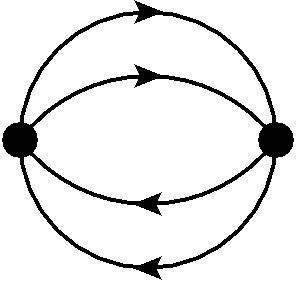}}}
\end{equation*}
\begin{equation*}
 Z^{ab}_{ac}Z^{cd}_{bd}=\vcenter{\hbox{\includegraphics[scale=0.2]{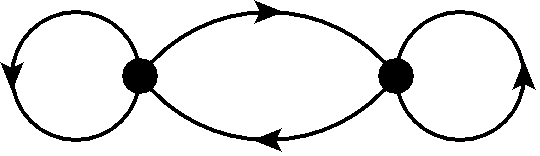}}}
\end{equation*}
\caption{Example diagrams corresponding to small invariants.}
\label{diagram_examples}
\end{figure}

A few small example diagrams for small invariants mentioned earlier in the text are shown in Figure~\ref{diagram_examples}. One can see there that for each of them, inverting the direction of the arrows produces the same diagram and thus the same invariant. This is the case for all of the small examples in Eqs.~(\ref{1Y_invariant})-(\ref{2Z_invariants}). The simplest invariant, $Y^a_a$ produces the diagram in Eq.~(\ref{loop_diagram}).

All invariants discussed so far were CP-even. The smallest CP-odd invariant was already found in~\cite{Branco:2005em} and is given by the difference $\mathcal{I}_1=I_1-I_1^{*}$ of
\begin{equation}
 I_1\equiv Z^{ab}_{ae}Z^{cd}_{bf}Y^e_cY^f_d=\vcenter{\hbox{\includegraphics[scale=0.2]{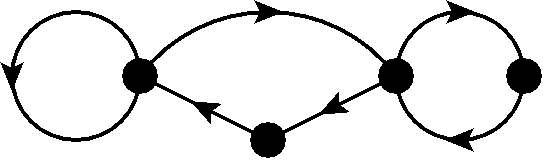}}}
 \label{I21}
\end{equation}
and its CP conjugate
\begin{equation}
I_1^{*}\equiv Z^{ae}_{ab}Z^{bf}_{cd}Y^c_eY^d_f=\vcenter{\hbox{\includegraphics[scale=0.2]{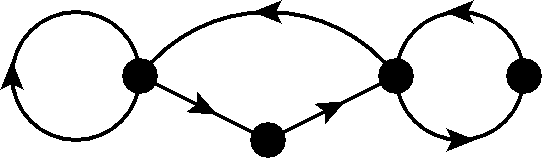}}}.
\label{I21CP}
\end{equation}
In whatever ways one tries to interchange the positions of vertices and arrows, it is impossible to make the diagrams equivalent. 

Additionally, out of all possible contractions of coupling tensors, many will
be related by interchanging the positions of tensors. The symmetries of the
diagrams can be used to classify invariants and search for CPIs in a
systematic way as will be explained in Appendix~\ref{sym_invs}. The results of
this systematic search are listed in the following.  

\subsection[CP-odd invariants only built from $Z$ tensors]{CP-odd invariants only built from $\boldsymbol{Z}$ tensors}
It is interesting to consider invariants that are only built from $Z$ tensors, as these indicate CP violation that is mediated purely through the interaction of fields and does not e.g.\ depend on a mass splitting.  
One could now wonder if for a non-diagonal $Y$ tensor CP violating effects could be shifted between $Y$ and $Z$ by diagonalising $Y$. However, because this is just another basis change, it drops out in any basis invariants, including also CPIs.

In Appendix~\ref{invariant_lists}, we list the representative CPIs with up to
$n_Z=6$ $Z$ tensors. All other CPIs are related to these representatives by
symmetries or CP conjugation.

An important first result is that \emph{all invariants up to $n_Z=4$ are CP-even}.
For $n_Z=5$, three different CPIs exist:
\begin{equation}
I^{(5)}_1\equiv  Z^{a_1a_2}_{a_7a_9}Z^{a_3a_4}_{a_5a_{10}}Z^{a_5a_6}_{a_3a_6}Z^{a_7a_8}_{a_4a_8}Z^{a_9a_{10}}_{a_1a_2}=\vcenter{\hbox{\includegraphics[scale=0.2]{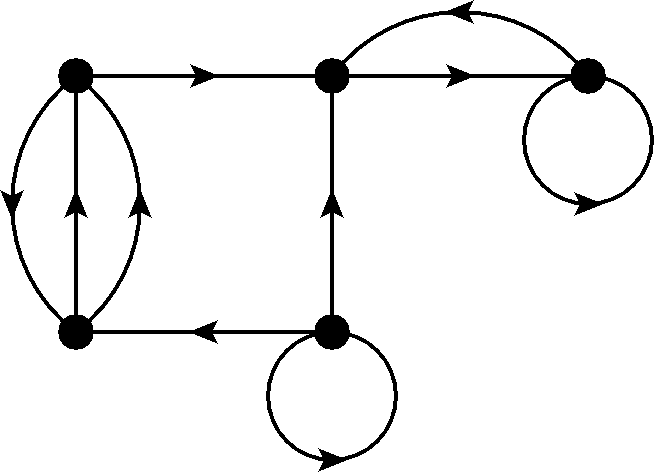}}}
\label{I5_1}
\end{equation}
\begin{equation}
 I^{(5)}_2\equiv Z^{a_1a_2}_{a_5a_7}Z^{a_3a_4}_{a_8a_9}Z^{a_5a_6}_{a_3a_6}Z^{a_7a_8}_{a_4a_{10}}Z^{a_9a_{10}}_{a_1a_2}=\vcenter{\hbox{\includegraphics[scale=0.2]{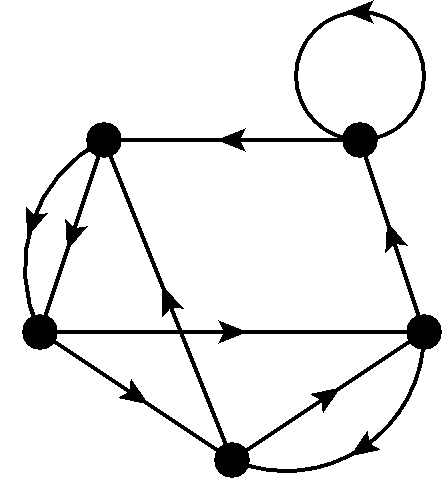}}}
\end{equation}
\begin{equation}
 I^{(5)}_3\equiv Z^{a_1a_2}_{a_5a_9}Z^{a_3a_4}_{a_3a_7}Z^{a_5a_6}_{a_6a_8}Z^{a_7a_8}_{a_1a_{10}}Z^{a_9a_{10}}_{a_2a_4}=\vcenter{\hbox{\includegraphics[scale=0.2]{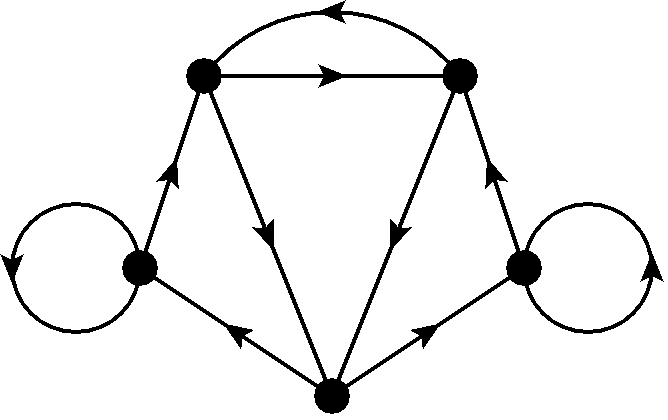}}}
 \label{I5_3}
\end{equation}
For $n_Z=6$, in total 56 different invariants exist out of which three are
products of the $n_Z=5$ invariants with a completely self-contracted $Z$
tensor, $Z_{ab}^{ab}$. These will not provide any new information. Next, of
particular interest are those invariants that contain no self-loops, as we
found that invariants with self-loops, i.e.\ $Z^{a.}_{a.}$ often vanish for
the example potentials considered in this work. With $n_Z=6$, only 5 invariants without self-loops remain:
\begin{equation}
I^{(6)}_1\equiv Z^{a_1 a_2}_{a_{11} a_{10}} Z^{a_3 a_4}_{a_5 a_8} Z^{a_5
  a_6}_{a_7 a_{12}} Z^{a_7 a_8}_{a_9 a_6} Z^{a_9 a_{10}}_{a_3 a_4} Z^{a_{11}
  a_{12}}_{a_1 a_2}\ ,
\end{equation}

\begin{equation}
\label{I62}
I^{(6)}_2\equiv Z^{a_1 a_2}_{a_7 a_{10}} Z^{a_3 a_4}_{a_{11} a_6} Z^{a_5 a_6}_{a_9 a_8} Z^{a_7 a_8}_{a_3 a_{12}} Z^{a_9 a_{10}}_{a_5 a_4} Z^{a_{11} a_{12}}_{a_1 a_2}\ ,
\end{equation}

\begin{equation}
\label{I63}
 I^{(6)}_3\equiv Z^{a_1 a_2}_{a_7 a_{10}} Z^{a_3 a_4}_{a_9 a_6} Z^{a_5 a_6}_{a_{11} a_8} Z^{a_7 a_8}_{a_3 a_{12}} Z^{a_9 a_{10}}_{a_5 a_4} Z^{a_{11} a_{12}}_{a_1 a_2}\ ,
\end{equation}

\begin{equation}
\label{I64}
 I^{(6)}_4\equiv Z^{a_1 a_2}_{a_{11} a_{10}} Z^{a_3 a_4}_{a_5 a_8} Z^{a_5 a_6}_{a_7 a_{12}} Z^{a_7 a_8}_{a_9 a_6} Z^{a_9 a_{10}}_{a_1 a_4} Z^{a_{11} a_{12}}_{a_3 a_2}\ ,
\end{equation}

\begin{equation}
\label{I65}
 I^{(6)}_5\equiv Z^{a_1 a_2}_{a_7 a_{12}} Z^{a_3 a_4}_{a_5 a_{10}} Z^{a_5 a_6}_{a_9 a_8} Z^{a_7 a_8}_{a_{11} a_4} Z^{a_9 a_{10}}_{a_1 a_6} Z^{a_{11} a_{12}}_{a_3 a_2}\ .
\end{equation}
The diagrams that correspond to the above invariants with $n_Z=6$ and the remaining representative CPIs with up to $n_Z=6$ are listed in Appendix~\ref{invariant_lists}.

\subsection[CP-odd invariants built from $Y$ and $Z$ tensors]{CP-odd
  invariants built from $\bs{Y}$ and $\bs{Z}$ tensors}
Mixed invariants consisting of $Y$ and $Z$ tensors can be CP-odd at lower
numbers of $Z$ tensors than $n_Z=5$. The reason for this is that additional
asymmetries can be introduced in the diagrams. The smallest CPI found
in~\cite{Branco:2005em}, Eqs.~(\ref{I21}) and (\ref{I21CP}), is of this type
with $m_Y=2,n_Z=2$ and will not be repeated here. There are no other CPIs for
$m_Y=2,n_Z=2$ that are not equivalent to the aforementioned one. The next
smallest CPIs are found for $n_Z=3$ and $m_Y=1$. There are two different
classes with the following representatives:  
\begin{equation}
 I^{(3,1)}_1\equiv Y^{a_1}_{a_4} Z^{a_2 a_3}_{a_6 a_7}Z^{a_4 a_5}_{a_2 a_5}Z^{a_6 a_7}_{a_1 a_3}=\vcenter{\hbox{\includegraphics[scale=0.2]{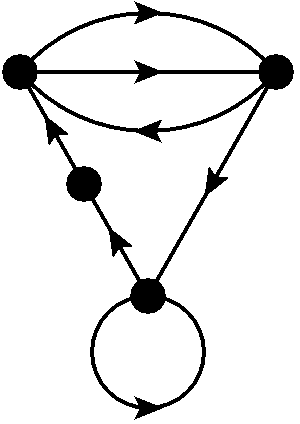}}}
 \label{I31_1_definition}
\end{equation}
\begin{equation}
 I^{(3,1)}_2\equiv Y^{a_1}_{a_4} Z^{a_2 a_3}_{a_2 a_6}Z^{a_4 a_5}_{a_5 a_7}Z^{a_6 a_7}_{a_1 a_3}=\vcenter{\hbox{\includegraphics[scale=0.2]{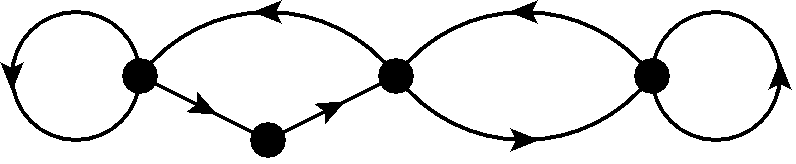}}}
 \label{I31_2_invariant}
\end{equation}
However, both invariants in Eqs.~(\ref{I31_1_definition}) and
(\ref{I31_2_invariant}) contain self-loops. As these often vanish in examples,
one would preferably like to find invariants without self-loops. Such invariants can already be found for $n_Z=3, m_Y=2$. There are in total 13 invariants with this number of $Y$ and $Z$ tensors, out of which 2 have no $Z$-self-loops. In one of them, the $Y$ tensor is inserted in a $Z$ loop and will only make a difference if $Y$ is not proportional to the identity, while the other invariant has genuinely no $Z$-self-loops. These invariants and diagrams are
\begin{equation}
 I^{(3,2)}_1\equiv Y^{a_1}_{a_7}Y^{a_2}_{a_5} Z^{a_3 a_4}_{a_6 a_7}Z^{a_5 a_6}_{a_3 a_4}Z^{a_7 a_8}_{a_1 a_2}=\vcenter{\hbox{\includegraphics[scale=0.2]{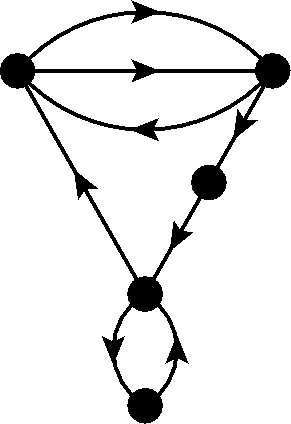}}}
\end{equation}
\begin{equation}
 I^{(3,2)}_2\equiv Y^{a_1}_{a_5}Y^{a_2}_{a_3} Z^{a_3 a_4}_{a_6 a_7}Z^{a_5 a_6}_{a_4 a_8}Z^{a_7 a_8}_{a_1 a_2}=\vcenter{\hbox{\includegraphics[scale=0.2]{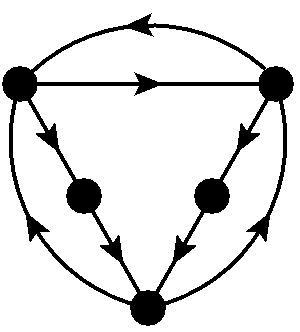}}}.
\end{equation}

Naively, there are 53 classes of invariants with $m_Y=3,n_Z=3$. However, many of these will be products of smaller CPIs with small CP-even invariants. Eventually, there are 10 invariants without $Z$-self loops which are not products of smaller invariants, the representatives of which are listed in the following:
\begin{equation}
 I^{(3,3)}_1\equiv Y^{a_1}_{a_8}Y^{a_2}_{a_6}Y^{a_3}_{a_4}Z^{a_4 a_5}_{a_7 a_9}Z^{a_6 a_7}_{a_3 a_5}Z^{a_8 a_9}_{a_1 a_2}\ ,
\end{equation}
\begin{equation}
 I^{(3,3)}_2\equiv Y^{a_1}_{a_6}Y^{a_2}_{a_7}Y^{a_3}_{a_4}Z^{a_4 a_5}_{a_8 a_9}Z^{a_6 a_7}_{a_3 a_5}Z^{a_8 a_9}_{a_1 a_2}\ ,
\end{equation}
\begin{equation}
 I^{(3,3)}_3\equiv Y^{a_1}_{a_8}Y^{a_2}_{a_4}Y^{a_3}_{a_6}Z^{a_4 a_5}_{a_7 a_9}Z^{a_6 a_7}_{a_3 a_5}Z^{a_8 a_9}_{a_1 a_2}\ ,
\end{equation}
\begin{equation}
 I^{(3,3)}_4\equiv Y^{a_1}_{a_6}Y^{a_2}_{a_4}Y^{a_3}_{a_8}Z^{a_4 a_5}_{a_7 a_9}Z^{a_6 a_7}_{a_3 a_5}Z^{a_8 a_9}_{a_1 a_2}\ ,
\end{equation}
\begin{equation}
 I^{(3,3)}_5\equiv Y^{a_1}_{a_6}Y^{a_2}_{a_4}Y^{a_3}_{a_7}Z^{a_4 a_5}_{a_8 a_9}Z^{a_6 a_7}_{a_3 a_5}Z^{a_8 a_9}_{a_1 a_2}\ ,
\end{equation}
\begin{equation}
 I^{(3,3)}_6\equiv Y^{a_1}_{a_8}Y^{a_2}_{a_3}Y^{a_3}_{a_6}Z^{a_4 a_5}_{a_7 a_9}Z^{a_6 a_7}_{a_4 a_5}Z^{a_8 a_9}_{a_1 a_2}\ ,
\end{equation}
\begin{equation}
 I^{(3,3)}_7\equiv Y^{a_1}_{a_6}Y^{a_2}_{a_3}Y^{a_3}_{a_8}Z^{a_4 a_5}_{a_7 a_9}Z^{a_6 a_7}_{a_4 a_5}Z^{a_8 a_9}_{a_1 a_2}\ ,
\end{equation}
\begin{equation}
 I^{(3,3)}_8\equiv Y^{a_1}_{a_6}Y^{a_2}_{a_3}Y^{a_3}_{a_4}Z^{a_4 a_5}_{a_7 a_8}Z^{a_6 a_7}_{a_5 a_9}Z^{a_8 a_9}_{a_1 a_2}\ ,
\end{equation}
\begin{equation}
 I^{(3,3)}_9\equiv Y^{a_1}_{a_6}Y^{a_2}_{a_3}Y^{a_3}_{a_4}Z^{a_4 a_5}_{a_7 a_8}Z^{a_6 a_7}_{a_1 a_9}Z^{a_8 a_9}_{a_2 a_5}\ ,
\end{equation}
\begin{equation}
 I^{(3,3)}_{10}\equiv Y^{a_1}_{a_6}Y^{a_2}_{a_3}Y^{a_3}_{a_4}Z^{a_4 a_5}_{a_8 a_9}Z^{a_6 a_7}_{a_1 a_5}Z^{a_8 a_9}_{a_2 a_7}\ .
\end{equation}
Finally, we have also analysed invariants with $n_Z=4$ and $m_Y=1$. Naively,
there are 18 different invariants with this number of $Z$ and $Y$
tensors. However, there is only one invariant with this number of coupling
tensors without self-loops,  
\begin{equation}
 I^{(4,1)}_1\equiv Y^{a_1}_{a_6}Z^{a_2 a_3}_{a_4 a_7}Z^{a_4 a_5}_{a_8 a_9}Z^{a_6 a_7}_{a_2 a_5}Z^{a_8 a_9}_{a_1a_3} =\vcenter{\hbox{\includegraphics[scale=0.2]{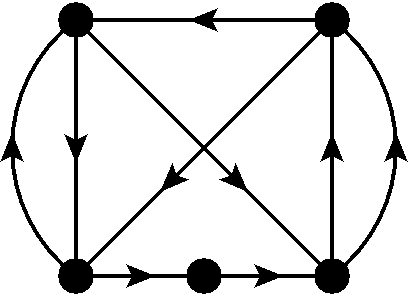}}}.
\end{equation}
This concludes our list of CPIs for explicit CP violation used in the main
text of this paper. Following our systematic approach, we have also calculated
larger invariants and the obtained CPIs are collected in Appendix~\ref{invariant_lists}. 

\section{Two Higgs doublet model potential \label{sec:2HDM}}
\cleqn

As a first example for an application of CPIs that is well known in the
literature we consider the most general potential of two copies of SM Higgs
bosons. For this potential, a complete basis of CPIs is
known~\cite{Gunion:2005ja}. All of these four CPIs have also been produced in our
systematic search. 
Using a slightly modified version of the notation
in~\cite{Branco:2005em}, the general 2HDM potential takes the form
\begin{eqnarray}
V (H_{1},H_{2})&=&m_{1}^2 \ H _{1}^{\dagger }H _{1}+ m_{12}^2\ e^{i\theta_0 }\ H
_{1}^{\dagger }H _{2}+ m_{12}^2 \ e^{-i\theta_0 }\ \ H _{2}^{\dagger }H
_{1}+m_{2}^2\ H _{2}^{\dagger }H _{2}+ \nonumber \\[2mm]
&&+a_{1}\ \left( H _{1}^{\dagger }H _{1}\right) ^{2}+a_{2}\ \left( H
_{2}^{\dagger }H _{2}\right) ^{2}\nonumber \\[2mm]
&&+b\ \left( H _{1}^{\dagger }H
_{1}\right) \left( H _{2}^{\dagger }H _{2}\right) +b^{\prime }\ \left(
H _{1}^{\dagger }H _{2}\right) \left( H _{2}^{\dagger }H
_{1}\right) + \nonumber \\[2mm]
&&+c_{1}\ e^{i\theta _{1}}\ \left( H _{1}^{\dagger }H _{1}\right) \left(
H _{2}^{\dagger }H _{1}\right) +c_{1}\ e^{-i\theta _{1}}\ \left( H
_{1}^{\dagger }H _{1}\right) \left( H _{1}^{\dagger }H _{2}\right) +
\nonumber \\[2mm] 
&&+c_{2}\ e^{i\theta _{2}}\ \left( H _{2}^{\dagger }H _{2}\right) \left(
H _{2}^{\dagger }H _{1}\right) +c_{2}\ e^{-i\theta _{2}}\ \left( H
_{2}^{\dagger }H _{2}\right) \left( H _{1}^{\dagger }H _{2}\right) +
\nonumber \\[2mm]
&&+d\ e^{i\theta_3 }\ \left( H _{1}^{\dagger }H _{2}\right) ^{2}+d\
e^{-i\theta_3 }\ \left( H _{2}^{\dagger }H _{1}\right) ^{2}.
\label{2SU2}
\end{eqnarray}
Here $H_1 = (h_{1,1},h_{1,2})$ and $H_2 = (h_{2,1},h_{2,2})$ and the $SU(2)_L$
invariant contractions are indicated by the brackets e.g. $(H_1^\dagger H_1)^2 = (h_{1,1}^\dagger h_{1,1} + h_{1,2}^\dagger h_{1,2})^2$.
Eq.~\eqref{eq:phi_to_h}
becomes
\begin{equation}
\phi=(\varphi_1,\varphi_2, \varphi_3, \varphi_4)=(h_{1,1},h_{1,2},h_{2,1},h_{2,2}) \ ,
\end{equation}
such that the $Z$ tensor corresponding to the quartic terms of the scalar
potential has $4^4=256$ components. It is straigtforward to determine these
explicitly for the potential of Eq.~\eqref{2SU2}. In the following, we display the
non-vanishing ones, 
\begin{align}
Z^{11}_{11}= Z^{22}_{22}= 2Z^{12}_{12}&= a_1\ ,\\
Z^{33}_{33}=Z^{44}_{44}=  2Z^{34}_{34}&= a_2\ ,\\
4Z^{14}_{14}=4Z^{23}_{23}&=  b \ ,\\
4Z^{14}_{23}=4Z^{23}_{14}&=  b'\ , \\
4Z^{13}_{13}=4Z^{24}_{24}&=  b+b'\ ,\\
4Z^{12}_{14}=4Z^{12}_{23}=2Z^{11}_{13}=2Z^{22}_{24}&=c_1 e^{i \theta_1}\ ,\\
4Z^{14}_{34}=4Z^{23}_{34}=2Z^{13}_{33}=2Z^{24}_{44}&=c_2 e^{i \theta_2}\ ,\\
2Z^{34}_{12}=Z^{33}_{11}=Z^{44}_{22}&=de^{i\theta_3}\ ,
\end{align}
and remind the reader of the general relations
$Z^{ab}_{cd}=Z^{ba}_{cd}=Z^{ab}_{dc}=Z^{ba}_{dc}$ 
and $Z^{cd}_{ab}=(Z^{ab}_{cd})^\ast$.  
Having determined the $Z$ tensor in terms of the parameters of the potential,
we can calculate CPIs explicitly. 

As a first illustration, we show the results of CPIs for the two Higgs doublet
potential of Eq.~\eqref{2SU2}.
In our notation, the smallest one becomes 
\begin{align}
\mathcal{I}_1=&-9 i m_{12}^2 \left(m_1^2-m_2^2\right) \Big[ c_2 (2 a_1-b-b') \sin (\theta_0+\theta_2)+c_1 (2 a_2-b-b') \sin (\theta_0+\theta_1)\nonumber\\[2mm]
 &\qquad+2 d (c_1 \sin (\theta_0-\theta_1-\theta_3)+c_2 \sin (\theta_0-\theta_2-\theta_3))\Big] \nonumber\\[2mm]
 &-9 i m_{12}^4 \Big[ 4 d (a_2-a_1) \sin (2 \theta_0-\theta_3)+c_1^2 \sin (2 (\theta_0+\theta_1))\nonumber\\[2mm]
 &\qquad-c_2 (2 c_1 \sin (\theta_1-\theta_2)+c_2 \sin (2 (\theta_0+\theta_2)))\Big]\nonumber\\[2mm]
 &-9 i c_1 c_2 \left(m_1^2-m_2^2\right)^2 \sin (\theta_1-\theta_2).
\end{align}
There are many ways of setting this expression to zero, the simpler ones
involve $m_{12}^2=0$ which leaves only the last line in the expression, which
vanishes either for $m_1^2-m_2^2$ or $\sin(\theta_1-\theta_2)$. Alternatively,
if $m_1^2-m_2^2=0$ there are other combinations of constraints that make this
CPI vanish, including $\sin(2 \theta_0 - \theta_3)= \sin(2 (\theta_0 +
\theta_1))=0$. 
However, at this stage it is not clear if any of these constraints are
sufficient to guarantee conservation of CP, as other CPIs could still be
non-zero. The invariant $\mathcal{I}_1$ being non-zero always requires
$m_{12}\neq0$ or $m_1^2\neq m_2^2$. 

As already mentioned, a complete basis of CPIs is known for the 2HDM
potential, cf.~\cite{Gunion:2005ja}. Of the four invariants given in that
paper, three are equivalent to invariants given in Section~\ref{CPoddI} of our
paper (GH denotes the invariant in~\cite{Gunion:2005ja}): 
\begin{equation}
 I_{2Y2Z}^\text{(GH)}=I_1=I^{(2,2)}_1\ ,
\end{equation}
\begin{equation}
 I_{Y3Z}^\text{(GH)}=I^{(3,1)}_2\ ,
\end{equation}
and
\begin{equation}
 I_{3Y3Z}^\text{(GH)}=(I^{(3,3)}_5)^{*}\ .
\end{equation}
The fourth invariant listed in~\cite{Gunion:2005ja} has $n_Z=6$ and has not
been given here yet as it contains $Z$-self-loops. For completeness, we
present it here following our general notation as well as diagrammatically:
\begin{equation}
I_{6Z}^{\mathrm{(GH)}}\equiv Z^{ac}_{bd}Z^{be}_{fe}Z^{dg}_{hg}Z^{fj}_{ak}Z^{km}_{jn}Z^{nh}_{mc}=\vcenter{\hbox{\includegraphics[scale=0.2]{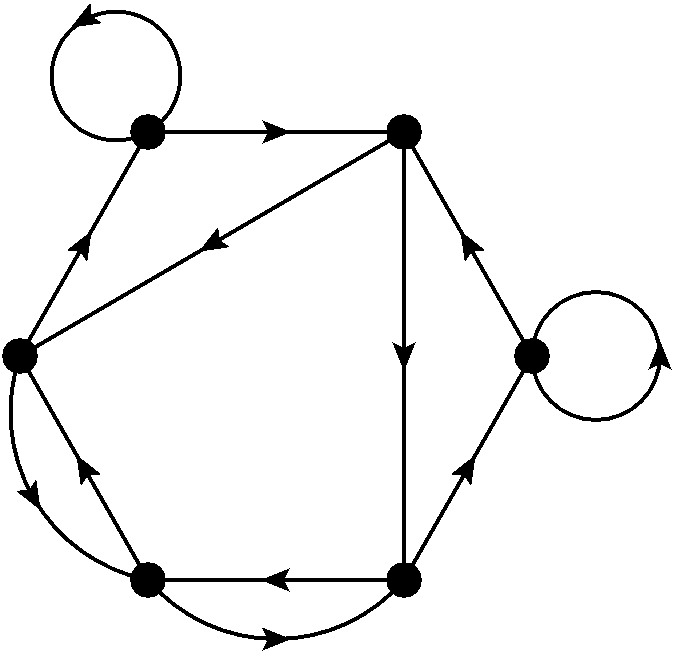}}}
\end{equation}
All CP-odd invariants with 5 $Z$ tensors, cf.\ Eqs.~(\ref{I5_1})-(\ref{I5_3}), vanish for this potential.
For this potential, $Z^{ab}_{ac}$ is non-diagonal, which is why the CPIs found
in~\cite{Gunion:2005ja} produce interesting results. While the invariants
from~\cite{Gunion:2005ja} form a complete basis of CPIs for the 2HDM, all
of them are zero for the potentials considered in the remainder of this
paper. It is our systematic search that reveals new non-zero CPIs in those situations.

\section{\label{sec:A4}$\bs{A_4=\Delta(12)}$ invariant potentials}
\cleqn

In this section we study potentials invariant under the discrete group $A_4$. We start with a field content of a single triplet of SM singlets, then consider a triplet of $SU(2)_L$ doublets, two triplets of SM singlets and two triplets of $SU(2)_L$ doublets.

$A_4$ contains a real triplet and three one-dimensional representations. The
product of two triplets decomposes as
\be
{\bf 3\otimes 3} ~=~ ({\bf 1}_0 + {\bf 1}_1 + {\bf 1}_2 + {\bf 3})_s + {\bf 3}_a\ . 
\label{eq:kronecker12}
\ee
Symmetric and antisymmetric combinations are denoted by subscripts $s$ and
$a$, respectively. Throughout this section we work in the basis
of~\cite{Ma:2001dn} which can be easily generalised to $\Delta(27)$ and the
complete $\Delta(3n^2)$ series~\cite{Luhn:2007uq,Ishimori:2010au} studied in
the Sections~\ref{sec:D27} and~\ref{sec:D3n2}.

\subsection{One flavour triplet}
With one triplet field, only the symmetric contribution in
Eq.~\eqref{eq:kronecker12} matters. It is convenient to define

\bea
V_0 (\varphi) =
 - ~m^2_{\varphi}\sum_i   \varphi_i \varphi^{*i}
+ r \left( \sum_i   \varphi_i \varphi^{*i}  \right)^2
+ s \sum_i ( \varphi_i \varphi^{*i})^2 \ ,
\label{eq:potV0}
\eea
where one notes that the first two terms are $SU(3)$ invariant. We consider
$\varphi$ to be additionally charged under some $U(1)$ symmetry (or an
appropriate discrete subgroup) such that terms of the form 
$\varphi_i \varphi_i$ or $\varphi_i \varphi_i \varphi_i$, for example, are not allowed. This leads to a more direct generalisation of
the case where the SM gauge group applies.

The resulting renormalisable scalar potential for $A_4$ reads
\bea\label{eq:A4ini}
V_{A_4} (\varphi) = V_0 (\varphi) &+&c \left(
\varphi_1 \varphi_1 \varphi^{*3} \varphi^{*3} +  \varphi_2 \varphi_2 \varphi^{*1} \varphi^{*1} + \varphi_3 \varphi_3 \varphi^{*2} \varphi^{*2} \right) \nonumber\\
&+&c^\ast \left(
\varphi^{*1} \varphi^{*1} \varphi_3 \varphi_3 + \varphi^{*2} \varphi^{*2} \varphi_1 \varphi_1 + \varphi^{*3} \varphi^{*3} \varphi_2 \varphi_2
 \right)
 \ ,
\eea
noting that this includes, as expected, four independent quartic terms.
Henceforth we use the convenient abbreviations cycl.\ to denote the cyclic
permutations, and h.c.\ to indicate the hermitian conjugate. We thus write the
$A_4$ invariant potential of Eq.~\eqref{eq:A4ini} in the compact form:
\bea
V_{A_4} (\varphi) = V_0 (\varphi) +~ \left[ c \left(
\varphi_1 \varphi_1 \varphi^{*3} \varphi^{*3} + 
\text{cycl.} \right) ~+~ \text{h.c.}
 \right]
 \ .
\label{eq:potA4}
\eea

The $A_4$ symmetric potential respects
the general CP symmetry with a 2-3 swap, namely the CP symmetry with unitary
matrix $U_{23}$ 
\bea
U_{23} = \begin{pmatrix}
1 & 0 & 0 \\
0 & 0 & 1 \\
0 & 1 & 0 
\end{pmatrix},
\label{U23}
\eea
for arbitrary coefficients $r,s \in \mathbb R$ and $c\in \mathbb C$.
Hence, despite the occurrence of the complex coupling $c$ the $A_4$ symmetric
potential of one triplet is invariant under this general CP symmetry. For this
reason, all possible CPIs for this potential will be zero. 

\subsection{One flavour triplet of Higgs doublets}

If each component of the $A_4$ triplet is an $SU(2)_L$ doublet,
\be
H = (h_{1\alpha},h_{2\beta},h_{3\gamma})\ ,
\ee
 the potential
is rather similar to the previous case. Indeed there is one additional
invariant, due to the two different ways to perform the $SU(2)_L$ contraction
on the $A_4$ invariant $\left( \sum_i   \varphi_i \varphi^{*i}  \right)^2$,
when the $\varphi$ are replaced by Higgs doublets\footnote{Since the doublet 
  ${\bf{2}}$ of $SU(2)_L$ is a pseudoreal representation, it is also possible to
combine  $( h_{i \alpha} h_{j \beta} \epsilon^{\alpha\beta})( h^{*i\gamma}
h^{*j\delta} \epsilon_{\gamma\delta} )$  using the antisymmetric $\epsilon$
tensor. However, such a term is not linearly independent of the two terms in
Eq.~\eqref{eq:doubletcontractions} as can be easily seen in an explicit
calculation or by noting that ${\bf{2\times 2=1+3}}$ which entails only two
independent  $SU(2)_L$ invariant quartic terms.}
\be
\sum_{i, j, \alpha, \beta} \left[
r_1 ( h_{i \alpha}  h^{*i\alpha})( h_{j \beta}  h^{*j\beta}) + r_2 ( h_{i
  \alpha}  h^{*i\beta})( h_{j \beta}  h^{*j\alpha}) \right]\ .
\label{eq:doubletcontractions}
\ee
Here 
we highlight the $SU(2)_L$ indices to clarify the distinct $SU(2)_L$ contractions.
We define $V_{0} (H)$ in analogy with Eq.~(\ref{eq:potV0}):

\begin{eqnarray}
 V_{0} ( H ) &=&
 - ~m^2_{h}\sum_{i, \alpha}   h_{i \alpha}  h^{*i\alpha}
+\sum_{i, j, \alpha, \beta} \left[ r_1 ( h_{i \alpha}  h^{*i\alpha})( h_{j \beta}  h^{*j \beta}) + r_2 ( h_{i \alpha}  h^{*i \beta})( h_{j \beta}  h^{*j\alpha}) \right]\notag \\
&&+ s \sum_{i, \alpha, \beta}  ( h_{i \alpha}  h^{*i\alpha})( h_{i \beta}
h^{*i\beta}) \  ,
\label{eq:potV0H}
\end{eqnarray}
and the $A_4$ potential is then
\begin{eqnarray}
 V_{A_4} ( H ) =
 V_{0} (H)
+ \sum_{\alpha, \beta} \left[ c \left(
h_{1 \alpha} h_{1 \beta}  h^{*3\alpha}  h^{*3\beta} + 
\text{cycl.} \right) + \text{h.c.} \right].
\label{eq:potA4SU2}
\end{eqnarray}
This potential is also invariant under a CP transformation that involves swapping the second and third component in flavour space while keeping $SU(2)_L$ contractions unchanged, i.e.\ $h_{2\alpha}\rightarrow h^{\ast 3\alpha}$ etc.:
\bea
U_{23}^H = \begin{pmatrix}
1 & 0 & 0 \\
0 & 0 & 1 \\
0 & 1 & 0 
\end{pmatrix} \otimes \delta^{\alpha}_{\beta}\ .
\label{U23H}
\eea
Therefore, CP is conserved automatically for this potential and all possible CPIs necessarily vanish. 

\subsection{Two flavour triplets}

Typically, realistic models of flavour require more than just one triplet 
flavon. We therefore consider the potential involving two physically different flavon
fields $\varphi$ and $\varphi'$ which both transform in the triplet representation of
$A_4$. 
In the case of two $A_4$ triplets distinguished by additional symmetries so
that the total symmetry is $A_4\times U(1) \times U(1)'$, the potential
includes a total of seven independent mixed quartic invariants of the form
$\varphi\,\varphi' \, \varphi^* \, \varphi'^*$. It is convenient to define:

\bea
V_1 (\varphi,\varphi') &=&
+~ \tilde r_1 \left( \sum_i \varphi_i \varphi^{*i} \right)
\left( \sum_j \varphi'_j \varphi'^{*j} \right) 
+ \tilde r_2\left( \sum_i \varphi_i \varphi'^{*i} \right)
\left( \sum_j \varphi'_j \varphi^{*j} \right) \notag \\[2mm]
&& +~ \tilde s_1\sum_i \left(\varphi_i \varphi^{*i} \varphi'_i \varphi'^{*i}
\right) \notag \\[2mm]
&& +~ \tilde s_2 \left(
\varphi_1 \varphi^{*1} \varphi'_2 \varphi'^{*2} + 
\varphi_2 \varphi^{*2} \varphi'_3 \varphi'^{*3} + 
\varphi_3 \varphi^{*3} \varphi'_1 \varphi'^{*1} 
\right)  \notag \\[2mm]
&& +~ i \, \tilde s_3 
\Big[
(\varphi_1 \varphi'^{*1} \varphi'_2 \varphi^{*2} + \text{cycl.}
) 
- 
( \varphi^{*1}\varphi'_1  \varphi'^{*2} \varphi_2 +\text{cycl.}
)
\Big] .
\label{eq:potV1}
\eea
Note that in this definition, the term multiplied by $\tilde{r}_1$ contains
the term multiplied by $\tilde{s}_2$ as well as the term obtained from the
latter by interchanging $\varphi$ with $\varphi'$:
\be
\left(
\varphi'_1 \varphi'^{*1} \varphi_2 \varphi^{*2} + 
\varphi'_2 \varphi'^{*2} \varphi_3 \varphi^{*3} + 
\varphi'_3 \varphi'^{*3} \varphi_1 \varphi^{*1} 
\right) ,
\ee
which is not included separately in $\tilde{s}_2$.

The $A_4$ symmetric renormalisable
potential takes the following explicit form, with $V_0$ as defined in Eq.~(\ref{eq:potV0}),
\bea\notag
V_{A_4} (\varphi,\varphi') &=&
V_0 (\varphi) + V'_0 (\varphi') + V_1 (\varphi,\varphi') +  \\[2mm]
&&+ \left[
c \left(
\varphi_1 \varphi_1 \varphi^{*3} \varphi^{*3} + 
\text{cycl.} \right) +\text{h.c.}\right]
 + \left[ c' \left(
\varphi'_1 \varphi'_1 \varphi'^{*3} \varphi'^{*3} + 
\text{cycl.} \right)+\text{h.c.}\right]\notag\\[2mm]
 &&+ \left[\tilde c \left(
\varphi_1 \varphi'_1 \varphi^{*3} \varphi'^{*3} + 
\text{cycl.} \right) + \text{h.c.}
\right] ,\label{eq:potA4-2}
\eea
where $V_0' (\varphi')$ has the same functional form as $V_0 (\varphi)$  with
different coefficients $m'_{\varphi'}$, $r'$, $s'$ and depends on $\varphi'$. 

Unlike the previous $A_4$ invariant potentials, this potential in general
violates CP, as confirmed by the non-zero CPIs listed in
Table~\ref{ta:summary} of Section~\ref{sec:table}.
The expressions are cumbersome and we do not reproduce them here. The
non-vanishing CPIs $\mathcal{I}^{(6)}_2,\mathcal{I}^{(6)}_3,\mathcal{I}^{(6)}_4,\mathcal{I}^{(6)}_5$ (Eqs.~(\ref{I62},\ref{I63},\ref{I64},\ref{I65})) all factorise as a product of
$\tilde s_2$ with different complicated functions of the remaining parameters,
for example, $\mathcal{I}^{(6)}_{2}$ takes the form: 
\bea
\mathcal{I}^{(6)}_{2} = \tilde{s}_2 f(...)\ , \label{eq:s2f}
\eea
where $f$ is a complicated function of the other parameters. 
Such a dependence on $\tilde s_2$ is expected because it corresponds to a CP
symmetry, where one imposes $U_{23}$ of Eq.~(\ref{U23}) on both triplets,
corresponding to the block matrix: 
\bea
U_{23}^{\varphi \varphi'} = \begin{pmatrix}
U_{23} & 0  \\
0 & U_{23} 
\end{pmatrix}.
\label{U23phiphi}
\eea
This CP symmetry constrains the potential such that $\tilde s_2=0$, which
forces all CPIs to vanish as expected from the presence of a CP symmetry. 
Furthermore, applying instead the trivial CP symmetry $CP_0$ forces $\tilde
s_3=0$ and all other complex parameters ($c,c',\tilde{c}$) to be real. 
As expected, this renders $f(...)=0$ in Eq.~\eqref{eq:s2f}, an makes all
other CPIs vanish as well.

\subsection{Two flavour triplets of Higgs doublets}
Earlier, when considering a potential of an $A_4$ triplet of $SU(2)_L$ doublets, the only difference was that the term with coefficient $r$ split into two different invariants corresponding to two different possible $SU(2)_L$ contractions, cf.\ Eq.~(\ref{eq:doubletcontractions}). Similarly, the potential of two triplets of SM doublets:
\be
H = (h_{1\alpha},h_{2\beta},h_{3\gamma}) \,, \quad H' =
(h'_{1\alpha},h'_{2\beta},h'_{3\gamma})\ ,
\ee
  can be obtained from the corresponding potential of singlets,
  Eq.~(\ref{eq:potA4-2}). In the first two parts of the potential,
  $V_0(\varphi)$ and $V_0(\varphi')$, as earlier, there are two different ways
  of $SU(2)_L$-contracting the invariants with coefficients $r$ and $r'$. In
  the part of the potential with $A_4$ contractions as in $V_1(\varphi,\varphi')$, for all $A_4$ invariants two possible ways of $SU(2)_L$ contracting the fields exists and this part of the potential becomes
\begin{align}
 V_1(H,H')&=
\sum_{i,j,\alpha,\beta} \left[
\tilde{r}_{11}h_{i\alpha}h^{\ast i \alpha}h_{j \beta}'h'^{\ast j
  \beta}+\tilde{r}_{12}h_{i \alpha}h'^{\ast j \alpha} h'_{j \beta}h^{\ast i
  \beta}
\right] \notag \\[2mm]
 &+\sum_{i,j,\alpha,\beta} \left[
\tilde{r}_{21}h_{i\alpha}h'^{\ast i \alpha}h'_{j \beta}h^{\ast j
  \beta}+\tilde{r}_{22}h_{i \alpha}h^{\ast j \alpha}h'_{j \beta}h'^{\ast i
  \beta}
\right]\notag \\[2mm]
 &+\sum_{i,\alpha,\beta} \left[
\tilde{s}_{11}h_{i \alpha}h^{\ast i \alpha}h'_{i \beta}h'^{\ast i
  \beta}+\tilde{s}_{12}h_{i \alpha}h'^{\ast i \alpha}h'_{i \beta}h^{\ast i
  \beta}
\right]\notag \\[2mm]
 &+\sum_{\alpha,\beta} \left[
\tilde{s}_{21}(h_{1\alpha}h^{\ast1\alpha}h'_{2\beta}h'^{\ast2\beta}+\text{cycl.})+\tilde{s}_{22}(h_{1\alpha}h'^{\ast
  2 \alpha}h'_{2\beta}h^{\ast 1\beta}+\text{cycl.})
\right]\notag \\[2mm]
 &+i\tilde{s}_{31}\sum_{\alpha,\beta} 
[(h_{1\alpha}h'^{\ast1\alpha}h'_{2\beta}h^{\ast2\beta}+\text{cycl.}) - (h^{\ast1\alpha}h'_{1\alpha}h'^{\ast2\beta}h_{2\beta}+\text{cycl.})]\notag \\[2mm]
 &+i\tilde{s}_{32}\sum_{\alpha,\beta} [(h_{1\alpha}h^{\ast 2 \alpha}h'_{2\beta}h'^{\ast1\beta}+\text{cycl.}) - (h^{\ast1\alpha}h_{2 \alpha}h'^{\ast2\beta}h'_{1\beta}+\text{cycl.})].
\label{eq:potV1H}
 \end{align}
Finally, of the remainder of the potential, only the invariant with coefficient $\tilde{c}$ from Eq.~\eqref{eq:potA4-2} needs to be doubled:
\begin{equation}\sum_{\alpha,\beta}
\left[ \tilde{c}_1(h_{1\alpha}h^{\ast3 \alpha}h'_{1\beta}h'^{\ast3\beta}+\text{cycl.})+\tilde{c}_2(h_{1\alpha}h'^{\ast 3\alpha}h'_{1\beta}h^{\ast3\beta}+\text{cycl.}) + \text{h.c.}\right].
\end{equation}
We therefore write
\begin{align}
V_{A_4} (H,H') &= V_0(H) + V_0'(H')+ V_1(H,H')\\
&+\sum_{\alpha,\beta}\left[ c \left(
h_{1\alpha} h_{1\beta} h^{\ast3\alpha} h^{\ast3\beta} + 
\text{cycl.} \right)
+ c' \left(
h'_{1 \alpha} h'_{1 \beta} h'^{\ast3\alpha} h'^{\ast3\beta} + 
\text{cycl.} \right) + \text{h.c.} \right]
\notag\\[2mm]
&+\sum_{\alpha,\beta}\left[ \tilde{c}_1(h_{1\alpha}h^{\ast3 \alpha}h'_{1\beta}h'^{\ast3\beta}+ \text{cycl.})+\tilde{c}_2(h_{1\alpha}h'^{\ast 3\alpha}h'_{1\beta}h^{\ast3\beta}+\text{cycl.})+\text{h.c.} \right] \notag.
\end{align}
We note that due to $SU(2)_L$ not allowing cubic invariants of $H$ and/or
$H'$, it is sufficient to use a $Z_3$ symmetry to distinguish the $A_4$
triplets.\footnote{The potential invariant under a
  $Z_2$~\cite{Varzielas:2015joa} would additionally allow for invariants of the form 
$h_{i\alpha}h'^{\ast i \alpha}h_{j \beta}h'^{\ast j   \beta}$ and 
$h_{i\alpha}h'^{\ast i \beta}h_{j \beta}h'^{\ast j   \alpha}$ 
where the conjugated fields are both related to $H'$. } 

This potential generally violates CP. This can be seen from the CP-odd invariants calculated, as $\mathcal{I}^{(6)}_2,\mathcal{I}^{(6)}_3,\mathcal{I}^{(6)}_4,\mathcal{I}^{(6)}_5$ (Eqs.~(\ref{I62},\ref{I63},\ref{I64},\ref{I65})) are non-zero (see Table \ref{ta:summary}) but with too large expressions to display here.
However, it is possible to impose a CP symmetry with 
\bea
U_{23}^{HH'} = \begin{pmatrix}
U_{23} & 0  \\
0 & U_{23} 
\end{pmatrix} \otimes \delta^{\alpha}_{\beta}\ ,
\label{U23HH}
\eea
which, similarly to previous examples, restricts the coefficients in the potential, namely
\begin{equation}
\tilde{s}_{21}=\tilde{s}_{22}=0\ ,
\end{equation}
thereby forcing all CPIs to vanish. 
Imposing, alternatively, the canonical CP symmetry $CP_0$ leads to $\tilde
s_{31}=\tilde s_{32}=0$ as well as $c,c',\tilde{c}_1,\tilde{c}_2\in \mathbb R$.

\subsection[$S_4$ invariant potentials]{$\boldsymbol{S_4}$ invariant potentials}

The transition from $\Delta(3n^2)$ invariant potentials with arbitrary $n\in
N$ to potentials which are symmetric under the larger group $\Delta(6n^2)$ is
discussed in Appendix~\ref{6n2_from_3n2}. The corresponding basis of
$S_4=\Delta(6\times 2^2)$ can be found
in~\cite{Ma:2005pd,Escobar:2008vc,Ishimori:2010au}. 
For the $A_4$ potential with one triplet of singlets as well the $A_4$ potential with a triplet of doublets, the corresponding $S_4$ invariant potentials are obtained by setting
\begin{equation}
 c^\ast =c \ ,
\end{equation}
so that
\begin{equation}
\begin{array}{rl}
 V_{S_4} (\varphi) = V_0 (\varphi) 
+c \left[ \left(\varphi_1 \varphi_1 \varphi^{*3} \varphi^{*3}
  +\text{cycl.}\right) + \text{h.c.} \right],
 \end{array}
\end{equation}
and
\begin{equation}
  V_{S_4} ( H ) =
 V_{0} (H)
+  \sum_{\alpha, \beta} c \left[\left(
h_{1 \alpha} h_{1 \beta}  h^{*3\alpha}  h^{*3\beta} + 
\text{cycl.}\right)  + \text{h.c.} \right],
\label{eq:potS4SU2}
\end{equation}
where the potentials $V_0$ were defined in Eq.~(\ref{eq:potV0}) and Eq.~(\ref{eq:potV0H}).
For the potential of two triplets of $A_4$, the $S_4$ invariant potential arises via setting
\bea
\tilde{s}_2 = \tilde{s}_3 = 0\ ,
\label{3n2_to_6n2}
\eea
and additionally
\bea
c^{\ast} = c \ ,\qquad c'^\ast = c' \ ,\qquad  \tilde{c}^\ast = \tilde{c} \ .
\eea
Defining the following abbreviation,
\bea\label{eq:potV2}
V_2 (\varphi,\varphi') &=&
 \tilde r_1 \left( \sum_i \varphi_i \varphi^{*i} \right)
\left( \sum_j \varphi'_j \varphi'^{*j} \right) 
+ \tilde r_2\left( \sum_i \varphi_i \varphi'^{*i} \right)
\left( \sum_j \varphi'_j \varphi^{*j} \right) \notag \\[2mm]
&&
 +~ \tilde s_1\sum_i \left(\varphi_i \varphi^{*i} \varphi'_i \varphi'^{*i}
\right), 
\eea
the full potential of two $S_4$ triplets  becomes
\bea
V_{S_4} (\varphi,\varphi') &=&
V_0 (\varphi) + V'_0 (\varphi') + V_2 (\varphi,\varphi') +  \notag\\[2mm]
&&+~
c \left[\left(
\varphi_1 \varphi_1 \varphi^{*3} \varphi^{*3} + 
\text{cycl.}\right)+\text{h.c.}\right]
 + c' \left[\left(
\varphi'_1 \varphi'_1 \varphi'^{*3} \varphi'^{*3} + 
\text{cycl.} \right)+ \text{h.c.}\right]\notag\\[2mm]
& &+ ~\tilde c \left[ \left(
\varphi_1 \varphi'_1 \varphi^{*3} \varphi'^{*3} + 
\text{cycl.}\right) + \text{h.c.}\right] .
\label{eq:potS4-2}
\eea
The $S_4$
potential with two triplets generally conserves CP. 
This can be understood 
from the non-vanishing CPIs obtained for $A_4$,  which were proportional to $\tilde{s}_2$ (see Eq.~(\ref{eq:s2f})) which is zero in the case of $S_4$.
Indeed, one CP symmetry
present in $V_{S_4} (\varphi, \varphi')$ is $U_{23}^{\varphi \varphi'}$ in
Eq.~(\ref{U23phiphi}), because $S_4$ enforces $\tilde{s}_2=0$ and therefore
the $V_{S_4} (\varphi, \varphi')$ potential is invariant under simultaneous
CP transformations with 2-3-swap on $\varphi$ and $\varphi'$. 

Turning to the case of Higgs doublets of $SU(2)_L$, for $V_{A_4}(H,H')$, enlarging the symmetry to $S_4$ constrains the
potential parameters as follows: 
\begin{equation}
c^\ast=c,\quad c'^\ast=c', \quad
\tilde{c}_1^\ast=\tilde{c}_1,\quad
\tilde{c}_2^\ast=\tilde{c}_2,
\end{equation}
and
\begin{equation}
 \tilde{s}_{21}=\tilde{s}_{22}=\tilde{s}_{31}=\tilde{s}_{32}=0\ .
\end{equation}
Again, introducing an abbreviation,
\bea
 V_2(H,H')&=&\sum_{i,j,\alpha,\beta}\left[
\tilde{r}_{11}h_{i\alpha}h^{\ast i \alpha}h_{j \beta}'h'^{\ast j
  \beta}+\tilde{r}_{12}h_{i \alpha}h'^{\ast j \alpha} h'_{j \beta}h^{\ast i
  \beta} \right]\notag \\
& &+\sum_{i,j,\alpha,\beta}\left[
\tilde{r}_{21}h_{i\alpha}h'^{\ast i \alpha}h'_{j \beta}h^{\ast j \beta}+\tilde{r}_{22}h_{i \alpha}h^{\ast j \alpha}h'_{j \beta}h'^{\ast i \beta}\right]\notag \\
 &&+\sum_{i,\alpha,\beta}\left[
\tilde{s}_{11}h_{i \alpha}h^{\ast i \alpha}h'_{i \beta}h'^{\ast i \beta}+\tilde{s}_{12}h_{i \alpha}h'^{\ast i \alpha}h'_{i \beta}h^{\ast i \beta}\right],
\label{eq:potV2H}
 \eea
 the $S_4$ invariant potential of two triplets of doublets becomes
\begin{align}
V_{S_4} (H,H') =~& V_0(H) + V_0'(H')+ V_2(H,H')\notag\\
&+\sum_{\alpha,\beta} c\left[  \left(
h_{1\alpha} h_{1\beta} h^{\ast3\alpha} h^{\ast3\beta} + 
\text{cycl.} \right)+ \text{h.c.}\right]\notag\\
&+\sum_{\alpha,\beta}  c' \left[\left(
h'_{1 \alpha} h'_{1 \beta} h'^{\ast3\alpha} h'^{\ast3\beta} + 
\text{cycl.}\right) + \text{h.c.} \right]
\notag\\
&+\sum_{\alpha,\beta} \tilde{c}_1\left[ \left(h_{1\alpha}h^{\ast3
    \alpha}h'_{1\beta}h'^{\ast3\beta}+ \text{cycl.}\right)+\text{h.c.}\right]\notag\\
&+\sum_{\alpha,\beta} \tilde{c}_2\left[\left(h_{1\alpha}h'^{\ast 3\alpha}h'_{1\beta}h^{\ast3\beta}+\text{cycl.}\right)+\text{h.c.} \right] .
\end{align}
As in Eq.~(\ref{eq:potS4-2}), the potential $V_{S_4} (H,H')$ conserves CP.
As all parameters of this potential are real, it is not surprising, that it is invariant under trivial CP, $CP_0$.

\section{$\bs{\Delta(27)}$ invariant potentials \label{sec:D27}}
\cleqn

In this section we concern ourselves with potentials invariant under $\Delta(27)$. As in the $A_4$ case, we consider the field content of a single triplet of SM singlets, then a single triplet which is also an $SU(2)_L$ doublet,
then two triplets of SM singlets, and finally two $\Delta(27)$ triplets of $SU(2)_L$ doublets.

The group $\Delta(27)$ has one irreducible triplet representation ${\bf 3}$, its
conjugate $\bar{\bf 3}$, and nine one-dimensional representations. The product of two
triplets decomposes as
\be
{\bf 3 \otimes 3} ~=~ ({\bf \ol 3 + \ol 3})_s ~+~ {\bf \ol 3}_a \ ,
\label{eq:Kronecker27} 
\ee
where the subscripts $s$ and $a$ denote symmetric and antisymmetric
combinations, respectively. In the following we adopt the basis of~\cite{Ma:2006ip,Luhn:2007uq,Ishimori:2010au}.

\subsection{One flavour triplet}

Having only one triplet field, the antisymmetric contribution in
Eq.~\eqref{eq:Kronecker27} vanishes identically. As a consequence there are
four independent quartic $\Delta(27)$ invariants of type ${\bf 3 \otimes 3
  \otimes \ol 3   \otimes \ol 3}$. Writing the components of the triplet field
as $\varphi_i$, with $i=1,2,3$, we can easily derive the renormalisable scalar
potential, 
\bea
V_{\Delta(27)} (\varphi) &=&
V_0(\varphi)
+~ \left[d \left(
\varphi_1 \varphi_1 \varphi^{*2} \varphi^{*3} + 
\text{cycl.} \right) +\text{h.c.}\right]
 \ .
\label{eq:pot27}
\eea
The coefficients inside $V_0 (\varphi)$ (cf.\ Eq.~(\ref{eq:potV0})) are real
but $d\in \mathbb C$. The number of independent real parameters is therefore
four. $V_{\Delta(27)} (\varphi)$ is accidentally also the potential for a
single $\Delta(54)$ triplet~\cite{deMedeirosVarzielas:2011zw}, as discussed
also in  Appendix~\ref{6n2_from_3n2}.

The potential of Eq.~\eqref{eq:pot27} in its most general form violates CP as
can be seen from the construction of CPIs which do not vanish for
general choices of the coefficients in the potential (see Table \ref{ta:summary}).
Calculating the CPIs $\mathcal{I}^{(6)}_{4,5}$ (Eqs.~(\ref{I64},\ref{I65})) explicitly yields the same
non-zero result for this potential: 
\bea
\mathcal{I}^{(6)}_{4,5} =  -\frac{3}{32} \left(d^3-d^{*3} \right) \left(d^3+6 d d^* s+ d^{*3}-8 s^3\right),
\label{D27CPI}
\eea
while the other explicit CPIs that are listed throughout Section~\ref{CPoddI}
are zero for this potential. 
The potential in Eq.~(\ref{eq:pot27}) is known to be CP conserving in the
cases $\mathrm{Arg}(d) =0, 2\pi/3, 4\pi/3$. 
Indeed this is reflected in the CPIs which are proportional to
\be
(d^3 - d^{*3}) \ .
\ee
This factor vanishes for $\mathrm{Arg}(d) =0, 2\pi/3, 4\pi/3$, where each case
corresponds to a distinct CP symmetry, defined by a $3\times 3$ matrix $U$. 
In the following, we explicitly list the CP transformations that enforce
various parameter relations. The $U_i$-notation we use in our work matches the
indices of the CP transformations listed in~\cite{Nishi:2013jqa},
\bea
 \text{Arg}(d)=0&\Longleftrightarrow&
 U_0=\begin{pmatrix}1&0&0\\0&1&0\\0&0&1\end{pmatrix}\text{ or
 }~~U_1=\begin{pmatrix}1&0&0\\0&0&1\\0&1&0\end{pmatrix}, \\
 \text{Arg}(d)=4\pi/3&\Longleftrightarrow& U_2=\begin{pmatrix}1&0&0\\0&1&0\\0&0&\omega\end{pmatrix}\text{ or }~~U_8=\begin{pmatrix}\omega&0&0\\0&0&1\\0&1&0\end{pmatrix},\\
 \text{Arg}(d)=2\pi/3&\Longleftrightarrow& U_3=U_2^*=\begin{pmatrix}1&0&0\\0&1&0\\0&0&\omega^2\end{pmatrix}\text{ or }~~U_9=U_8^*=\begin{pmatrix}\omega^2&0&0\\0&0&1\\0&1&0\end{pmatrix}.~~~~~
\eea
We recall that for each CP transformation an equivalent one can be obtained by multiplying it by an element of 
$\Delta(27)$. 
Note also that $U_1 = U_{23}$ from Eq.~(\ref{U23}).
Focusing on the other factor of Eq.~\eqref{D27CPI}, all CPIs we have
identified vanish if we set
\be
\left(d^3+6 d d^* s+ d^{*3}-8 s^3\right) = 0\ .
\label{D27CPC2}
\ee
This is a strong hint that there are CP symmetries that make the potential CP
conserving, not by fixing the phase of $d$ but by imposing specific relations
between the parameters $d$ and $s$. Indeed, there are three solutions to
Eq.~(\ref{D27CPC2}) which are listed with the corresponding CP
transformations from~\cite{Nishi:2013jqa}, 

\bea
\!\!\!&&\!\!\!\!\!\!\!2 s=(d+d^*)=2 \mathrm{Re} (d)\notag\\[2mm] &&\!\!\Longleftrightarrow U_4=\frac{1}{\sqrt{3}}\begin{pmatrix}1&1&1\\1&\omega&\omega^2\\1&\omega^2&\omega\end{pmatrix} \text{ or }~
 U_5=U_4 U_1=U_4^\ast=\frac{1}{\sqrt{3}}
\left(
\begin{array}{ccc}
 1 & 1 & 1 \\
 1 & \omega^2 & \omega \\
 1 & \omega & \omega^2 \\
\end{array}
\right)\!,
\label{eq:notsomagic}\\[7mm]
\!\!\!&&\!\!\!\!\!\!\!2 s=-\mathrm{Re} (d) - \sqrt{3} \mathrm{Im} (d)\notag\\[2mm]  &&\!\!\Longleftrightarrow U_6=\frac{-i}{\sqrt{3}}
\left(
\begin{array}{ccc}
 1 & \omega & \omega \\
 \omega & \omega & 1 \\
 \omega & 1 & \omega \\
\end{array}
\right)
\text{ or }~U_{10}=U_6 U_1=\frac{-i}{\sqrt{3}}\left(
\begin{array}{ccc}
 1 & \omega & \omega \\
 \omega & 1 & \omega \\
 \omega & \omega & 1 \\
\end{array}\right)\!, \\[7mm]
 \!\!\!&& \!\!\!\!\!\!\!2 s=-\mathrm{Re} (d) + \sqrt{3} \mathrm{Im} (d)\notag\\[2mm] &&\!\!\Longleftrightarrow U_7=U_6^\ast=\frac{i}{\sqrt{3}}\left(
\begin{array}{ccc}
 1 & \omega^2 & \omega^2 \\
 \omega^2 & \omega^2 & 1 \\
 \omega^2 & 1 & \omega^2 \\
\end{array}
\right)\text{ or }~U_{11}=U_7 U_1=\frac{i}{\sqrt{3}}\left(
\begin{array}{ccc}
 1 & \omega^2 & \omega^2 \\
 \omega^2 & 1 & \omega^2 \\
 \omega^2 & \omega^2 & 1 \\
\end{array}
\right)\!.~~~~~~~~~
\eea

We conclude that there exist 12 CP symmetries, listed in~\cite{Nishi:2013jqa},
which correspond to two CP symmetries for each of the 6 CP conserving
conditions that make either $(d^3 - d^{*3})=0$ or $\left(d^3+6 d d^* s+
d^{*3}-8 s^3\right) = 0$. 
The fact that there are two distinct classes of CP symmetries, unrelated by $\Delta(27)$
transformations, for each of the 6 CP conserving conditions is due to the
$\Delta(27)$ potential being accidentally invariant under
$\Delta(54)$~\cite{deMedeirosVarzielas:2011zw}.
The two classes of CP symmetries in each case 
are related to each other by a $\Delta(54)$ transformation.

\subsection{One flavour triplet of Higgs doublets}

If each component of the $\Delta(27)$ triplet is an $SU(2)_L$ doublet, the
potential is rather similar to the previous case, and in analogy with the
$A_4$ potential there is one additional invariant which is contained in
$V_0(H)$. The resulting potential reads

\bea
V_{\Delta(27)} (H) = V_0(H) 
~+~ \sum_{\alpha, \beta} \left[d \left(
h_{1 \alpha} h_{1 \beta}  h^{*2 \alpha}  h^{*3 \beta} + 
\text{cycl.} \right) +
\text{h.c.}\right]
.
\label{eq:pot27SU2}
\eea
In general the potential explicitly violates CP. It is possible to impose CP
conservation as in the previous case, as follows.

Calculating the CPIs, we see that up to a prefactor, $\mathcal{I}^{(6)}_{4,5}$ have the same form as in
Eq.~\eqref{D27CPI} for the previous potential:
\begin{equation}
 \frac{512}{315}\mathcal{I}^{(6)}_{4}=\frac{1024}{495}\mathcal{I}^{(6)}_{5}= - \left(d^3-d^{*3} \right) \left(d^3+6 d d^* s+ d^{*3}-8 s^3\right).
 \label{D27CPIHiggs}
\end{equation}
This means that the same conditions ensure CP conservation
as in the previous $\Delta(27)$ invariant potential. They are associated to CP
symmetries with the $U_i$ matrices discussed in the previous subsection, simply multiplied by
$\delta^\alpha_\beta$ acting on $SU(2)_L$ indices (similarly to Eq.~(\ref{U23H})). 

\subsection{Two flavour triplets}
\label{subsec:27-2fields}

As for the $A_4$ case, we consider the potential involving two physically different flavon
fields $\varphi$ and $\varphi'$ which both transform in the triplet representation of
$\Delta(27)$. Note that the triplet representation of $\Delta(27)$ is unique
up to complex conjugation. In addition to the invariants of each field, the
full potential contains also mixed terms. Confining ourselves to quartic
terms of the form $\varphi\,\varphi' \, \varphi^* \, \varphi'^*$ (which can
be enforced e.g.\ by $U(1)$ symmetries, such that the imposed symmetry is really $\Delta(27)\times U(1) \times U(1)'$), we obtain nine independent mixed
invariants. The resulting renormalisable potential is then given by
\bea\label{eq:pot27-2}
V_{\Delta(27)} (\varphi,\varphi') &\!\!=\!\!&
V_0 (\varphi) + V_0'(\varphi') + V_1 (\varphi, \varphi') 
  \\[2mm]
&& + \left[d \left(
\varphi_1 \varphi_1 \varphi^{*2} \varphi^{*3} + 
\text{cycl.} \right) + \text{h.c.}\right] + \left[d' \left(
\varphi'_1 \varphi'_1 \varphi'^{*2} \varphi'^{*3} + 
\text{cycl.} \right) +\text{h.c.}\right]
\notag\\[2mm]
&& +\left[ \tilde d_1  \left(
\varphi_1 \varphi'_1 \varphi^{*2} \varphi'^{*3} + 
\text{cycl.} \right) +\text{h.c.}\right] 
+ \left[ \tilde d_2  \left(
\varphi_1 \varphi'_1 \varphi^{*3} \varphi'^{*2} + 
\text{cycl.} \right) +\text{h.c.}\right]
\notag 
 \! .
\eea
Here the masses as well as the coupling constants inside $V_0$, $V_0'$ and
$V_1$ are all real (note the explicit factor of $i$ multiplying
$\tilde{s}_3$), while the couplings $d$, $d'$, $\tilde{d}_1$ and $\tilde{d}_2$
are generally complex.  

This potential explicitly violates CP, since 
several of the CPIs are non-zero as can be seen in Table \ref{ta:summary} of Section~\ref{sec:table}, but the expressions are cumbersome.
However it is possible to impose CP conservation.
For example, imposing trivial CP ($CP_0$) enforces the 4 complex coefficients $d, d',
\tilde{d}_1, \tilde{d}_2$ to be real and $\tilde s_3=0$. We have verified
explicitly that all CPIs vanish in this case.
Alternatively, imposing $U_{23}^{\varphi \varphi'}$ Eq.~(\ref{U23phiphi}) enforces $\tilde s_2=0$
and relates $\tilde{d}_1 = \tilde{d}_2^*$ as well as $d^*= d$, $d'^* = d'$,
which implies that all the CPIs vanish as expected.

\subsection{Two flavour triplets of Higgs doublets}
As earlier, a potential for two triplets of $SU(2)_L$ doublets can be obtained
by including all possible $SU(2)_L$ contractions of the fields in the
$\Delta(27)$ invariants. The only difference of this potential to earlier
Higgs potentials lies in the invariants with $d$-coefficients, out of which
only the invariants corresponding to $\tilde{d}_1$ and $\tilde{d}_2$ 
in Eq.~(\ref{eq:pot27-2}) need to
be doubled.
Therefore the potential is in this case:
\bea
 V_{\Delta(27)} (H,H') \!\!&\!\!\!\!=\!\!\!\!& \!\!V_0(H) + V_0'(H')+ V_1(H,H')+ \\
& +&\!\!\sum_{\alpha,\beta}\left[ d \left(
h_{1\alpha} h_{1\beta} h^{\ast 2 \alpha} h^{\ast 3 \beta}+
\text{cycl.} \right) + 
d' \left(
h'_{1\alpha} h'_{1\beta} h'^{\ast 2 \alpha} h'^{\ast 3 \beta} + 
\text{cycl.} \right)
+ \text{h.c.}\right]
\notag\\
&+& \!\!\sum_{\alpha,\beta} \left[
 \tilde{d}_{11}(h_{1\alpha}h^{\ast 2 \alpha}h'_{1\beta}h'^{\ast 3 \beta} +
\text{cycl.})+\tilde{d}_{12}(h_{1\alpha}h'^{\ast 3 \alpha}h'_{1\beta}h^{\ast 2 \beta} +
\text{cycl.}) + \text{h.c.} \right] \notag \\
&+&\!\! \sum_{\alpha,\beta} \left[  \tilde{d}_{21}(h_{1\alpha}h^{\ast 3 \alpha}h'_{1\beta}h'^{\ast 2 \beta}+
\text{cycl.})+\tilde{d}_{22}(h_{1\alpha}h'^{\ast 2 \alpha}h'_{1\beta}h^{\ast 3 \beta}+
\text{cycl.}) + \text{h.c.} \right] \notag
 \! .
\eea
The potential $V_{\Delta(27)}(H,H')$ is CP violating in general. Of the CPIs
calculated, cf.\ Table~\ref{ta:summary}, $\mathcal{I}^{(6)}_2, \mathcal{I}^{(6)}_3,\mathcal{I}^{(6)}_4,\mathcal{I}^{(6)}_5$ (Eqs.~(\ref{I62},\ref{I63},\ref{I64},\ref{I65})) are non-zero, but the expressions are too large to display here.

\subsection[$\Delta(54)$ invariant potentials]{$\boldsymbol{\Delta(54)}$
  invariant potentials\label{subsec:54}}
Working in the basis of~\cite{Escobar:2008vc,Ishimori:2008uc,Ishimori:2010au},
the potentials of one triplet of singlets or $SU(2)_L$ doublets are both
identical for $\Delta(27)$ and $\Delta(54)$.
The $\Delta(54)$ symmetric potential of two triplets of $SU(2)_L$ singlets is obtained
from the corresponding $\Delta(27)$ potential by imposing the constraint of
Eq.~\eqref{3n2_to_6n2}, 
\begin{equation}
 \tilde{s}_2 = \tilde{s}_3 = 0\ ,
\end{equation}
as well as
\bea
\tilde{d}_1 = \tilde{d}_2 \ ,
\eea
from which $V_{\Delta(27)} (\varphi,\varphi')$ becomes
\begin{align}
 V_{\Delta(54)} (\varphi,\varphi')=&V_0(\varphi)+V'_0(\varphi')+V_2(\varphi,\varphi')\notag\\&+\left[d \left(
\varphi_1 \varphi_1 \varphi^{*2} \varphi^{*3} + 
\text{cycl.} \right) + \text{h.c.}\right] + \left[d' \left(
\varphi'_1 \varphi'_1 \varphi'^{*2} \varphi'^{*3} + 
\text{cycl.} \right) +\text{h.c.}\right]
 \notag\\&+\tilde d_1\left[   \left(
\varphi_1 \varphi'_1 \varphi^{*2} \varphi'^{*3} + 
\text{cycl.} \right) +\left(
\varphi_1 \varphi'_1 \varphi^{*3} \varphi'^{*2} + 
\text{cycl.} \right) \right]+\text{h.c.}.
\label{11}
\end{align}
 The $\Delta(54)$ potential with two triplets does not conserve CP in general,
 as seen in Table~\ref{ta:summary}. The potential in Eq.~\eqref{11} is obtained from Eq.~\eqref{eq:pot27-2} in the
 $\tilde{s}_2 = \tilde{s}_3 = 0$, $\tilde{d}_1 = \tilde{d}_2$ limit, which
 makes it rather similar to the CP conserving
 $V_{\Delta(27)}(\varphi,\varphi')$ after imposing the $U_{23}^{\varphi
   \varphi'}$ (cf.\ Eq.~(\ref{U23phiphi})), but note that
 $V_{\Delta(54)}(\varphi,\varphi')$ does not have $\tilde{d}_1^* =
 \tilde{d}_1$, ${d}^* = {d}$ nor ${d'}^* = {d'}$. Therefore, even though CPI
 $\mathcal{I}_2^{(6)}$ vanishes,  $\mathcal{I}^{(6)}_3,\mathcal{I}^{(6)}_4,\mathcal{I}^{(6)}_5$ are non-zero. 

For the potential of two triplets of Higgs doublets, the following conditions
on the parameters arise when enlarging the symmetry to $\Delta(54)$: 
\begin{equation}
  \tilde{d}_{21}=\tilde{d}_{11} \text{ , } \quad 
\tilde{d}_{22}=\tilde{d}_{12}\text{ , }\quad
\tilde{s}_{21}=\tilde{s}_{22}=\tilde{s}_{31}=\tilde{s}_{32}=0.
\end{equation}
The potential becomes
\begin{align} 
 V_{\Delta(54)} (H,H') &=V_0(H) + V_0'(H')+ V_2(H,H')\\&+ 
  \sum_{\alpha,\beta}\left[ d \left(
h_{1\alpha} h_{1\beta} h^{\ast 2 \alpha} h^{\ast 3 \beta}+
\text{cycl.} \right) + 
d' \left(
h'_{1\alpha} h'_{1\beta} h'^{\ast 2 \alpha} h'^{\ast 3 \beta} + 
\text{cycl.} \right)
+ \text{h.c.}\right] \notag
 \\&+  \sum_{\alpha,\beta} \left[
 \tilde{d}_{11}(h_{1\alpha}h^{\ast 2 \alpha}h'_{1\beta}h'^{\ast 3 \beta} +
\text{cycl.})+\tilde{d}_{12}(h_{1\alpha}h'^{\ast 3 \alpha}h'_{1\beta}h^{\ast 2 \beta} +
\text{cycl.}) + \text{h.c.} \right]  \notag
 \\&+  \sum_{\alpha,\beta} \left[  \tilde{d}_{11}(h_{1\alpha}h^{\ast 3 \alpha}h'_{1\beta}h'^{\ast 2 \beta}+
\text{cycl.})+\tilde{d}_{12}(h_{1\alpha}h'^{\ast 2 \alpha}h'_{1\beta}h^{\ast 3 \beta}+
\text{cycl.}) + \text{h.c.} \right] \! . ~~ \notag
\end{align}
This potential is also generally CP violating and $\mathcal{I}^{(6)}_3,\mathcal{I}^{(6)}_4,\mathcal{I}^{(6)}_5$ are non-zero but too large to display here.

\section{$\bs{\Delta(3n^2)}$ invariant potentials with $\bs{n>3}$ \label{sec:D3n2}}
\cleqn
So far we have considered the finite groups $A_4=\Delta(3\cdot
2^2)$ and $\Delta(27)=\Delta(3\cdot 3^2)$ which correspond to the first two
non-Abelian members of the series $\Delta(3n^2)$ with  $n\in \mathbb N$. 
In this section we derive renormalisable potentials which are invariant
under $\Delta(3n^2)$ with $n>3$. The field contents considered are a single triplet of SM singlets, then one triplet of $SU(2)_L$ doublets, then two triplets of SM singlets and finally two triplets of $SU(2)_L$ doublets. Following~\cite{Luhn:2007uq}, a triplet of
$\Delta(3n^2)$ can be written as ${\bf 3}_{(k,l)}$, where
$k,l=0,1,...,n-1$. The complex conjugate of ${\bf 3}_{(k,l)}$ is given by ${\bf 3}_{(-k,-l)}$, which we sometimes
denote as~${\bf \ol 3}$, dropping the indices. The
cyclic permutation symmetry included in $\Delta(3n^2)$ entails an ambiguity in 
labelling the {\it {same}} triplet representation such that 
${\bf 3}_{(k,l)}  = {\bf 3}_{(l,-k-l)} = {\bf 3}_{(-k-l,k)}$.
With these preliminary remarks, we can determine the product of two identical
triplet representations~\cite{Luhn:2007uq}
\be
{\bf 3}_{(k,l)} \otimes {\bf 3}_{(k,l)} ~=~ 
[{\bf 3}_{(2k,2l)} + {\bf 3}_{(-k,-l)}]_s + [{\bf 3}_{(-k,-l)}]_a \ .
\label{eq:kronecker3n2}
\ee
Again the subscripts $s$ and $a$ denote symmetric and antisymmetric
combinations. Assuming the original triplet ${\bf 3}_{(k,l)}$ to be a faithful
(and thus irreducible) representation of $\Delta(3n^2)$, all representations
on the right-hand side are irreducible for $n\neq 2$. Excluding moreover the
case with $n=3$, the triplets ${\bf 3}_{(2k,2l)}$ and ${\bf 3}_{(-k,-l)}$
denote different representations.
Throughout this section we adopt the basis of~\cite{Luhn:2007uq,Ishimori:2010au}.

\subsection{One flavour triplet}
With one triplet field, only the symmetric part of
Eq.~\eqref{eq:kronecker3n2} is relevant for constructing quartic terms of the
form ${\bf 3\otimes 3 \otimes \ol 3 \otimes \ol 3}$. 
Considering $n>3$, the two triplets in the symmetric contraction of
Eq.~\eqref{eq:kronecker3n2} are distinct, so that only two independent quartic
invariants can be constructed. The renormalisable scalar potential, which is
additionally invariant under a $U(1)$ symmetry, thus takes the form
\bea
V_{\Delta(3n^2)} (\varphi) &=&V_0 (\varphi)
 \ ,
\label{eq:pot3n2}
\eea
where the explicit form of $V_0(\varphi)$ is given in Eq.~\eqref{eq:potV0}. 
This potential always explicitly conserves CP.
It is a reduced version of the $A_4$ symmetric potential $V_{A_4} (\varphi)$
of Eq.~\eqref{eq:potA4} which generally conserves CP. Therefore it is clear that
$V_{\Delta(3n^2)} (\varphi)$ is left invariant under the same CP symmetry,
i.e.~the one defined with a 2-3 swap, $U_{23}$. In addition, $V_{\Delta(3n^2)}
(\varphi)$ respects the trivial CP symmetry $CP_0$ (which $V_{A_4} (\varphi)$ in general does not).

\subsection{One flavour triplet of Higgs doublets}

If each component of the faithful $\Delta(3n^2)$ triplet transforms as an $SU(2)_L$
doublet, the corresponding renormalisable potential consists of four
independent terms. As described in Section~\ref{sec:A4}, the different ways of
contracting the $SU(2)_L$ indices entail a doubling of the $\Delta(3n^2)$
invariant term in Eq.~\eqref{eq:pot3n2} which is proportional to $r$. The
resulting Higgs potential then takes the form
\bea
V_{\Delta(3n^2)} (H) = V_{0} (H)  \ ,
\label{eq:pot3n2SU2}
\eea
with the right-hand side defined in Eq.~\eqref{eq:potV0H}. 
This potential always conserves CP explicitly (for any choice of parameters).
Similar to the
corresponding $A_4$ case, $V_{\Delta(3n^2)} (H)$ is left invariant under a CP
transformation with a 2-3 swap. Additionally, it also respects the trivial CP symmetry $CP_0$.

\subsection{Two flavour triplets}

We now turn to the case of two flavour multiplets, $\varphi$ and $\varphi'$,  in
the same faithful triplet representation. The potential can be simplified by
imposing individual $U(1)$ symmetries for each of the scalar fields, such that
the actual symmetry of the potential is given by $\Delta(3n^2)\times U(1)
\times U(1)'$. In addition to the potential of the individual
(non-interacting) fields, only mixed terms of the form  $\varphi\,\varphi' \, 
\varphi^* \, \varphi'^*$ are possible; in particular cubic terms are absent.
In order to construct the mixed quartic terms, we consider the Kronecker product
given in Eq.~\eqref{eq:kronecker3n2}, now also including the antisymmetric
combination. Multiplying the right-hand side with its complex conjugate, we
see that there are five independent mixed quartic $\Delta(3n^2)$ invariants if
$n>3$. The renormalisable potential can be written as follows,
\bea
V(\varphi,\varphi')_{\Delta(3n^2)} &=&
V_0 (\varphi) + V_0'(\varphi') + V_1 (\varphi, \varphi') \ ,
\label{eq:potDelta3n2-2b}
\eea
where the individual contributions to the right-hand side are defined in
Eqs.~(\ref{eq:potV0}) and (\ref{eq:potV1}). 

Unlike the previous $\Delta(3n^2)$ invariant potentials for $n>3$, this potential generally violates CP, as confirmed by the non-zero CPI $\mathcal{I}^{(6)}_2$ (Eq.~\eqref{I62}) which for this potential becomes
\be
\mathcal{I}^{(6)}_2 =  \frac{3}{512} i \tilde{s}_2 \tilde{s}_3 (-3 \tilde{r}_2^2 + \tilde{s}_3^2) (-\tilde{s}_1^2 + \tilde{s}_1 \tilde{s}_2 + 
   \tilde{r}_2 (-2 \tilde{s}_1 + \tilde{s}_2) + \tilde{s}_3^2)\ .
\label{eq:inv3n2-2flavons}
\ee
Imposing the trivial CP symmetry $CP_0$ entails $\tilde s_3=0$, whereas the $U_{23}^{\varphi \varphi'}$ 2-3 swap
CP symmetry constrains the potential such that $\tilde s_2=0$. 
As expected, both CP symmetries enforce $\mathcal{I}^{(6)}_2=0$ 
(and make any other CPIs vanish), but they are distinct CP symmetries with 
distinct effects on the potential.

Inspection of other CPIs reveals that also the factor $(-3 \tilde{r}_2^2 +
\tilde{s}_3^2)$ is present in each non-vanishing CPI we found.
This raises the question if there exists a CP symmetry which is associated
with setting this factor to zero. Such a symmetry must relate different terms
of the potential in Eq.~\eqref{eq:potDelta3n2-2b}, namely
$$
 \tilde r_2\Big( \sum_i \varphi_i \varphi'^{*i} \Big)
\Big( \sum_j \varphi'_j \varphi^{*j} \Big)
\,+~ i \, \tilde s_3 
\Big[
(\varphi_1 \varphi'^{*1} \varphi'_2 \varphi^{*2} + \text{cycl.}) 
- 
( \varphi^{*1}\varphi'_1  \varphi'^{*2} \varphi_2 +\text{cycl.})
\Big] \ .
$$
Clearly, the term proportional to $\tilde r_2$ is invariant under a
general CP transformation where the unitary matrix $U$ is block diagonal and the blocks are the same for both
triplets $\varphi$ and $\varphi'$. Hence, we are led to more general choices
with different $3\times 3$ blocks $U_{\varphi}$ and $U_{\varphi'}$ for
$\varphi$  and $\varphi'$, respectively.  Pursuing the simple ansatz 
\be
\label{eq:3n2-Uphiphi'}
U^{\varphi\varphi'} = \begin{pmatrix} U_\varphi & 0 \\0&U_{\varphi'}
\end{pmatrix}
, ~\quad \mathrm{with} ~\quad
U_{\varphi}=\begin{pmatrix} 1&0&0\\
0&\omega&0\\
0&0&\omega^2 \end{pmatrix}
 , \quad
U_{\varphi'}=\begin{pmatrix} 1&0&0\\
0&\omega^2&0\\
0&0&\omega \end{pmatrix},
\ee
we find that the potential remains invariant under the corresponding
general CP transformation if and only if
\be
\tilde s_3 ~=~ \tilde r_2 ~ i (\omega-\omega^2) \ .
\ee
Inserting $\omega=e^{2\pi i/3}$, we get $\tilde s_3 = -\sqrt{3} \tilde r_2$
which corresponds to one solution of the quadratic equation 
$3\tilde r_2^2 -\tilde s_3^2=0$. The other solution, $\tilde s_3 = \sqrt{3}
\tilde r_2$, is related to the CP transformation where the roles of the
explicit matrices in Eq.~\eqref{eq:3n2-Uphiphi'} are exchanged.
Imposing either of the two CP symmetries guarantees that all CPIs vanish.

An example of a larger non-trivial CPI is provided by $\mathcal{I}^{(7,2)}_1$, defined
in Eq.~\eqref{eq:deltaCPoddinv} of Appendix~\ref{app:larger}. Explicit
evaluation in the parametrisation of Eq.~\eqref{eq:potDelta3n2-2b} yields
\bea\nonumber
\mathcal I^{(7,2)}_1  &\!\!=\!\!&
\frac{9}{8192} i \tilde s_2 \tilde s_3
\big(3\tilde r_2^2 -\tilde s_3^2\big) \big(m_\varphi^2-m_{\varphi'}^2\big)^2 
\big(\tilde r_1 +\tilde r_2 +\tilde s_1\big)  \times  \Big[
16(s^2+s s'+ {s'}^2)+\\
&& 
+8r(2s+s')+8r'(2s'+s)
+\tilde s_1^2 +\tilde s_2^2-\tilde s_3^2
-\tilde s_1\tilde s_2
+\tilde r_2( 2 \tilde s_1 - \tilde s_2 )
\Big] \ . 
\eea
While this more complicated CPI vanishes for $m_\varphi=m_{\varphi'}$, we already know that such
a relation is not a consequence of any CP symmetry because the simpler CPI
derived above does not depend on the masses. In other words, any CP symmetry
that would relate the masses by $m_\varphi=m_{\varphi'}$ would have to impose additional
constraints on the other parameters of the potential. 

Having identified the CP symmetries corresponding to the zeros of 
$\tilde s_2 \tilde s_3 \big(3\tilde r_2^2 -\tilde s_3^2\big)$, one may wonder
about the consequences of imposing other CP symmetries on the
potential of Eq.~\eqref{eq:potDelta3n2-2b}. As an example, one could for
instance consider the situation where $U$ is given by the block matrix where $U_\varphi$ and $U_\varphi'$ are both given by one of the matrices of
Eq.~\eqref{eq:notsomagic}. A straightforward but somewhat tedious calculation
reveals that such a ``general'' CP symmetry would require vanishing
coefficients for all non-$SU(3)$ type terms. In other words $s=s'=\tilde
s_1=\tilde s_2=\tilde s_3=0$. The symmetry of the resulting potential would
therefore be enhanced from $\Delta(3n^2)$ to $SU(3)$ in addition to preserving CP.

\subsection{Two flavour triplets of Higgs doublets}

The potential of two triplets of $SU(2)_L$ doublets can be deduced from the
potential of two flavour triplets of $SU(2)_L$ singlets.
It is a particular case of the corresponding $A_4$ potential. 
We therefore write the potential in terms of the expressions 
defined in Eqs.~(\ref{eq:potV0H}) and (\ref{eq:potV1H}),
\bea
 V_{\Delta(3n^2)} (H,H') &= V_0(H) + V_0'(H')+ V_1(H,H')\ .\label{eq:pot3n2_2H}
\eea
We note again that due to the $SU(2)_L\times U(1)_Y$ gauge group, the
potential cannot contain any cubic terms. In fact, each term must have an
equal number of Higgs and complex conjugate Higgs fields. Hence it is
sufficient to impose e.g.\ a $Z_3$ symmetry with non-trivial charge for only one
of the two triplets of Higgs doublets in order to enforce the potential of Eq.~\eqref{eq:pot3n2_2H}.
This potential in Eq.~\eqref{eq:pot3n2_2H} generally violates CP explicitly.
Of the CP-odd invariants calculated, cf.\ Table \ref{ta:summary}, $\mathcal{I}^{(6)}_2,\mathcal{I}^{(6)}_3,\mathcal{I}^{(6)}_4,\mathcal{I}^{(6)}_5$ (Eqs.~(\ref{I62},\ref{I63},\ref{I64},\ref{I65})) are non-zero, but the expressions are too large to display here.

\subsection[${\Delta(6n^2)}$ invariant potentials with $n>3$]{$\bs{\Delta(6n^2)}$
 invariant potentials with $\bs{n>3}$}

Working in the basis of~\cite{Escobar:2008vc,Ishimori:2010au}, it is straightforward to
enhance the symmetry of $\Delta(3n^2)$ invariant potentials to
$\Delta(6n^2)$ by imposing extra constraints, see
Appendix~\ref{6n2_from_3n2}. With only one flavour triplet $\varphi$ or $H$,
the renormalisable potentials are automatically symmetric under
$\Delta(6n^2)$, i.e. 
\be
V_{\Delta(6n^2)} (\varphi)= V_{\Delta(3n^2)}(\varphi) = V_0(\varphi) \ ,\qquad~~
V_{\Delta(6n^2)} (H)= V_{\Delta(3n^2)}(H) = V_0(H) \ ,
\ee
where $V_0(\varphi)$ and $V_0(H)$ are defined in
Eqs.~(\ref{eq:potV0}) and (\ref{eq:potV0H}), respectively.
With two flavour triplets, it is necessary to impose $\tilde s_2=\tilde s_3=0$
for $V_{\Delta(6n^2)}(\varphi,\varphi')$ and $\tilde s_{21}=\tilde
s_{22}=\tilde s_{31}=\tilde s_{32}=0$ for $V_{\Delta(6n^2)}(H,H')$. Using the
definitions of Eqs.~(\ref{eq:potV2}) and (\ref{eq:potV2H}), we then have
\bea
V_{\Delta(6n^2)} (\varphi,\varphi')&=& V_0(\varphi)  + V'_0(\varphi')  + V_2(\varphi,\varphi')  \ ,\\[2mm]
V_{\Delta(6n^2)} (H,H')&=& V_0(H) + V'_0(H') + V_2(H,H')\ .
\eea
All of the above $\Delta(6n^2)$ invariant potentials (with $n>3$) conserve CP
explicitly. For instance, one can easily show that the respective trivial CP transformations $CP_0$ as
well as the respective CP transformations with a 2-3 swap ($U_{23}$, $U_{23}^H$, $U_{23}^{\varphi \varphi'}$, $U_{23}^{HH'}$) do not constrain the parameters of
the potentials as they are all real.

\section{Summary of CPIs for explicit CP violation\label{sec:table}}
 
\begin{table}[t]
\centering
\begin{tabular}{|c||c|c|c|c||c|}
\hline  & $\mathcal{I}^{(6)}_{2}$&$\mathcal{I}^{(6)}_{3}$&$\mathcal{I}^{(6)}_{4}$&$\mathcal{I}^{(6)}_{5}$ & CP \\\hline\hline
 
 $({\bf3}_{A_4},{\bf1}_{SU(2)_L})$  
& 0 & 0 & 0 & 0 & Eq.~(\ref{U23}) \\\hline
 $({\bf3}_{A_4},{\bf2}_{SU(2)_L})$ & 0 & 0 & 0 & 0 & Eq.~(\ref{U23H})\\\hline
$2\times({\bf3}_{A_4},{\bf1}_{SU(2)_L})$  
& * & * & * & * & NA\\\hline
 $2\times({\bf3}_{A_4},{\bf2}_{SU(2)_L})$ & * & * & * & * & NA\\\hline\hline
 
$({\bf3}_{\Delta(27)},{\bf1}_{SU(2)_L})$  
& 0 & 0 & Eq.~(\ref{D27CPI}) & Eq.~(\ref{D27CPI}) & NA\\\hline
 $({\bf3}_{\Delta(27)},{\bf2}_{SU(2)_L})$ & 0 & 0 & Eq.~(\ref{D27CPIHiggs}) & Eq.~(\ref{D27CPIHiggs}) & NA \\\hline
$2\times({\bf3}_{\Delta(27)},{\bf1}_{SU(2)_L})$ 
& * & * & * & * & NA \\\hline
 $2\times({\bf3}_{\Delta(27)},{\bf2}_{SU(2)_L})$ & * & * & * & * & NA\\\hline\hline
 
$({\bf3}_{\Delta(3n^2)},{\bf1}_{SU(2)_L})$
& 0 & 0 & 0 & 0 & Eq.~(\ref{U23}) \\\hline
 $({\bf3}_{\Delta(3n^2)},{\bf2}_{SU(2)_L})$ & 0 & 0 & 0 & 0 & Eq.~(\ref{U23H}) \\\hline
$2\times({\bf3}_{\Delta(3n^2)},{\bf1}_{SU(2)_L})$ 
& Eq.~(\ref{eq:inv3n2-2flavons})  & * & * & * & NA \\\hline
 $2\times({\bf3}_{\Delta(3n^2)},{\bf2}_{SU(2)_L})$ & * & * & * & * & NA \\\hline\hline
 
$({\bf3}_{S_4},{\bf1}_{SU(2)_L})$
& 0 & 0 & 0 & 0 & $CP_0$\:\& Eq.~(\ref{U23}) \\\hline
 $({\bf3}_{S_4},{\bf2}_{SU(2)_L})$ & 0 & 0 & 0 & 0 &  $CP_0$\:\& Eq.~(\ref{U23H}) \\\hline
$2\times({\bf3}_{S_4},{\bf1}_{SU(2)_L})$ 
& 0 & 0 & 0 & 0 &  $CP_0$\:\& Eq.~(\ref{U23phiphi}) \\\hline
 $2\times({\bf3}_{S_4},{\bf2}_{SU(2)_L})$ & 0 & 0 & 0 & 0 &  $CP_0$\:\& Eq.~\eqref{U23HH} \\\hline\hline
 
$({\bf3}_{\Delta(54)},{\bf1}_{SU(2)_L})$
& 0 & 0 & * & * & NA \\\hline
 $({\bf3}_{\Delta(54)},{\bf2}_{SU(2)_L})$ & 0 & 0 & * & * & NA \\\hline
$2\times({\bf3}_{\Delta(54)},{\bf1}_{SU(2)_L})$ 
& 0 & * & * & * &  NA \\\hline
 $2\times({\bf3}_{\Delta(54)},{\bf2}_{SU(2)_L})$ & 0 & * & * & * & NA \\\hline\hline
 
$({\bf3}_{\Delta(6n^2)},{\bf1}_{SU(2)_L})$
& 0 & 0 & 0 & 0 &  $CP_0$\:\& Eq.~(\ref{U23}) \\\hline
 $({\bf3}_{\Delta(6n^2)},{\bf2}_{SU(2)_L})$ & 0 & 0 & 0 & 0 &  $CP_0$\:\& Eq.~(\ref{U23H}) \\\hline
$2\times({\bf3}_{\Delta(6n^2)},{\bf1}_{SU(2)_L})$
& 0 & 0 & 0 & 0 &  $CP_0$ \:\& Eq.~(\ref{U23phiphi})\\\hline
 $2\times({\bf3}_{\Delta(6n^2)},{\bf2}_{SU(2)_L})$ & 0 & 0 & 0 & 0 &  $CP_0$ \:\& Eq.~\eqref{U23HH} \\\hline
\end{tabular}
\caption{Summary of CPIs and (if applicable) CP symmetry transformations
    for scalar potentials with discrete symmetry.}  \label{ta:summary}
\end{table}
 
In this section, we collect our results of Sections~\ref{sec:A4},~\ref{sec:D27}
and~\ref{sec:D3n2}. We have calculated CPIs for a number of different
potentials which are invariant under either of the following discrete
symmetries $A_4$, $S_4$, $\Delta(27)$, $\Delta(54)$, $\Delta(3n^2)$ and
$\Delta(6n^2)$ with $n>3$.  All these symmetries have irreducible triplet
representations. Choosing Higgs fields in a faithful triplet, we have
determined the potential for one triplet of $SU(2)_L$ singlets, one triplet of
$SU(2)_L$ doublets, two triplets of $SU(2)_L$ singlets and finally two triplets of
$SU(2)_L$ doublets. The (scalar) particle content for each of these 
$6\times 4$ cases is listed intuitively in the leftmost column of Table~\ref{ta:summary}. 

Many of the CPIs defined in Section~\ref{CPoddI} vanish for all of these 24
potentials. We have checked explicitly that $\mathcal{I}^{(2,2)}_{1}$,
$\mathcal{I}^{(3,1)}_{1}$, $\mathcal{I}^{(3,1)}_{2}$,
$\mathcal{I}^{(3,2)}_{1}$, $\mathcal{I}^{(3,2)}_{2}$,
$\mathcal{I}^{(4,1)}_{1}$, $\mathcal{I}^{(5)}_{1}$, $\mathcal{I}^{(5)}_{2}$,
$\mathcal{I}^{(5)}_{3}$, $\mathcal{I}^{(6)}_{1}$ vanish in all
cases. Table~\ref{ta:summary} shows the relevant invariants 
$\mathcal{I}^{(6)}_{2}$, 
$\mathcal{I}^{(6)}_{3}$, 
$\mathcal{I}^{(6)}_{4}$,
$\mathcal{I}^{(6)}_{5}$, evaluated for each potential. 
A 0-entry means that the corresponding CPI was found to be zero. A
non-vanishing CPI is indicated by either an asterisk or an equation number,
where the latter refers to the position in our paper where the corresponding
expression for the CPI is given. The asterisk is used for non-zero CPIs which we
have calculated analytically but whose expressions are too large to display in the text.

We observe from  Table~\ref{ta:summary} that 12 potentials feature explicit CP
violation. On the other hand, all four CPIs shown in the table vanish for
the other 12 potentials, which suggests CP is conserved in those cases. Indeed,
as listed in the rightmost column, one can easily identify CP transformations
which leave the potential unchanged, thereby explicitly proving that CP is
conserved. We recall that trivial CP ($CP_0$) means complex conjugation on all
scalar fields, cf.\ Eq.~\eqref{CP0}. 
``NA'' stands for ``Not Applicable'' and is used for CP violating cases. 
 
\section{CP-odd invariants for spontaneous CP violation}
\label{spontCPI}
\cleqn

So far we have discussed CPIs that signal explicit CP violation in scalar
potentials. It is also useful to consider CPIs that indicate the presence of
spontaneous CP violation. In order to extend our formalism  (which is
applicable to any potentials once translated into the standard form) we need to
include also VEVs. 

Recall that VEVs transform as vectors under basis transformations,
cf. Eqs.~(\ref{eq:basi}) and~(\ref{eq:basi*}):
\begin{equation}
 v_a~\mapsto~ V_a^{a'}v_{a'}\ ,
\end{equation}
\begin{equation}
 {v^\ast}^a~\mapsto ~v^{\ast a'} V^{\dagger a}_{\ a'} \ .
\end{equation}
When used in invariants, first, if the potential does not contain trilinear couplings, VEVs can only appear in pairs of $v$ and corresponding $v^\ast$ because otherwise indices would remain uncontracted. Furthermore, all VEVs commute and thus can be combined into one large tensor,
\begin{equation}
 W^{w_1\ldots w_{n_v}}_{w'_1\ldots w'_{n_v}}=v_{w'_1}\ldots v_{w'_{n_v}}v^{\ast w_1}\ldots v^{\ast w_{n_v}}.
 \label{W_definition}
\end{equation}
where $n_v$ is the number of $v$, $v^\ast$  pairs.\footnote{In~\cite{Davidson:2005cw}, VEVs are always assigned in pairs to matrices $V^a_b=v^{\ast a}v_b$, however, since all VEVs commute, even for four or more VEVs, also all $V^a_b$ commute and can be summarised in one large totally symmetric tensor.} Using $W$, all invariants with $n_v$ pairs of VEV and conjugated VEV can be written using
\begin{align}
J_\sigma^{(n_v,m_Y,n_Z)}&\equiv W^{w_1\ldots w_{n_v}}_{\sigma(w_1)\ldots \sigma(w_{n_v})}Y^{a_1}_{\sigma(a_1)}\ldots Y^{a_{m_Y}}_{\sigma(a_{m_Y})}Z^{b_1 b_2}_{\sigma(b_1)\sigma(b_2)}\ldots Z^{b_{2n_Z-1} b_{2n_Z}}_{\sigma(b_{2n_Z-1})\sigma(b_{2n_Z})} \nonumber\\&\mapsto 
W^{\sigma(w_1)\ldots \sigma(w_{n_v})}_{w_1\ldots
  w_{n_v}}Y^{\sigma(a_1)}_{a_1}\ldots
Y^{\sigma(a_{m_Y})}_{a_{m_Y}}Z^{\sigma(b_1)\sigma(b_2)}_{b_1 b_2}\ldots
Z^{\sigma(b_{2n_Z-1})\sigma(b_{2n_Z})}_{b_{2n_Z-1} b_{2n_Z}}\equiv  (J_\sigma^{(n_v,m_Y,n_Z)})^{\ast}\ ,
\label{spont_CP_of_invariant}
\end{align}
with $\sigma \in S_{n_{v}+m_Y+2n_Z}$.
When drawing diagrams, there are additional rules for contractions with VEVs, again with $X=Y,Z$:
\begin{equation}
 X^{a..}_{..}v_a=\vcenter{\hbox{\includegraphics[scale=0.2]{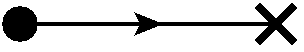}}}
\end{equation}
and
\begin{equation}
 X_{a..}^{..}v^{\ast a}=\vcenter{\hbox{\includegraphics[scale=0.2]{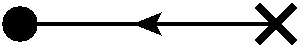}}}
\end{equation}
Invariants containing only $Y$ tensors and VEVs are always CP-even. The smallest examples of CPIs for spontaneous CP violation built from $Z$ tensors and VEVs are
\begin{equation}
J_{1}^{(2,2)}\equiv Z^{a_1a_2}_{a_1a_3}Z^{a_3a_4}_{a_5a_6}v_{a_2}v_{a_4}v^{\ast a_5}v^{\ast a_6}=\vcenter{\hbox{\includegraphics[scale=0.2]{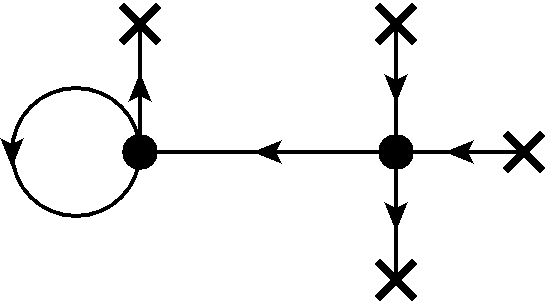}}}
\label{SI2Z2v1_short_definition}
\end{equation}
\begin{equation}
J_{1}^{(3,1)}\equiv Z^{a_1a_2}_{a_5a_6}Z^{a_3a_4}_{a_1a_3}Z^{a_5a_6}_{a_2 a_7}v_{a_4}v^{\ast a_7}=\vcenter{\hbox{\includegraphics[scale=0.2]{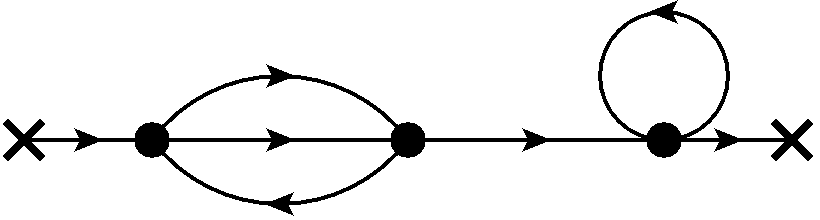}}}
\end{equation}
\begin{equation}
J_{2}^{(3,1)}\equiv Z^{a_1a_2}_{a_1a_5}Z^{a_3a_4}_{a_3a_6}Z^{a_5a_6}_{a_2 a_7}v_{a_4}v^{\ast a_7}=\vcenter{\hbox{\includegraphics[scale=0.2]{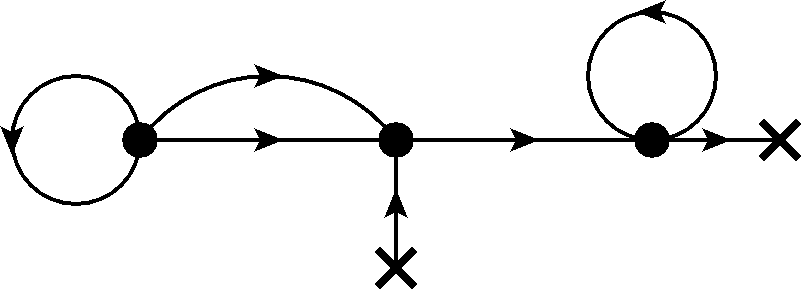}}}
\end{equation}
where the superscripts on  $J$ indicate the number of $Z$ tensors and pairs of VEVs in
the invariant. A complete search for invariants with $(n_Z,n_v)=(2,2),(3,1),(3,2),(4,1)$ was performed. The method is explained in Appendix \ref{sym_invs} and the invariants not given in the main text are listed in Appendix \ref{SCPI_list}.

\subsection{Minimisation condition in terms of diagrams}
The minima of the a potential written as in Eq.~(\ref{eq:potCPgeneral}) fulfil 
\begin{equation}
 0=\frac{\partial V }{\partial \phi_e}=\phi^{\ast a}Y^e_a+2 \phi^{\ast
   a}\phi^{\ast c}Z^{ed}_{ac}\phi_d\ ,
\end{equation}
and
\begin{equation}
 0=\frac{\partial V}{\partial \phi^{\ast e}}=Y^b_e\phi_b+2 \phi^{\ast
   c}Z^{bd}_{ec}\phi_b\phi_d\ ,
\end{equation}
where the factor of 2 appears because of the symmetry of $Z^{bd}_{ac}$ under $b\leftrightarrow d$ and $a\leftrightarrow c$. Replacing the fields by their VEVs, these minimisation conditions can be expressed in terms of diagrams:
\begin{equation}
 0=\vcenter{\hbox{\includegraphics[scale=0.2]{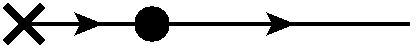}}}+2\text{ } \vcenter{\hbox{\includegraphics[scale=0.2]{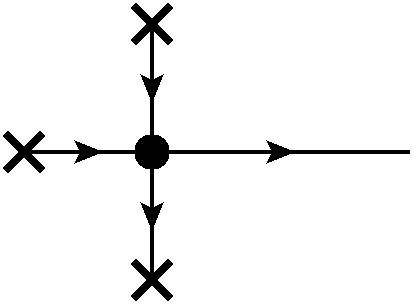}}}
 \label{minimization_condition}
\end{equation}
and
\begin{equation}
 0=\vcenter{\hbox{\includegraphics[scale=0.2]{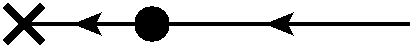}}}+2\text{ } \vcenter{\hbox{\includegraphics[scale=0.2]{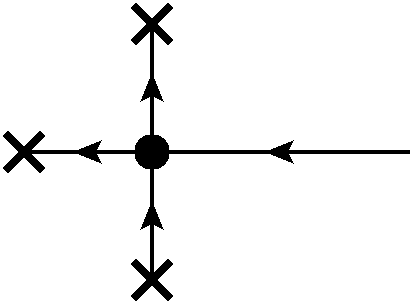}}}
\end{equation}
This can be used later to simplify CPIs, as can be seen by applying
\begin{equation}
 \vcenter{\hbox{\includegraphics[scale=0.2]{EOM1.png}}}=-2\text{ } \vcenter{\hbox{\includegraphics[scale=0.2]{EOM2.png}}}
 \label{EOM3}
\end{equation}
in Eq.~\eqref{SI2Z2v1_short_definition}.
Using the minimisation condition Eq.~(\ref{EOM3}), the invariant
$J_{1}^{(2,2)}$ can be simplified to\footnote{The resulting expression
  corresponds to the invariant $J_3$ in Eq.~(26) of~\cite{Botella:1994cs}.}
\begin{equation}
 J_{1}^{(2,2)}\equiv -\frac{1}{2}\vcenter{\hbox{\includegraphics[scale=0.2]{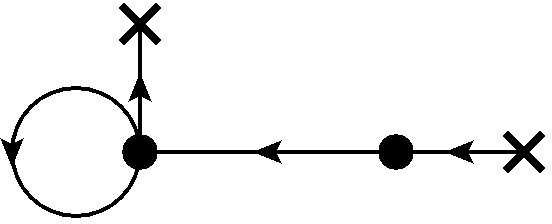}}}
\end{equation}
This can only be CP-odd if $Y$ is not proportional to the identity. One can now search for more complicated invariants built from $Z$ tensors and VEVs that will not simplify like this. 
The smallest CPIs for spontaneous CP violation without self-loops which also cannot be simplified using the minimisation condition for $n_Z=3,4$ respectively are
\begin{equation}
J_{1}^{(3,2)}\equiv Z^{a_1a_2}_{a_4a_5}Z^{a_3a_4}_{a_2a_6}Z^{a_5a_6}_{a_7a_8}v_{a_1}v_{a_3}v^{\ast a_7}v^{\ast a_8}=\vcenter{\hbox{\includegraphics[scale=0.2]{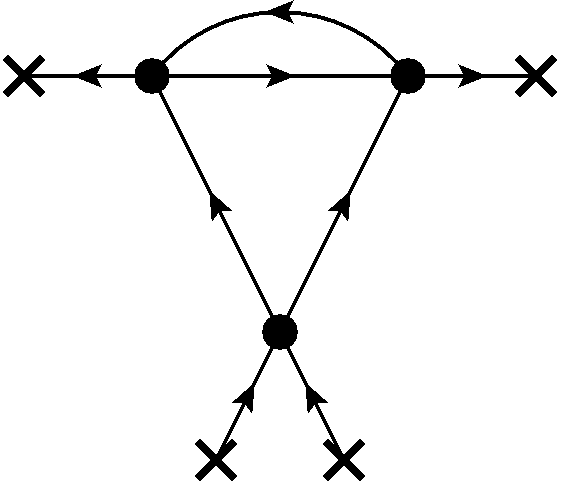}}}
\label{J322}
\end{equation}
and
\begin{equation}
J_{1}^{(4,1)}\equiv Z^{a_1a_2}_{a_3a_5}Z^{a_3a_4}_{a_7a_8}Z^{a_5a_6}_{a_1 a_4}Z^{a_7a_8}_{a_2a_9}v_{a_6}v^{\ast a_9}=\vcenter{\hbox{\includegraphics[scale=0.2]{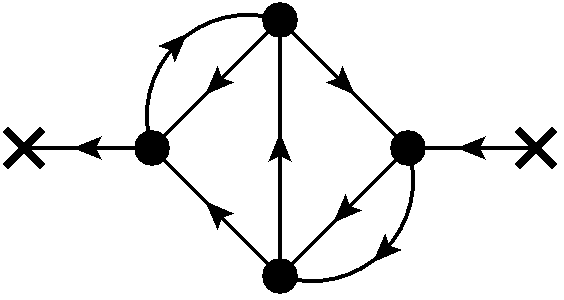}}}.
\end{equation}

\subsection{Example applications}
\subsubsection[One triplet of $A_4$]{One triplet of $\bs{A_4}$}
As we have seen, the potential in Eq.~\eqref{eq:potA4SU2} conserves CP explicitly.
By an analysis of all VEVs, it has been shown~\cite{Ivanov:2014doa} that CP cannot be spontaneously broken.
Using our approach we have verified that the low order invariants vanish.
In particular, 
all spontaneous invariants up to $n_Z=3,n_{v}=2$ are found to vanish for this potential.
\subsubsection[One triplet of $\Delta(27)$]{One triplet of $\bs{\Delta(27)}$}
One can now calculate SCPIs for this potential for arbitrary VEVs and the smallest 
 non-zero SCPI found is $\mathcal{J}_1^{(3,2)}$, as defined via Eq.\ (\ref{J322}).
For the general potential $V_{\Delta(27)}(\varphi)$ (which we note is CP
violating), it takes the value
\bea\nonumber 
  \mathcal{J}_1^{(3,2)}& =& \frac{1}{4}(d^{\ast 3}-d^3)(|v_1|^4+|v_2|^4+|v_3|^4-2|v_1|^2|v_2|^2-2|v_1|^2|v_3|^2-2|v_2|^2|v_3|^2)
  \\&&+\frac{1}{2}(dd^{\ast 2}-2d^\ast s^2+d^{2}s)(v_2 v_3 v_1^{\ast 2}+v_1
  v_3 v_2^{\ast2}+v_1 v_2 v_3^{\ast2}) \nonumber\\
&&-\frac{1}{2}(d^2d^\ast-2ds^2+d^{\ast 2}s)(v_2^\ast v_3^\ast
 v_1^2+v_1^\ast v_3^\ast v_2^2+v_1^\ast v_2^\ast v_3^2)   \ .\label{eq:J27ex2}
   \eea
In order to demonstrate the usefulness of SCPIs, let us consider the following special cases of $V_{\Delta(27)} (\varphi)$ where we impose different CP symmetries.
We start by considering trivial CP ($CP_0$), which in this case is the $U_0$ matrix, forcing $\text{Arg}(d)=0$ which simplifies the SCPI expression to
\begin{equation}
  \mathcal{J}_1^{(3,2)} = \frac{1}{2}(d^3 -2d s^2+d^{2}s) \left[(v_2 v_3 v_1^{\ast 2}+v_1 v_3 v_2^{\ast2}+v_1 v_2 v_3^{\ast2})-(v_2^\ast v_3^\ast v_1^2+v_1^\ast v_3^\ast v_2^2+v_1^\ast v_2^\ast v_3^2) \right]  \!.\,
   \end{equation}
It is known~\cite{Branco:1983tn, deMedeirosVarzielas:2011zw, Varzielas:2012nn, Bhattacharyya:2012pi}  that the complex VEV $(1,\omega,\omega^2)$ is not CP violating when starting with trivial CP. This can be confirmed easily by using the SCPI above. Instead, the geometrically CP violating VEV $(\omega,1,1)$ does give non-zero when plugged into the SCPI.
Let us consider now the CP symmetry $U_3$, forcing $\text{Arg}(d)=2 \pi/3$. Because $d$ remains complex, even a real VEV like $(1,1,1)$ spontaneously violates CP~\cite{Fallbacher:2015rea} and this is shown by the SCPI:
\begin{equation}
  \mathcal{J}_1^{(3,2)} = \frac{1}{2} \mathrm{Im} (dd^{\ast 2}-2d^\ast s^2+d^{2}s) \left[(3 v_1^4) \right]  .
   \end{equation}
Another interesting case is the CP symmetry $U_4$, forcing $2 s=(d+d^*)=2
\mathrm{Re} (d)$. This simplifies Eq.~\eqref{eq:J27ex2} to
\begin{equation}
  \mathcal{J}_1^{(3,2)} = \frac{1}{4}(d^{\ast
    3}-d^3)(|v_1|^4+|v_2|^4+|v_3|^4-2|v_1|^2|v_2|^2-2|v_1|^2|v_3|^2-2|v_2|^2|v_3|^2)\ .
   \end{equation}
It is interesting that in this case the SCPI indicates that  spontaneous CP violation is independent of the phases of the VEV. Indeed, the known VEVs for the $U_0$ symmetric potential, such as $(0,0,1)$, $(1,1,1)$ (which are real) and $(\omega,1,1)$ are still candidate VEVs of the $U_4$ symmetric potential and all violate CP spontaneously, as indicated by the SCPI.

\section{Conclusions \label{sec:conc}}
\cleqn

In this paper we have been concerned with CPV arising from 
scalar potentials which go beyond the one Higgs doublet of the SM.
After reviewing the well-known general technique of constructing CPIs, we
    have introduced and developed powerful tools based on diagrams and
    contraction matrices that allow to systematically identify CPIs in an
    efficient and straightforward way. 
Such CPIs, which are valid for any scalar potential, provide a reliable
indicator for whether CP is explicitly violated by the parameters of the
potential. We have also extended our formalism to construct the spontaneous
CPIs involving the VEVs, in order to reliably determine whether CP is
spontaneously violated.

In order to illustrate the usefulness of the CPI approach, 
we then applied our results to multi-Higgs scalar potentials of physical interest.
We first considered the general 2HDM case which was known to be CP violating, with a complete basis of CPIs known, with several small CPIs being non-zero.
We then considered 3HDM and 6HDM which are symmetric under $\Delta(3n^2)$ and 
$\Delta(6n^2)$ groups. Many of these potentials had not been studied before and the new CPIs we found with our systematic search were needed as the previously known ones vanish even for potentials where the new CPIs reveal the presence of explicit CP violation.

For each potential, we either determined the lowest order non-zero CPIs (thereby proving that potential is CP violating) or, in cases where all the considered 
CPIs vanish, we derived the explicit CP symmetries that leave the potential invariant (thereby proving that potential is CP conserving). Since the potentials considered were very symmetric, 
we found that most of the smaller CPIs vanish.
Although the CPIs apply to any potential, they take different expressions as functions of the parameters of the potential, as clearly illustrated in the 2HDM example.
Furthermore, CPIs that are useful for one potential can vanish for other CP violating potentials.

We found that the $A_4$ potentials, although generally CP conserving for one triplet of Higgs 
doublets or singlets, are no longer CP conserving in general when two $A_4$ triplets are present (either doublets or singlets). By contrast we find that $\Delta(27)$ potentials are all CP violating in general. Although the 
$\Delta(27)$ potentials with a single triplet (whether the scalars are Higgs doublets or not) had previously been studied extensively, by using the calculated expression for a CPI we completely mapped specific CP symmetries to different ways to make the CPI expression vanish. For such potentials, we further analysed spontaneous CP violation when considering different CP symmetries by using a non-trivial SCPI.
The potentials with $\Delta(3n^2)$ with $n>3$ turn out to be particular cases of $A_4$ potentials.
For such cases it is notable that the expressions for the non-zero CPI become manageable for the case with two triplets (non-Higgs), which allowed us to find a CP symmetry that relates two of the real parameters of the potential.
Moreover, we found that 
all of the $\Delta(6n^2)$ potentials are special cases of the respective $\Delta(3n^2)$ potentials. In the $S_4$ case, this makes even the potentials with two triplets automatically CP conserving. Although the $\Delta(54)$ potential for one triplet (whether the scalars are Higgs doublets or not) coincides with the $\Delta(27)$ potential, this is no longer the case when two triplets are present, but they still generally violate CP. $\Delta(6n^2)$ with $n>3$ is a particular case of $S_4$ and therefore the potentials considered are again automatically CP conserving.

Finally, we briefly showed how our approach may also be applied to spontaneous CPV.
As an illustration of this we calculated the SCPIs
which are relevant for a $\Delta(27)$ potential showing how it reveals the CP properties of candidate VEVs.

In conclusion, the invariant approach to CP violation provides a reliable method for 
studying the CP properties of multi-Higgs potentials. We have developed a systematic
formalism for determing the CPIs for multi-Higgs potentials in general,
and have extensively applied this formalism to both the familiar general 2HDM as well as
many examples in which the Higgs fields fall into irreducible triplet representations 
of a discrete symmetry. We considered not only 
SM Higgs doublets, but also SM singlets which play the role of flavons in flavour models.
In each case we catalogued all the lowest order CPIs, many of which previously unknown,
thereby elucidating the CP properties of the considered potentials
and finding the relevant CP symmetry transformations where applicable.

\section*{Acknowledgements}

This project has received funding from the European Union's Seventh Framework Programme for research, technological development and demonstration under grant agreement no PIEF-GA-2012-327195 SIFT.
SFK and TN acknowledge partial support from the STFC Consolidated
ST/J000396/1 grant and the European Union FP7 ITN-INVISIBLES 
(Marie Curie Actions, PITN-GA-2011-289442). 
The work of CL is supported by the Deutsche Forschungsgemeinschaft (DFG) within
the Research Unit FOR 1873 ``Quark Flavour Physics and Effective Field Theories''.
The authors thank Celso Nishi for bringing to their attention an error in Section \ref{sec:2HDM} of the previous version of the manuscript.

\begin{appendix}

\section*{Appendix}
\section{Symmetries of invariants \label{sym_invs}}
In the main text, invariants both without and with VEVs were defined via permutations of indices, cf.\ Eqs.~(\ref{invariant_definition}) and (\ref{spont_CP_of_invariant}). Firstly, it might seem as if there is a huge number of invariants, one for each possible permutation of indices, whose number grows as the factorial of the number of indices. But luckily, as already hinted at in subsection~\ref{CPIs}, invariants have symmetries which will reduce the number of inequivalent invariants. Secondly, one still has to find those index permutations which correspond to CPIs. This appendix concerns itself with these two issues.

Invariants were defined in the following way via index permutations $\sigma \in S_n$ where $n$ is the total number of upper indices coming from all involved tensors and VEVs: for invariants without VEVs,
\begin{equation}
I_\sigma^{(m_Y n_Z)}=Y^{a_1}_{\sigma(a_1)}\ldots Y^{a_{m_Y}}_{\sigma(a_{m_Y})}Z^{b_1 b_2}_{\sigma(b_1)\sigma(b_2)}\ldots Z^{b_{2n_Z-1} b_{2n_Z}}_{\sigma(b_{2n_Z-1})\sigma(b_{2n_Z})} \text{ with } \sigma \in S_{m_Y+2n_Z},
\end{equation}
where $n=m_Y+2n_Z$, and for invariants containing VEVs,
\begin{align}
J_\sigma^{(n_v,m_Y,n_Z)}&=W^{w_1\ldots w_{n_v}}_{\sigma(w_1)\ldots \sigma(w_{n_v})}Y^{a_1}_{\sigma(a_1)}\ldots Y^{a_{m_Y}}_{\sigma(a_{m_Y})}Z^{b_1 b_2}_{\sigma(b_1)\sigma(b_2)}\ldots Z^{b_{2n_Z-1} b_{2n_Z}}_{\sigma(b_{2n_Z-1})\sigma(b_{2n_Z})} \nonumber\\&\mapsto 
W^{\sigma(w_1)\ldots \sigma(w_{n_v})}_{w_1\ldots w_{n_v}}Y^{\sigma(a_1)}_{a_1}\ldots Y^{\sigma(a_{m_Y})}_{a_{m_Y}}Z^{\sigma(b_1)\sigma(b_2)}_{b_1 b_2}\ldots Z^{\sigma(b_{2n_Z-1})\sigma(b_{2n_Z})}_{b_{2n_Z-1} b_{2n_Z}}= (J_\sigma^{(n_v,m_Y,n_Z)})^{\ast},
\end{align}
where now $n=n_v+m_Y+2n_Z$ and $W$ as defined in Eq.~\eqref{W_definition}.

There are the following sources of symmetries of invariants: renaming of indices, permutations of tensors of the same type, and internal symmetries of tensors. Internal symmetries of tensors can refer to symmetries under exchanging indices on the tensor, and symmetries induced by the symmetry of the Lagrangian. 
Except for the latter, which are not discussed here,
all of these symmetries exist for arbitrary invariants corresponding to arbitrary potentials. These sources of symmetries will now be discussed. To streamline notation, write all indices into a multi-index,
\begin{equation}
\alpha=(a_1,\ldots,a_n)\ ,
\end{equation}
where now permutations act on $\alpha$ by acting on each index as usual:
\begin{equation}
 \sigma(\alpha)=(\sigma(a_1),\ldots,\sigma(a_n)).
\end{equation}
Also, let $\mathcal{Z}$ stand for the product of tensors (both $Y$ and $Z$) and VEVs appropriate to the invariant in discussion, then any invariant can be written as
\begin{equation}
 I_\sigma=\mathcal{Z}^\alpha_{\sigma(\alpha)}.
 \label{multiindex_notation_definition}
\end{equation}
Renaming indices into each other corresponds to another permutation of all indices. For $a_i\mapsto \pi(a_i)$ with $\pi\in S_n$, the invariant becomes
\begin{equation}
 I_\sigma \xrightarrow{\pi} \mathcal{Z}^{\pi(\alpha)}_{\sigma(\pi(\alpha))}.
\end{equation}
The original invariant and the invariant with indices renamed into each other have the same value.

Next, some elements of invariants are symmetric under \emph{independent} permutations of upper and lower indices. For example, as discussed in section \ref{CPIs}, the following four versions of the $Z$ tensor are equal,
\begin{equation}
 Z^{ab}_{cd}=Z^{ba}_{cd}=Z^{ab}_{dc}=Z^{ba}_{dc},
\end{equation}
because $Z^{ab}_{cd}$ is symmetric under $a\leftrightarrow b$ and/or $c\leftrightarrow d$. This means that for each $Z$ tensor in the invariant there are 4 equivalent ways of connecting it to the rest of the invariant and thus for $n_Z$ $Z$ tensors, there would be $4^{n_Z}$ $\sigma$ matrices producing the same invariant and diagram. 
Similarly, in the tensor $W$ that summaries the product of all VEVs and complex conjugates of VEVs, all upper and lower indices can be permuted independently of each other. Denoting any such permutation of indices that is allowed by internal symmetries of tensors by $\tau$, then this condition becomes
\begin{equation}
\mathcal{Z}^\alpha_{\sigma(\alpha)}=\mathcal{Z}^{\tau(\alpha)}_{\sigma(\alpha)}=\mathcal{Z}^\alpha_{\tau(\sigma(\alpha))}=\mathcal{Z}^{\tau(\alpha)}_{\tau(\sigma(\alpha))}=I_\sigma.
\label{tau_definition}
\end{equation}
These internal symmetries can be taken into account in the actual search for CPIs by defining a new matrix that is produced from one of the equivalent $\sigma$ matrices, which maps all invariants that are related by transformations of the type $\tau$ onto a single matrix that also uniquely corresponds to the diagram corresponding to all these invariants. (In the diagram the symmetries are taken into account automatically.) This new matrix will be called contraction matrix and denoted by $m$. Define the following submatrices of $\sigma$ and $m$:
\begin{equation}
 \sigma=\begin{pmatrix}
         \sigma_{vv}&\sigma_{vY}&\sigma_{vZ}\\
         \sigma_{Yv}&\sigma_{YY}&\sigma_{YZ}\\
         \sigma_{Zv}&\sigma_{ZY}&\sigma_{ZZ}
        \end{pmatrix} \text{ , }
        m=\begin{pmatrix}
         m_{vv}&m_{vY}&m_{vZ}\\
         m_{Yv}&m_{YY}&m_{YZ}\\
         m_{Zv}&m_{ZY}&m_{ZZ}
        \end{pmatrix},
\end{equation}
where now the $vv$ parts correspond to contractions between VEVs, $vY$ between VEVs and $Y$ tensors, and so on, until $ZZ$, which corresponds to contractions between $Z$ tensors. While $\sigma$ is a $n\times n$ matrix with $n$ the number of indices, $m$ will be a $N\times N$ matrix where $N$ is the number of tensors in the invariant. $W$ would only be counted once. The relations between the submatrices of $\sigma$ and $m$ are as follows:
\begin{equation}
\begin{aligned}
 m_{vv}&=\sum_{i,j} (\sigma_{vv})_{ij}\ , \\
 (m_{vY})_{j}&=\sum_{i} (\sigma_{vY})_{ij}\ , \\
 (m_{vZ})_{j}&=\sum_{i} (\sigma_{vZ})_{2i-1,j}+\sum_{i} (\sigma_{vZ})_{2i,j}\ , \\
 (m_{Yv})_{i}&=\sum_{j} (\sigma_{Yv})_{ij}\ , \\
 (m_{Zv})_{i}&=\sum_{j} (\sigma_{Zv})_{i,2j-1}+\sum_{j} (\sigma_{Zv})_{i,2j}\ , \\
 (m_{YY})_{ij}&=\sigma_{ij}\ , \\
 (m_{YZ})_{ij}&=\sigma_{i,2j-1}+\sigma_{i,2j}\ , \\
 (m_{ZY})_{ij}&=\sigma_{2i-1,j}+\sigma_{2i,j}\ , \\
 (m_{ZZ})_{ij}&=\sigma_{2i-1,2j-1}+\sigma_{2i-1,2j}+\sigma_{2i,2j-1}+\sigma_{2i,2j}\ .
 \label{sigma_to_m}
\end{aligned}
\end{equation}

The element $m_{ij}$ denotes how many arrows are pointing from the $i$-th tensor in the invariant to the $j$-th tensor. What is happening in Eq.~\eqref{sigma_to_m} is that all equivalent ways of contracting the $i$-th and $j$-th tensor are summarised in $m_{ij}$ which means that e.g.\ for a contraction from a $Y$ tensor to a $Z$ tensor, one has to add the two elements corresponding to the two possible permutations of the lower index of $Z$, out of which only one can be non-zero in $\sigma$. Similarly, for contractions of a $Z$ tensor with another $Z$ tensor (or itself), one has to add all entries in the $2\times 2$ submatrix that corresponds to the four involved indices, out of which only two can be non-zero in $\sigma$.

For an invariant that only consists of $Y$ tensors, the contraction matrix $m$ is identical to $\sigma$. For invariants only consisting of $Z$ tensors, the situation also becomes a little simpler, as the full contraction matrix is given by the last line of Eq.~\eqref{sigma_to_m}. As $\sigma$ is a permutation matrix, in a $2\times2$ submatrix only either the two diagonal or the two off-diagonal elements can be non-zero at the same time and the contraction matrix decays into the sum of two smaller permutation matrices of only $n_Z$ elements, i.e.\ with $\sigma^{Z}_1,\sigma^Z_2\in S_{n_Z}$:
\begin{equation}
 m_{ij}=(\sigma^{Z}_1)_{ij}+(\sigma^{Z}_2)_{ij}.
 \label{m_only_Z}
\end{equation}

Now, one can discuss the last source of symmetry, namely permutations of tensors of the same type. Interchanging the position of two $Z$ tensors in an invariant,
\begin{equation}
 Z^{a_1a_2}_{\sigma(a_1)\sigma(a_2)}Z^{a_3a_4}_{\sigma(a_3)\sigma(a_4)}\ldots
 \rightarrow
 Z^{a_3a_4}_{\sigma(a_3)\sigma(a_4)}Z^{a_1a_2}_{\sigma(a_1)\sigma(a_2)}\ldots
 \ ,
\end{equation}
induces simultaneous permutations of both upper and lower indices of the form
\begin{equation}
\tilde{\tau}=\begin{pmatrix}
 0&1&&\\
 1&0&&\\
 &&1&\\
 &&&\ddots
\end{pmatrix}\otimes
\begin{pmatrix}
 1&0\\
 0&1
\end{pmatrix},
\end{equation}
such that an invariant transforms as
\begin{equation}
I_\sigma \xrightarrow{\tilde{\tau}}\mathcal{Z}^{\tilde{\tau}(\alpha)}_{\tilde{\tau}(\sigma(\alpha))}.
\end{equation}
For larger invariants that also contain $Y$ tensors and VEVs the index transformation induced by permutations of tensors of equal type works similarly. 
Now one can rename $\tilde{\tau}(\alpha)\equiv \alpha'$ such that the invariant becomes
\begin{equation}
 I_\sigma \rightarrow \mathcal{Z}^{\alpha'}_{\tilde{\tau}(\sigma(\tilde{\tau}^{-1}(\alpha')))}
 \ ,\label{tautilde_definiton}
\end{equation}
which shows that permutations of tensors relate different $\sigma$ matrices in a way similar to conjugacy class transformations, except that the index permutations induced by tensor permutations do not generate the full permutation group $S_n$ of the $n$ indices.
To summarise the symmetries of $\sigma$, all permutation matrices that are related to $\sigma$ by conjugation with transformations of type $\tau$, Eq.~\eqref{tau_definition} and transformations of type $\tilde{\tau}$, Eq.~\eqref{tautilde_definiton},
\begin{equation}
\tilde{\tau} \circ \tau \circ \sigma \circ \tau' \circ \tilde{\tau}^{-1},
\end{equation}
where $\tau$ and $\tau'$ can be two different transformations, produce the same invariant as $\sigma$.

On a contraction matrix $m$, the permutations on tensors act in a simpler way. For all $\sigma^Y\in S_{m_Y}$ and $\sigma^Z \in S_{n_Z}$, all of the contraction matrices, first for invariants without VEVs,
\begin{equation}
\begin{pmatrix}
 \sigma_Y&0\\0&\sigma_Z
\end{pmatrix}
m
\begin{pmatrix}
 \sigma_Y&0\\0&\sigma_Z
\end{pmatrix}^T,
\end{equation}
and for invariants with VEVs,
\begin{equation}
 \begin{pmatrix}
 1&0&0\\0&\sigma_Y&0\\0&0&\sigma_Z
\end{pmatrix}
m
\begin{pmatrix}
 1&0&0\\0&\sigma_Y&0\\0&0&\sigma_Z
\end{pmatrix}^T,
\label{invariant_classes}
\end{equation}
will produce equivalent invariants. Similarly, for invariants only involving $Z$ tensors, all $\sigma^Z m (\sigma^Z)^T$ will produce equivalent invariants. Applying this to Eq.~(\ref{m_only_Z}), this means that one of the two summands can be chosen to be a conjugacy class representative of $S_{n_Z}$ which reduces the number of invariants that need to be considered.

Finally, all pieces are in place to discuss the CP properties first of
$\sigma$ and after that of $m$. The CP conjugate of an invariant can be obtained by interchanging upper and lower indices, or in the shorthand notation introduced in Eq.~\eqref{multiindex_notation_definition},
\begin{equation}
 I_\sigma=\mathcal{Z}^\alpha_{\sigma(\alpha)}\xrightarrow{CP}\mathcal{Z}_\alpha^{\sigma(\alpha)}.
\end{equation}
One can now rename $\sigma(\alpha)=\alpha'$ and subsequently drop the prime to obtain
\begin{equation}
 I_\sigma \xrightarrow{CP}\mathcal{Z}^\alpha_{\sigma^{-1}(\alpha)}.
\end{equation}
Naively, an invariant is CP-even if it equals its CP conjugate which leads to the condition 
\begin{equation}
 \sigma^2=1.
\end{equation}
However, one has to take into account also all permutation matrices that are equivalent to $\sigma$ such that the condition becomes
\begin{equation}
 \sigma^{-1}=\tilde{\tau} \circ \tau \circ \sigma \circ \tau' \circ \tilde{\tau}^{-1},
\end{equation}
which means that as soon as any $\tau, \tau', \tilde{\tau}$ exist such that the above condition can be fulfilled, $\sigma$ produces a CP-even invariant. 

\begin{figure}[tbp]
\begin{equation*}
 Y^a_a=\includegraphics[scale=0.2]{FR_3.png}=\begin{pmatrix}1\end{pmatrix}
\end{equation*}
\begin{equation*}
 Y^a_aY^b_b=\includegraphics[scale=0.2]{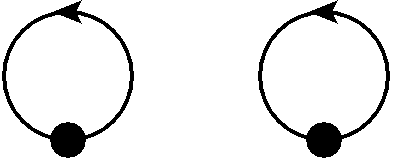}=\begin{pmatrix}1&0\\0&1\end{pmatrix}
\end{equation*}
\begin{equation*}
 Y^a_b Y^b_a=\includegraphics[scale=0.2]{YabYba.png}=\begin{pmatrix}0&1\\1&0\end{pmatrix}
\end{equation*}
\begin{equation*}
 Z^{ab}_{ab}=\includegraphics[scale=0.2]{Zabab.png}=\begin{pmatrix}2\end{pmatrix}
\end{equation*}
\begin{equation*}
 Z^{ac}_{bc}Y^b_a=\includegraphics[scale=0.2]{ZacbcYab.png}=\begin{pmatrix}0&1\\1&1\end{pmatrix}
\end{equation*}
\begin{equation*}
 Z^{ab}_{cd}Z^{cd}_{ab}=\includegraphics[scale=0.2]{ZabcdZcdab.png}=\begin{pmatrix}0&2\\2&0\end{pmatrix}
\end{equation*}
\begin{equation*}
 Z^{ab}_{ac}Z^{cd}_{bd}=\includegraphics[scale=0.2]{ZabacZcdbd.png}=\begin{pmatrix}1&1\\1&1\end{pmatrix}
\end{equation*}
\begin{equation*}
 I_1\equiv Z^{ab}_{ae}Z^{cd}_{bf}Y^e_cY^f_d=\includegraphics[scale=0.2]{I1.png}=\begin{pmatrix}0&0&1&0\\0&0&0&1\\0&0&1&1\\1&1&0&0\end{pmatrix}
\end{equation*}
\caption{Examples of contraction matrices for small invariants. All contraction matrices are symmetric except for the CPI.}
\label{contraction_matrix_examples}
\end{figure}

\begin{figure}[tbp]
\begin{equation*}
J^{(3,1)}_1=Z^{a_1a_2}_{a_5a_6}Z^{a_3a_4}_{a_1a_3}Z^{a_5a_6}_{a_2 a_7}v_{a_4}v^{\ast a_7}=\vcenter{\hbox{\includegraphics[scale=0.2]{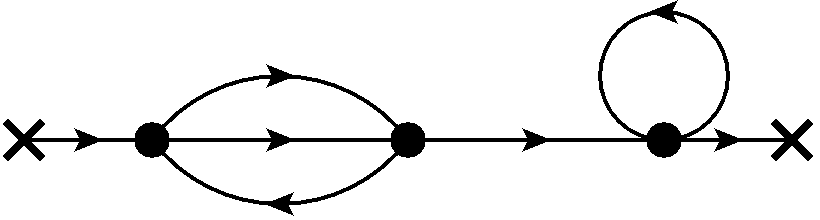}}}=\begin{pmatrix}0&0&0&1\\0&0&1&1\\1&0&1&0\\0&2&0&0\end{pmatrix}
\end{equation*}
 \begin{equation*}
J_{1}^{(3,2)}\equiv Z^{a_1a_2}_{a_4a_5}Z^{a_3a_4}_{a_2a_6}Z^{a_5a_6}_{a_7a_8}v_{a_1}v_{a_3}v^{\ast a_7}v^{\ast a_8}=\vcenter{\hbox{\includegraphics[scale=0.2]{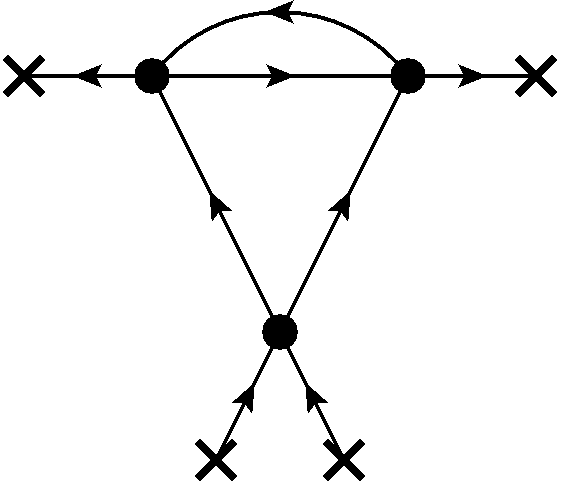}}}=\begin{pmatrix}0&0&0&2\\1&0&1&0\\1&1&0&0\\0&1&1&0\end{pmatrix}
\end{equation*}
\caption{Examples for contraction matrices of CPIs for spontaneous CP violation. We draw each of the VEVs (as opposed to a single vertex for the whole $W$ tensor).}
\label{spont_cont_mat_examples}
\end{figure}

For contraction matrices, the condition testing if an invariant is CP-even simplifies. With $\sigma^{-1}=\sigma^T$, from which follows that $m\xrightarrow{CP}m^T$ and if $\sigma_Y$ and $\sigma_Z$ exist, such that the right-hand side is fulfilled, then the condition for invariants without VEVs to be CP-even becomes
\begin{equation}
 \text{Invariant CP-even}\Leftrightarrow m^T=\begin{pmatrix}
 \sigma_Y&0\\0&\sigma_Z
\end{pmatrix}
m
\begin{pmatrix}
 \sigma_Y&0\\0&\sigma_Z
\end{pmatrix}^T,
\end{equation}
and for invariants with VEVs
\begin{equation}
 \text{spont.\ Invariant CP-even}\Leftrightarrow m^T=\begin{pmatrix}
 1&0&0\\0&\sigma_Y&0\\0&0&\sigma_Z
\end{pmatrix}
m
\begin{pmatrix}
 1&0&0\\0&\sigma_Y&0\\0&0&\sigma_Z
\end{pmatrix}^T
\label{invariant_CP_classes},
\end{equation}
where the actions of $\tau$ and $\tau'$ are absorbed in $m$. 
Figures~\ref{contraction_matrix_examples} and \ref{spont_cont_mat_examples} contain examples of contraction matrices for small diagrams. There, all contraction matrices happen to be trivially symmetric except for the CPIs.

It is condition Eq.~(\ref{invariant_CP_classes}) that was used to find CP-odd invariants. In the actual search, first all $\sigma$ matrices for a certain number of $Y$ and $Z$ tensors, and VEVs was generated. This list of $\sigma$ matrices was then reduced to a list of contraction matrices, which was condensed using Eq.~(\ref{invariant_classes}) to classes of equivalent contraction matrices, out of which a representative was tested for CP-oddness using Eq.~(\ref{invariant_CP_classes}). This search was performed for invariants without VEVs for $m_Y=0$ up to $n_Z=6$, where it was found that all invariants without $Y$ tensors until $n_Z=4$ are CP even. Furthermore, CP-odd invariants were found for $(m_Y,n_Z)=(1,3),(1,4),(2,2),(2,3),(3,3)$. For invariants with VEVs, only a search for invariants with $m_Y=0$ was performed, where CP-odd invariants were found for $(n_v,n_Z)=(1,3),(2,3),(1,4)$. All inequivalent invariants from these classes are listed in section \ref{CPoddI} or in appendix \ref{invariant_lists}.

As one progresses to more complicated invariants, one has to make sure not to count invariants that are products or powers of smaller invariants. An invariant that is a product of two smaller invariants will correspond to a diagram that decays into two separate graphs.  As this means that some vertices are only connected among each other while being unconnected to the rest of the diagram, such a reducible invariant will be described by a contraction matrix that can be brought to block-diagonal form only using permutation matrices. This means in particular, as for invariants with VEVs, $m_{vv}$ denotes the number of VEVs that are only connected to other VEVs, that $m_{vv}\neq 0$ would mean that the diagram would contain graphs for $v_a v^{\ast a}$ that are unconnected to the rest of the diagram.

Finally, there is one last condition that relates invariants, namely the minimisation condition Eq.~\eqref{minimization_condition}. In the contraction matrix for an invariant with VEVs this can be used if there is an $i$ such that
\begin{equation}
 m_{1i}=1\text{ and }m_{i1}=2\ ,
\end{equation}
or 
\begin{equation}
 m_{1i}=2\text{ and }m_{i1}=1.
\end{equation}
In both cases, the Z tensor at position $i$ in the invariant is connected to three VEVs.

\section{List of invariants \label{invariant_lists}}
\cleqn
\subsection[Contraction matrices of $n_Z=5$ invariants]{Contraction matrices of $\bs{n_Z=5}$ invariants}

\begin{align}
I^{(5)}_1= Z^{a_1a_2}_{a_7a_9}Z^{a_3a_4}_{a_5a_{10}}Z^{a_5a_6}_{a_3a_6}Z^{a_7a_8}_{a_4a_8}Z^{a_9a_{10}}_{a_1a_2}=\vcenter{\hbox{\includegraphics[scale=0.2]{I5_1.png}}}=\begin{pmatrix}0&0&0&0&2\\0&0&1&1&0\\0&1&1&0&0\\1&0&0&1&0\\1&1&0&0&0\end{pmatrix}\\
 I^{(5)}_2=Z^{a_1a_2}_{a_5a_7}Z^{a_3a_4}_{a_8a_9}Z^{a_5a_6}_{a_3a_6}Z^{a_7a_8}_{a_4a_{10}}Z^{a_9a_{10}}_{a_1a_2}=\vcenter{\hbox{\includegraphics[scale=0.2]{I5_2.png}}}=\begin{pmatrix}0&0&0&0&2\\0&0&1&1&0\\1&0&1&0&0\\1&1&0&0&0\\0&1&0&1&0\end{pmatrix}\\
 I^{(5)}_3=Z^{a_1a_2}_{a_5a_9}Z^{a_3a_4}_{a_3a_7}Z^{a_5a_6}_{a_6a_8}Z^{a_7a_8}_{a_1a_{10}}Z^{a_9a_{10}}_{a_2a_4}=\vcenter{\hbox{\includegraphics[scale=0.2]{I5_3.png}}}=\begin{pmatrix}0&0&0&1&1\\0&1&0&0&1\\1&0&1&0&0\\0&1&1&0&0\\1&0&0&1&0\end{pmatrix}
\end{align}
\subsection[Contraction matrices of $n_Z=6$ invariants without $Z$-self-loops]{Contraction matrices of $\bs{n_Z=6}$ invariants without $\bs{Z}$-self-loops}
\begin{equation}
I_1^{(6)}=Z^{a_1 a_2}_{a_{11} a_{10}} Z^{a_3 a_4}_{a_5 a_8} Z^{a_5 a_6}_{a_7 a_{12}} Z^{a_7 a_8}_{a_9 a_6} Z^{a_9 a_{10}}_{a_3 a_4} Z^{a_{11} a_{12}}_{a_1 a_2}=\vcenter{\hbox{\includegraphics[scale=0.2]{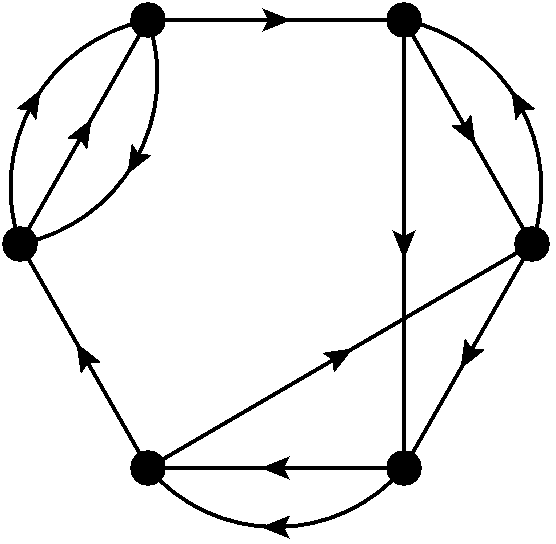}}}=\left(
\begin{array}{cccccc}
 0 & 0 & 0 & 0 & 0 & 2 \\
 0 & 0 & 0 & 0 & 2 & 0 \\
 0 & 1 & 0 & 1 & 0 & 0 \\
 0 & 1 & 1 & 0 & 0 & 0 \\
 1 & 0 & 0 & 1 & 0 & 0 \\
 1 & 0 & 1 & 0 & 0 & 0 \\
\end{array}
\right)
\end{equation}
\begin{equation}
I_2^{(6)}=Z^{a_1 a_2}_{a_7 a_{10}} Z^{a_3 a_4}_{a_{11} a_6} Z^{a_5 a_6}_{a_9 a_8} Z^{a_7 a_8}_{a_3 a_{12}} Z^{a_9 a_{10}}_{a_5 a_4} Z^{a_{11} a_{12}}_{a_1 a_2}=\vcenter{\hbox{\includegraphics[scale=0.2]{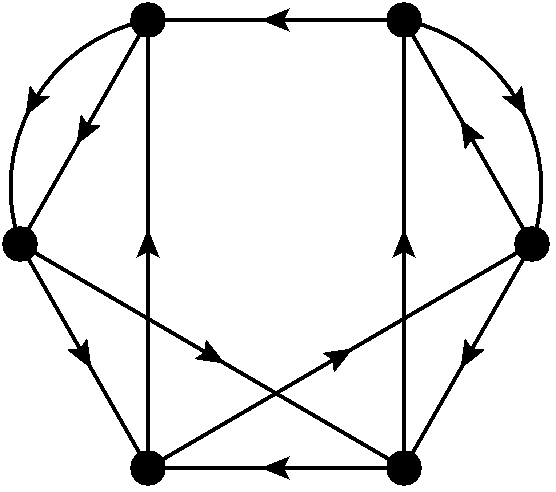}}}=\left(
\begin{array}{cccccc}
 0 & 0 & 0 & 0 & 0 & 2 \\
 0 & 0 & 0 & 1 & 1 & 0 \\
 0 & 1 & 0 & 0 & 1 & 0 \\
 1 & 0 & 1 & 0 & 0 & 0 \\
 1 & 0 & 1 & 0 & 0 & 0 \\
 0 & 1 & 0 & 1 & 0 & 0 \\
\end{array}
\right)
\end{equation}
\begin{equation}
 I_3^{(6)}=Z^{a_1 a_2}_{a_7 a_{10}} Z^{a_3 a_4}_{a_9 a_6} Z^{a_5 a_6}_{a_{11} a_8} Z^{a_7 a_8}_{a_3 a_{12}} Z^{a_9 a_{10}}_{a_5 a_4} Z^{a_{11} a_{12}}_{a_1 a_2}=\vcenter{\hbox{\includegraphics[scale=0.2]{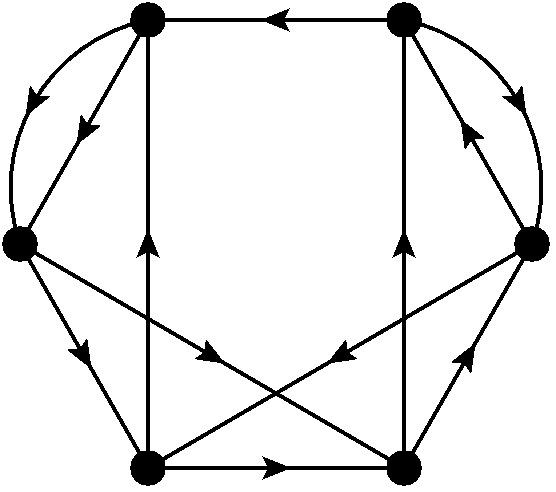}}}=\left(
\begin{array}{cccccc}
 0 & 0 & 0 & 0 & 0 & 2 \\
 0 & 0 & 0 & 1 & 1 & 0 \\
 0 & 1 & 0 & 0 & 1 & 0 \\
 1 & 0 & 1 & 0 & 0 & 0 \\
 1 & 1 & 0 & 0 & 0 & 0 \\
 0 & 0 & 1 & 1 & 0 & 0 \\
\end{array}
\right)
\end{equation}
\begin{equation}
 I_4^{(6)}=Z^{a_1 a_2}_{a_{11} a_{10}} Z^{a_3 a_4}_{a_5 a_8} Z^{a_5 a_6}_{a_7 a_{12}} Z^{a_7 a_8}_{a_9 a_6} Z^{a_9 a_{10}}_{a_1 a_4} Z^{a_{11} a_{12}}_{a_3 a_2}=\vcenter{\hbox{\includegraphics[scale=0.2]{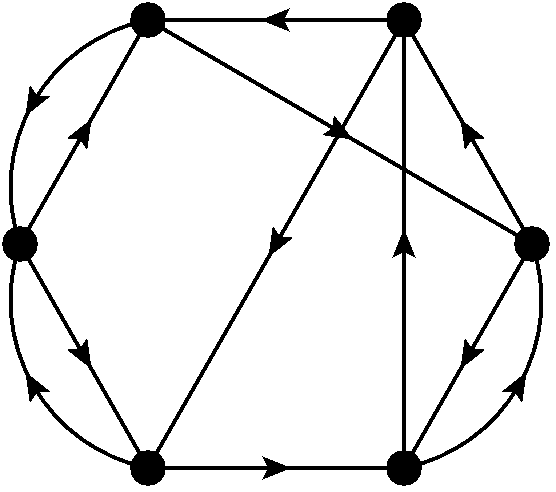}}}=\left(
\begin{array}{cccccc}
 0 & 0 & 0 & 0 & 1 & 1 \\
 0 & 0 & 0 & 0 & 1 & 1 \\
 0 & 1 & 0 & 1 & 0 & 0 \\
 0 & 1 & 1 & 0 & 0 & 0 \\
 1 & 0 & 0 & 1 & 0 & 0 \\
 1 & 0 & 1 & 0 & 0 & 0 \\
\end{array}
\right)
\end{equation}
\begin{equation}
 I_5^{(6)}=Z^{a_1 a_2}_{a_7 a_{12}} Z^{a_3 a_4}_{a_5 a_{10}} Z^{a_5 a_6}_{a_9 a_8} Z^{a_7 a_8}_{a_{11} a_4} Z^{a_9 a_{10}}_{a_1 a_6} Z^{a_{11} a_{12}}_{a_3 a_2}=\vcenter{\hbox{\includegraphics[scale=0.2]{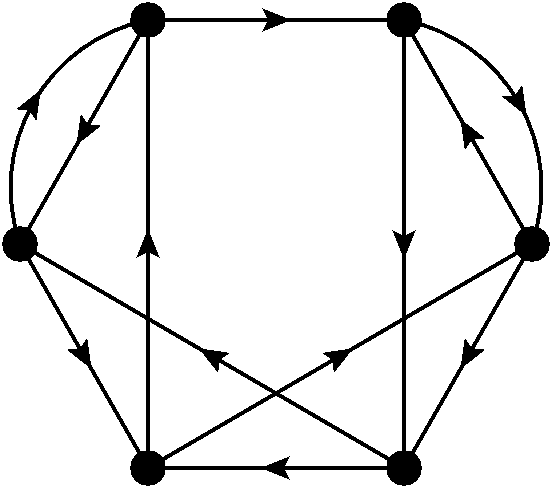}}}=\left(
\begin{array}{cccccc}
 0 & 0 & 0 & 0 & 1 & 1 \\
 0 & 0 & 0 & 1 & 0 & 1 \\
 0 & 1 & 0 & 0 & 1 & 0 \\
 1 & 0 & 1 & 0 & 0 & 0 \\
 0 & 1 & 1 & 0 & 0 & 0 \\
 1 & 0 & 0 & 1 & 0 & 0 \\
\end{array}
\right)
\end{equation}
\subsection[$n_Z=6$ invariants with self-loops]{$\bs{n_Z=6}$ invariants with self-loops}
\begin{align}
Z^{a_1 a_2}_{a_9 a_{12}}	Z^{a_3 a_4}_{a_5 a_8}		Z^{a_5 a_6}_{a_{11} a_6}	Z^{a_7 a_8}_{a_7 a_{10}}	Z^{a_9 a_{10}}_{a_3 a_4}	Z^{a_{11} a_{12}}_{a_1 a_2}\\	
Z^{a_1 a_2}_{a_7 a_{12}}	Z^{a_3 a_4}_{a_{11} a_8}	Z^{a_5 a_6}_{a_5 a_{10}}	Z^{a_7 a_8}_{a_9 a_6}	Z^{a_9 a_{10}}_{a_3 a_4}	Z^{a_{11} a_{12}}_{a_1 a_2}\\	
Z^{a_1 a_2}_{a_{11} a_{10}}	Z^{a_3 a_4}_{a_7 a_8}		Z^{a_5 a_6}_{a_5 a_{12}}	Z^{a_7 a_8}_{a_9 a_6}	Z^{a_9 a_{10}}_{a_3 a_4}	Z^{a_{11} a_{12}}_{a_1 a_2}\\	
Z^{a_1 a_2}_{a_{11} a_8}	Z^{a_3 a_4}_{a_7 a_{10}}	Z^{a_5 a_6}_{a_5 a_{12}}	Z^{a_7 a_8}_{a_9 a_6}	Z^{a_9 a_{10}}_{a_3 a_4}	Z^{a_{11} a_{12}}_{a_1 a_2}\\	
Z^{a_1 a_2}_{a_7 a_{10}}	Z^{a_3 a_4}_{a_{11} a_8}	Z^{a_5 a_6}_{a_5 a_{12}}	Z^{a_7 a_8}_{a_9 a_6}	Z^{a_9 a_{10}}_{a_3 a_4}	Z^{a_{11} a_{12}}_{a_1 a_2}\\	
Z^{a_1 a_2}_{a_9 a_{12}}	Z^{a_3 a_4}_{a_7 a_{10}}	Z^{a_5 a_6}_{a_{11} a_6}	Z^{a_7 a_8}_{a_3 a_8}	Z^{a_9 a_{10}}_{a_5 a_4}	Z^{a_{11} a_{12}}_{a_1 a_2}\\	
Z^{a_1 a_2}_{a_{11} a_8}	Z^{a_3 a_4}_{a_9 a_{12}}	Z^{a_5 a_6}_{a_7 a_6}	Z^{a_7 a_8}_{a_3 a_{10}}	Z^{a_9 a_{10}}_{a_5 a_4}	Z^{a_{11} a_{12}}_{a_1 a_2}\\	
Z^{a_1 a_2}_{a_9 a_{12}}	Z^{a_3 a_4}_{a_{11} a_8}	Z^{a_5 a_6}_{a_7 a_6}	Z^{a_7 a_8}_{a_3 a_{10}}	Z^{a_9 a_{10}}_{a_5 a_4}	Z^{a_{11} a_{12}}_{a_1 a_2}\\	
Z^{a_1 a_2}_{a_9 a_8}		Z^{a_3 a_4}_{a_7 a_{12}}	Z^{a_5 a_6}_{a_{11} a_6}	Z^{a_7 a_8}_{a_3 a_{10}}	Z^{a_9 a_{10}}_{a_5 a_4}	Z^{a_{11} a_{12}}_{a_1 a_2}\\	
Z^{a_1 a_2}_{a_7 a_{12}}	Z^{a_3 a_4}_{a_9 a_8}		Z^{a_5 a_6}_{a_{11} a_6}	Z^{a_7 a_8}_{a_3 a_{10}}	Z^{a_9 a_{10}}_{a_5 a_4}	Z^{a_{11} a_{12}}_{a_1 a_2}\\	
Z^{a_1 a_2}_{a_9 a_{10}}	Z^{a_3 a_4}_{a_{11} a_8}	Z^{a_5 a_6}_{a_7 a_6}	Z^{a_7 a_8}_{a_3 a_{12}}	Z^{a_9 a_{10}}_{a_5 a_4}	Z^{a_{11} a_{12}}_{a_1 a_2}\\	
Z^{a_1 a_2}_{a_9 a_8}		Z^{a_3 a_4}_{a_{11} a_{10}}	Z^{a_5 a_6}_{a_7 a_6}	Z^{a_7 a_8}_{a_3 a_{12}}	Z^{a_9 a_{10}}_{a_5 a_4}	Z^{a_{11} a_{12}}_{a_1 a_2}\\	
Z^{a_1 a_2}_{a_7 a_{10}}	Z^{a_3 a_4}_{a_{11} a_8}	Z^{a_5 a_6}_{a_9 a_6}	Z^{a_7 a_8}_{a_3 a_{12}}	Z^{a_9 a_{10}}_{a_5 a_4}	Z^{a_{11} a_{12}}_{a_1 a_2}\\	
Z^{a_1 a_2}_{a_7 a_{10}}	Z^{a_3 a_4}_{a_9 a_8}		Z^{a_5 a_6}_{a_{11} a_6}	Z^{a_7 a_8}_{a_3 a_{12}}	Z^{a_9 a_{10}}_{a_5 a_4}	Z^{a_{11} a_{12}}_{a_1 a_2}\\	
Z^{a_1 a_2}_{a_9 a_{12}}	Z^{a_3 a_4}_{a_5 a_{10}}	Z^{a_5 a_6}_{a_{11} a_6}	Z^{a_7 a_8}_{a_3 a_8}	Z^{a_9 a_{10}}_{a_7 a_4}	Z^{a_{11} a_{12}}_{a_1 a_2}\\	
Z^{a_1 a_2}_{a_9 a_8}		Z^{a_3 a_4}_{a_5 a_{12}}	Z^{a_5 a_6}_{a_{11} a_6}	Z^{a_7 a_8}_{a_3 a_{10}}	Z^{a_9 a_{10}}_{a_7 a_4}	Z^{a_{11} a_{12}}_{a_1 a_2}\\	
Z^{a_1 a_2}_{a_9 a_8}		Z^{a_3 a_4}_{a_5 a_{10}}	Z^{a_5 a_6}_{a_{11} a_6}	Z^{a_7 a_8}_{a_3 a_{12}}	Z^{a_9 a_{10}}_{a_7 a_4}	Z^{a_{11} a_{12}}_{a_1 a_2}\\	
Z^{a_1 a_2}_{a_{11} a_{10}}	Z^{a_3 a_4}_{a_5 a_{12}}	Z^{a_5 a_6}_{a_7 a_6}	Z^{a_7 a_8}_{a_3 a_8}	Z^{a_9 a_{10}}_{a_9 a_4}	Z^{a_{11} a_{12}}_{a_1 a_2}\\	
Z^{a_1 a_2}_{a_5 a_{12}}	Z^{a_3 a_4}_{a_{11} a_{10}}	Z^{a_5 a_6}_{a_7 a_6}	Z^{a_7 a_8}_{a_3 a_8}	Z^{a_9 a_{10}}_{a_9 a_4}	Z^{a_{11} a_{12}}_{a_1 a_2}\\	
Z^{a_1 a_2}_{a_5 a_{10}}	Z^{a_3 a_4}_{a_{11} a_6}	Z^{a_5 a_6}_{a_7 a_{12}}	Z^{a_7 a_8}_{a_3 a_8}	Z^{a_9 a_{10}}_{a_9 a_4}	Z^{a_{11} a_{12}}_{a_1 a_2}\\	
Z^{a_1 a_2}_{a_7 a_{10}}	Z^{a_3 a_4}_{a_5 a_{12}}	Z^{a_5 a_6}_{a_{11} a_6}	Z^{a_7 a_8}_{a_3 a_8}	Z^{a_9 a_{10}}_{a_9 a_4}	Z^{a_{11} a_{12}}_{a_1 a_2}\\	
Z^{a_1 a_2}_{a_5 a_{10}}	Z^{a_3 a_4}_{a_7 a_{12}}	Z^{a_5 a_6}_{a_{11} a_6}	Z^{a_7 a_8}_{a_3 a_8}	Z^{a_9 a_{10}}_{a_9 a_4}	Z^{a_{11} a_{12}}_{a_1 a_2}\\	
Z^{a_1 a_2}_{a_5 a_{10}}	Z^{a_3 a_4}_{a_{11} a_8}	Z^{a_5 a_6}_{a_7 a_6}	Z^{a_7 a_8}_{a_3 a_{12}}	Z^{a_9 a_{10}}_{a_9 a_4}	Z^{a_{11} a_{12}}_{a_1 a_2}\\		
Z^{a_1 a_2}_{a_9 a_{12}}	Z^{a_3 a_4}_{a_7 a_6}		Z^{a_5 a_6}_{a_5 a_{10}}	Z^{a_7 a_8}_{a_3 a_8}	Z^{a_9 a_{10}}_{a_{11} a_4}	Z^{a_{11} a_{12}}_{a_1 a_2}\\	
Z^{a_1 a_2}_{a_9 a_{10}}	Z^{a_3 a_4}_{a_7 a_6}		Z^{a_5 a_6}_{a_5 a_{12}}	Z^{a_7 a_8}_{a_3 a_8}	Z^{a_9 a_{10}}_{a_{11} a_4}	Z^{a_{11} a_{12}}_{a_1 a_2}\\	
Z^{a_1 a_2}_{a_5 a_{10}}	Z^{a_3 a_4}_{a_9 a_{12}}	Z^{a_5 a_6}_{a_7 a_6}	Z^{a_7 a_8}_{a_3 a_8}	Z^{a_9 a_{10}}_{a_{11} a_4}	Z^{a_{11} a_{12}}_{a_1 a_2}\\	
Z^{a_1 a_2}_{a_7 a_{10}}	Z^{a_3 a_4}_{a_5 a_{12}}	Z^{a_5 a_6}_{a_9 a_6}	Z^{a_7 a_8}_{a_3 a_8}	Z^{a_9 a_{10}}_{a_{11} a_4}	Z^{a_{11} a_{12}}_{a_1 a_2}\\	
Z^{a_1 a_2}_{a_5 a_{10}}	Z^{a_3 a_4}_{a_7 a_{12}}	Z^{a_5 a_6}_{a_9 a_6}	Z^{a_7 a_8}_{a_3 a_8}	Z^{a_9 a_{10}}_{a_{11} a_4}	Z^{a_{11} a_{12}}_{a_1 a_2}\\	
Z^{a_1 a_2}_{a_9 a_8}		Z^{a_3 a_4}_{a_5 a_{10}}	Z^{a_5 a_6}_{a_7 a_6}	Z^{a_7 a_8}_{a_3 a_{12}}	Z^{a_9 a_{10}}_{a_{11} a_4}	Z^{a_{11} a_{12}}_{a_1 a_2}\\	
Z^{a_1 a_2}_{a_9 a_8}		Z^{a_3 a_4}_{a_3 a_{12}}	Z^{a_5 a_6}_{a_5 a_{10}}	Z^{a_7 a_8}_{a_7 a_6}	Z^{a_9 a_{10}}_{a_{11} a_4}	Z^{a_{11} a_{12}}_{a_1 a_2}\\	
Z^{a_1 a_2}_{a_9 a_{12}}	Z^{a_3 a_4}_{a_5 a_8}		Z^{a_5 a_6}_{a_{11} a_6}	Z^{a_7 a_8}_{a_7 a_{10}}	Z^{a_9 a_{10}}_{a_1 a_4}	Z^{a_{11} a_{12}}_{a_3 a_2}\\	
Z^{a_1 a_2}_{a_{11} a_8}	Z^{a_3 a_4}_{a_7 a_{10}}	Z^{a_5 a_6}_{a_5 a_{12}}	Z^{a_7 a_8}_{a_9 a_6}	Z^{a_9 a_{10}}_{a_1 a_4}	Z^{a_{11} a_{12}}_{a_3 a_2}
\end{align}
\begin{align}
Z^{a_1 a_2}_{a_9 a_{12}}	Z^{a_3 a_4}_{a_7 a_{10}}	Z^{a_5 a_6}_{a_5 a_8}	Z^{a_7 a_8}_{a_{11} a_4}	Z^{a_9 a_{10}}_{a_1 a_6}	Z^{a_{11} a_{12}}_{a_3 a_2}\\	
Z^{a_1 a_2}_{a_7 a_{10}}	Z^{a_3 a_4}_{a_9 a_{12}}	Z^{a_5 a_6}_{a_5 a_8}	Z^{a_7 a_8}_{a_{11} a_4}	Z^{a_9 a_{10}}_{a_1 a_6}	Z^{a_{11} a_{12}}_{a_3 a_2}\\	
Z^{a_1 a_2}_{a_7 a_{12}}	Z^{a_3 a_4}_{a_5 a_{10}}	Z^{a_5 a_6}_{a_{11} a_6}	Z^{a_7 a_8}_{a_9 a_4}	Z^{a_9 a_{10}}_{a_1 a_8}	Z^{a_{11} a_{12}}_{a_3 a_2}\\	
Z^{a_1 a_2}_{a_7 a_{10}}	Z^{a_3 a_4}_{a_9 a_6}		Z^{a_5 a_6}_{a_5 a_{12}}	Z^{a_7 a_8}_{a_{11} a_4}	Z^{a_9 a_{10}}_{a_1 a_8}	Z^{a_{11} a_{12}}_{a_3 a_2}\\	
Z^{a_1 a_2}_{a_{11} a_8}	Z^{a_3 a_4}_{a_9 a_6}		Z^{a_5 a_6}_{a_5 a_{12}}	Z^{a_7 a_8}_{a_7 a_4}	Z^{a_9 a_{10}}_{a_1 a_{10}}	Z^{a_{11} a_{12}}_{a_3 a_2}\\	
Z^{a_1 a_2}_{a_9 a_{12}}	Z^{a_3 a_4}_{a_5 a_8}		Z^{a_5 a_6}_{a_{11} a_6}	Z^{a_7 a_8}_{a_7 a_4}	Z^{a_9 a_{10}}_{a_1 a_{10}}	Z^{a_{11} a_{12}}_{a_3 a_2}\\	
Z^{a_1 a_2}_{a_9 a_8}		Z^{a_3 a_4}_{a_5 a_{12}}	Z^{a_5 a_6}_{a_{11} a_6}	Z^{a_7 a_8}_{a_7 a_4}	Z^{a_9 a_{10}}_{a_1 a_{10}}	Z^{a_{11} a_{12}}_{a_3 a_2}\\	
Z^{a_1 a_2}_{a_9 a_8}		Z^{a_3 a_4}_{a_5 a_4}		Z^{a_5 a_6}_{a_{11} a_6}	Z^{a_7 a_8}_{a_7 a_{12}}	Z^{a_9 a_{10}}_{a_1 a_{10}}	Z^{a_{11} a_{12}}_{a_3 a_2}\\	
Z^{a_1 a_2}_{a_7 a_{12}}	Z^{a_3 a_4}_{a_9 a_6}		Z^{a_5 a_6}_{a_5 a_8}	Z^{a_7 a_8}_{a_{11} a_4}	Z^{a_9 a_{10}}_{a_1 a_{10}}	Z^{a_{11} a_{12}}_{a_3 a_2}\\	
Z^{a_1 a_2}_{a_9 a_8}		Z^{a_3 a_4}_{a_7 a_6}		Z^{a_5 a_6}_{a_5 a_{12}}	Z^{a_7 a_8}_{a_{11} a_4}	Z^{a_9 a_{10}}_{a_1 a_{10}}	Z^{a_{11} a_{12}}_{a_3 a_2}\\	
Z^{a_1 a_2}_{a_9 a_8}		Z^{a_3 a_4}_{a_5 a_{12}}	Z^{a_5 a_6}_{a_7 a_6}	Z^{a_7 a_8}_{a_{11} a_4}	Z^{a_9 a_{10}}_{a_1 a_{10}}	Z^{a_{11} a_{12}}_{a_3 a_2}\\	
Z^{a_1 a_2}_{a_7 a_{10}}	Z^{a_3 a_4}_{a_9 a_4}		Z^{a_5 a_6}_{a_{11} a_6}	Z^{a_7 a_8}_{a_5 a_8}	Z^{a_9 a_{10}}_{a_1 a_{12}}	Z^{a_{11} a_{12}}_{a_3 a_2}\\	
Z^{a_1 a_2}_{a_{11} a_{10}}	Z^{a_3 a_4}_{a_5 a_8}		Z^{a_5 a_6}_{a_9 a_6}	Z^{a_7 a_8}_{a_7 a_4}	Z^{a_9 a_{10}}_{a_1 a_{12}}	Z^{a_{11} a_{12}}_{a_3 a_2}\\	
Z^{a_1 a_2}_{a_{11} a_8}	Z^{a_3 a_4}_{a_5 a_{10}}	Z^{a_5 a_6}_{a_9 a_6}	Z^{a_7 a_8}_{a_7 a_4}	Z^{a_9 a_{10}}_{a_1 a_{12}}	Z^{a_{11} a_{12}}_{a_3 a_2}\\	
Z^{a_1 a_2}_{a_9 a_8}		Z^{a_3 a_4}_{a_5 a_{10}}	Z^{a_5 a_6}_{a_{11} a_6}	Z^{a_7 a_8}_{a_7 a_4}	Z^{a_9 a_{10}}_{a_1 a_{12}}	Z^{a_{11} a_{12}}_{a_3 a_2}\\	
Z^{a_1 a_2}_{a_{11} a_8}	Z^{a_3 a_4}_{a_9 a_4}		Z^{a_5 a_6}_{a_5 a_{10}}	Z^{a_7 a_8}_{a_7 a_6}	Z^{a_9 a_{10}}_{a_1 a_{12}}	Z^{a_{11} a_{12}}_{a_3 a_2}\\	
\end{align}
\subsection[Invariants with $n_Y\neq0$ not listed in the main text]{Invariants with $\bs{n_Y\neq0}$ not listed in the main text}
\label{SCPI_list}
$n_Y=2,n_Z=3$ invariants with self-loops
\begin{align}
Y^{a_1}_{a_7} Y^{a_2}_{a_5} Z^{a_3 a_4}_{a_3 a_8} Z^{a_5 a_6}_{a_4 a_6} Z^{a_7 a_8}_{a_1 a_2}\\
Y^{a_1}_{a_7} Y^{a_2}_{a_5} Z^{a_3 a_4}_{a_3 a_6} Z^{a_5 a_6}_{a_4 a_8} Z^{a_7 a_8}_{a_1 a_2}\\
Y^{a_1}_{a_5} Y^{a_2}_{a_3} Z^{a_3 a_4}_{a_7 a_8} Z^{a_5 a_6}_{a_4 a_6} Z^{a_7 a_8}_{a_1 a_2}\\
Y^{a_1}_{a_5} Y^{a_2}_{a_3} Z^{a_3 a_4}_{a_4 a_7} Z^{a_5 a_6}_{a_6 a_8} Z^{a_7 a_8}_{a_1 a_2}\\
Y^{a_1}_{a_7} Y^{a_2}_{a_3} Z^{a_3 a_4}_{a_5 a_8} Z^{a_5 a_6}_{a_2 a_6} Z^{a_7 a_8}_{a_1 a_4}\\
Y^{a_1}_{a_5} Y^{a_2}_{a_3} Z^{a_3 a_4}_{a_6 a_7} Z^{a_5 a_6}_{a_2 a_4} Z^{a_7 a_8}_{a_1 a_8}\\
Y^{a_1}_{a_2} Y^{a_2}_{a_5} Z^{a_3 a_4}_{a_7 a_8} Z^{a_5 a_6}_{a_3 a_6} Z^{a_7 a_8}_{a_1 a_4}\\
Y^{a_1}_{a_2} Y^{a_2}_{a_5} Z^{a_3 a_4}_{a_3 a_7} Z^{a_5 a_6}_{a_6 a_8} Z^{a_7 a_8}_{a_1 a_4} 
\end{align}

$n_Y=3,n_Z=3$ Invariants with self-loops
\begin{align}
Y^{a_1}_{a_8} Y^{a_2}_{a_6} Y^{a_3}_{a_4} Z^{a_4 a_5}_{a_5 a_9} Z^{a_6 a_7}_{a_3 a_7} Z^{a_8 a_9}_{a_1 a_2}\\ 
Y^{a_1}_{a_8} Y^{a_2}_{a_6} Y^{a_3}_{a_4} Z^{a_4 a_5}_{a_5 a_7} Z^{a_6 a_7}_{a_3 a_9} Z^{a_8 a_9}_{a_1 a_2}\\ 
Y^{a_1}_{a_6} Y^{a_2}_{a_7} Y^{a_3}_{a_4} Z^{a_4 a_5}_{a_5 a_8} Z^{a_6 a_7}_{a_3 a_9} Z^{a_8 a_9}_{a_1 a_2}\\ 
Y^{a_1}_{a_8} Y^{a_2}_{a_4} Y^{a_3}_{a_6} Z^{a_4 a_5}_{a_5 a_9} Z^{a_6 a_7}_{a_3 a_7} Z^{a_8 a_9}_{a_1 a_2}\\ 
Y^{a_1}_{a_8} Y^{a_2}_{a_4} Y^{a_3}_{a_6} Z^{a_4 a_5}_{a_5 a_7} Z^{a_6 a_7}_{a_3 a_9} Z^{a_8 a_9}_{a_1 a_2}\\ 
Y^{a_1}_{a_6} Y^{a_2}_{a_4} Y^{a_3}_{a_8} Z^{a_4 a_5}_{a_5 a_9} Z^{a_6 a_7}_{a_3 a_7} Z^{a_8 a_9}_{a_1 a_2}\\ 
Y^{a_1}_{a_6} Y^{a_2}_{a_4} Y^{a_3}_{a_8} Z^{a_4 a_5}_{a_5 a_7} Z^{a_6 a_7}_{a_3 a_9} Z^{a_8 a_9}_{a_1 a_2}\\ 
Y^{a_1}_{a_6} Y^{a_2}_{a_4} Y^{a_3}_{a_7} Z^{a_4 a_5}_{a_5 a_8} Z^{a_6 a_7}_{a_3 a_9} Z^{a_8 a_9}_{a_1 a_2}\\ 
Y^{a_1}_{a_8} Y^{a_2}_{a_3} Y^{a_3}_{a_6} Z^{a_4 a_5}_{a_4 a_9} Z^{a_6 a_7}_{a_5 a_7} Z^{a_8 a_9}_{a_1 a_2}\\ 
Y^{a_1}_{a_8} Y^{a_2}_{a_3} Y^{a_3}_{a_6} Z^{a_4 a_5}_{a_4 a_7} Z^{a_6 a_7}_{a_5 a_9} Z^{a_8 a_9}_{a_1 a_2}\\ 
Y^{a_1}_{a_6} Y^{a_2}_{a_3} Y^{a_3}_{a_8} Z^{a_4 a_5}_{a_4 a_9} Z^{a_6 a_7}_{a_5 a_7} Z^{a_8 a_9}_{a_1 a_2}\\ 
Y^{a_1}_{a_6} Y^{a_2}_{a_3} Y^{a_3}_{a_8} Z^{a_4 a_5}_{a_4 a_7} Z^{a_6 a_7}_{a_5 a_9} Z^{a_8 a_9}_{a_1 a_2}\\ 
Y^{a_1}_{a_6} Y^{a_2}_{a_3} Y^{a_3}_{a_4} Z^{a_4 a_5}_{a_8 a_9} Z^{a_6 a_7}_{a_5 a_7} Z^{a_8 a_9}_{a_1 a_2}\\ 
Y^{a_1}_{a_6} Y^{a_2}_{a_3} Y^{a_3}_{a_4} Z^{a_4 a_5}_{a_5 a_8} Z^{a_6 a_7}_{a_7 a_9} Z^{a_8 a_9}_{a_1 a_2}\\ 
Y^{a_1}_{a_6} Y^{a_2}_{a_3} Y^{a_3}_{a_4} Z^{a_4 a_5}_{a_5 a_7} Z^{a_6 a_7}_{a_8 a_9} Z^{a_8 a_9}_{a_1 a_2}\\ 
Y^{a_1}_{a_8} Y^{a_2}_{a_4} Y^{a_3}_{a_3} Z^{a_4 a_5}_{a_6 a_9} Z^{a_6 a_7}_{a_2 a_7} Z^{a_8 a_9}_{a_1 a_5}\\ 
Y^{a_1}_{a_6} Y^{a_2}_{a_4} Y^{a_3}_{a_3} Z^{a_4 a_5}_{a_7 a_8} Z^{a_6 a_7}_{a_2 a_5} Z^{a_8 a_9}_{a_1 a_9}\\ 
Y^{a_1}_{a_8} Y^{a_2}_{a_3} Y^{a_3}_{a_6} Z^{a_4 a_5}_{a_4 a_7} Z^{a_6 a_7}_{a_2 a_9} Z^{a_8 a_9}_{a_1 a_5}\\ 
Y^{a_1}_{a_6} Y^{a_2}_{a_3} Y^{a_3}_{a_8} Z^{a_4 a_5}_{a_4 a_9} Z^{a_6 a_7}_{a_2 a_7} Z^{a_8 a_9}_{a_1 a_5}\\ 
Y^{a_1}_{a_8} Y^{a_2}_{a_3} Y^{a_3}_{a_4} Z^{a_4 a_5}_{a_6 a_9} Z^{a_6 a_7}_{a_2 a_7} Z^{a_8 a_9}_{a_1 a_5}\\ 
Y^{a_1}_{a_6} Y^{a_2}_{a_3} Y^{a_3}_{a_4} Z^{a_4 a_5}_{a_8 a_9} Z^{a_6 a_7}_{a_2 a_7} Z^{a_8 a_9}_{a_1 a_5}\\ 
Y^{a_1}_{a_6} Y^{a_2}_{a_3} Y^{a_3}_{a_4} Z^{a_4 a_5}_{a_7 a_8} Z^{a_6 a_7}_{a_2 a_5} Z^{a_8 a_9}_{a_1 a_9}\\ 
Y^{a_1}_{a_6} Y^{a_2}_{a_3} Y^{a_3}_{a_4} Z^{a_4 a_5}_{a_5 a_8} Z^{a_6 a_7}_{a_2 a_9} Z^{a_8 a_9}_{a_1 a_7}\\ 
Y^{a_1}_{a_6} Y^{a_2}_{a_3} Y^{a_3}_{a_4} Z^{a_4 a_5}_{a_5 a_8} Z^{a_6 a_7}_{a_2 a_7} Z^{a_8 a_9}_{a_1 a_9}\\ 
Y^{a_1}_{a_6} Y^{a_2}_{a_3} Y^{a_3}_{a_4} Z^{a_4 a_5}_{a_5 a_7} Z^{a_6 a_7}_{a_2 a_8} Z^{a_8 a_9}_{a_1 a_9}\\ 
Y^{a_1}_{a_4} Y^{a_2}_{a_3} Y^{a_3}_{a_6} Z^{a_4 a_5}_{a_7 a_8} Z^{a_6 a_7}_{a_2 a_5} Z^{a_8 a_9}_{a_1 a_9}\\ 
Y^{a_1}_{a_6} Y^{a_2}_{a_3} Y^{a_3}_{a_2} Z^{a_4 a_5}_{a_8 a_9} Z^{a_6 a_7}_{a_4 a_7} Z^{a_8 a_9}_{a_1 a_5}\\ 
Y^{a_1}_{a_6} Y^{a_2}_{a_3} Y^{a_3}_{a_2} Z^{a_4 a_5}_{a_4 a_8} Z^{a_6 a_7}_{a_7 a_9} Z^{a_8 a_9}_{a_1 a_5}\\ 
Y^{a_1}_{a_3} Y^{a_2}_{a_6} Y^{a_3}_{a_2} Z^{a_4 a_5}_{a_8 a_9} Z^{a_6 a_7}_{a_4 a_7} Z^{a_8 a_9}_{a_1 a_5}\\ 
Y^{a_1}_{a_3} Y^{a_2}_{a_6} Y^{a_3}_{a_2} Z^{a_4 a_5}_{a_4 a_8} Z^{a_6 a_7}_{a_7 a_9} Z^{a_8 a_9}_{a_1 a_5} 
\end{align}

$n_y=1,n_Z=4$ Invariants with self-loops
\begin{align}
Y^{a_1}_{a_8} Z^{a_2 a_3}_{a_6 a_7} Z^{a_4 a_5}_{a_4 a_9} Z^{a_6 a_7}_{a_2 a_5} Z^{a_8 a_9}_{a_1 a_3}\\
Y^{a_1}_{a_6} Z^{a_2 a_3}_{a_8 a_9} Z^{a_4 a_5}_{a_4 a_7} Z^{a_6 a_7}_{a_2 a_5} Z^{a_8 a_9}_{a_1 a_3}\\
Y^{a_1}_{a_8} Z^{a_2 a_3}_{a_4 a_6} Z^{a_4 a_5}_{a_5 a_9} Z^{a_6 a_7}_{a_2 a_7} Z^{a_8 a_9}_{a_1 a_3}\\
Y^{a_1}_{a_6} Z^{a_2 a_3}_{a_4 a_8} Z^{a_4 a_5}_{a_5 a_9} Z^{a_6 a_7}_{a_2 a_7} Z^{a_8 a_9}_{a_1 a_3}\\
Y^{a_1}_{a_6} Z^{a_2 a_3}_{a_4 a_8} Z^{a_4 a_5}_{a_5 a_7} Z^{a_6 a_7}_{a_2 a_9} Z^{a_8 a_9}_{a_1 a_3}\\
Y^{a_1}_{a_6} Z^{a_2 a_3}_{a_4 a_7} Z^{a_4 a_5}_{a_5 a_8} Z^{a_6 a_7}_{a_2 a_9} Z^{a_8 a_9}_{a_1 a_3}\\
Y^{a_1}_{a_6} Z^{a_2 a_3}_{a_4 a_5} Z^{a_4 a_5}_{a_8 a_9} Z^{a_6 a_7}_{a_2 a_7} Z^{a_8 a_9}_{a_1 a_3}\\
Y^{a_1}_{a_4} Z^{a_2 a_3}_{a_8 a_9} Z^{a_4 a_5}_{a_5 a_6} Z^{a_6 a_7}_{a_2 a_7} Z^{a_8 a_9}_{a_1 a_3}\\
Y^{a_1}_{a_4} Z^{a_2 a_3}_{a_6 a_8} Z^{a_4 a_5}_{a_5 a_9} Z^{a_6 a_7}_{a_2 a_7} Z^{a_8 a_9}_{a_1 a_3}\\
Y^{a_1}_{a_4} Z^{a_2 a_3}_{a_6 a_8} Z^{a_4 a_5}_{a_5 a_7} Z^{a_6 a_7}_{a_2 a_9} Z^{a_8 a_9}_{a_1 a_3}\\
Y^{a_1}_{a_4} Z^{a_2 a_3}_{a_6 a_7} Z^{a_4 a_5}_{a_5 a_8} Z^{a_6 a_7}_{a_2 a_9} Z^{a_8 a_9}_{a_1 a_3}\\
Y^{a_1}_{a_6} Z^{a_2 a_3}_{a_4 a_8} Z^{a_4 a_5}_{a_5 a_9} Z^{a_6 a_7}_{a_2 a_3} Z^{a_8 a_9}_{a_1 a_7}\\
Y^{a_1}_{a_6} Z^{a_2 a_3}_{a_4 a_8} Z^{a_4 a_5}_{a_5 a_7} Z^{a_6 a_7}_{a_2 a_3} Z^{a_8 a_9}_{a_1 a_9}\\
Y^{a_1}_{a_6} Z^{a_2 a_3}_{a_2 a_8} Z^{a_4 a_5}_{a_4 a_9} Z^{a_6 a_7}_{a_5 a_7} Z^{a_8 a_9}_{a_1 a_3}\\
Y^{a_1}_{a_6} Z^{a_2 a_3}_{a_2 a_4} Z^{a_4 a_5}_{a_5 a_8} Z^{a_6 a_7}_{a_7 a_9} Z^{a_8 a_9}_{a_1 a_3}
\end{align}
\subsection{Lists of spontaneous CP-odd invariants}
$n_{v}=1,n_Z=3$
\begin{align}
v_{a_4} v^{\ast a_1} Z^{a_2 a_3}_{a_6 a_7} Z^{a_4 a_5}_{a_2 a_5} Z^{a_6 a_7}_{a_1 a_3}\\
v_{a_4} v^{\ast a_1} Z^{a_2 a_3}_{a_2 a_6} Z^{a_4 a_5}_{a_5 a_7} Z^{a_6 a_7}_{a_1 a_3}
\end{align}
$n_{v}=2,n_Z=3$
\begin{align}
v_{a_5} v_{a_7} v^{\ast a_1} v^{\ast a_2} Z^{a_3 a_4}_{a_6 a_8} Z^{a_5 a_6}_{a_3 a_4} Z^{a_7 a_8}_{a_1 a_2}\\
v_{a_5} v_{a_7} v^{\ast a_1} v^{\ast a_2} Z^{a_3 a_4}_{a_3 a_8} Z^{a_5 a_6}_{a_4 a_6} Z^{a_7 a_8}_{a_1 a_2}\\
v_{a_5} v_{a_7} v^{\ast a_1} v^{\ast a_2} Z^{a_3 a_4}_{a_3 a_6} Z^{a_5 a_6}_{a_4 a_8} Z^{a_7 a_8}_{a_1 a_2}\\
v_{a_5} v_{a_7} v^{\ast a_1} v^{\ast a_2} Z^{a_3 a_4}_{a_3 a_4} Z^{a_5 a_6}_{a_6 a_8} Z^{a_7 a_8}_{a_1 a_2}\\
v_{a_3} v_{a_5} v^{\ast a_1} v^{\ast a_2} Z^{a_3 a_4}_{a_7 a_8} Z^{a_5 a_6}_{a_4 a_6} Z^{a_7 a_8}_{a_1 a_2}\\
v_{a_3} v_{a_5} v^{\ast a_1} v^{\ast a_2} Z^{a_3 a_4}_{a_6 a_7} Z^{a_5 a_6}_{a_4 a_8} Z^{a_7 a_8}_{a_1 a_2}\\
v_{a_3} v_{a_5} v^{\ast a_1} v^{\ast a_2} Z^{a_3 a_4}_{a_4 a_7} Z^{a_5 a_6}_{a_6 a_8} Z^{a_7 a_8}_{a_1 a_2}\\
v_{a_3} v_{a_7} v^{\ast a_1} v^{\ast a_2} Z^{a_3 a_4}_{a_5 a_8} Z^{a_5 a_6}_{a_1 a_6} Z^{a_7 a_8}_{a_2 a_4} 
\end{align}
$n_{v}=1,n_Z=4$
\begin{align}
v_{a_8} v^{\ast a_1} Z^{a_2 a_3}_{a_6 a_7} Z^{a_4 a_5}_{a_4 a_9} Z^{a_6 a_7}_{a_2 a_5} Z^{a_8 a_9}_{a_1 a_3}\\
v_{a_6} v^{\ast a_1} Z^{a_2 a_3}_{a_8 a_9} Z^{a_4 a_5}_{a_4 a_7} Z^{a_6 a_7}_{a_2 a_5} Z^{a_8 a_9}_{a_1 a_3}\\
v_{a_6} v^{\ast a_1} Z^{a_2 a_3}_{a_4 a_7} Z^{a_4 a_5}_{a_8 a_9} Z^{a_6 a_7}_{a_2 a_5} Z^{a_8 a_9}_{a_1 a_3}\\
v_{a_8} v^{\ast a_1} Z^{a_2 a_3}_{a_4 a_6} Z^{a_4 a_5}_{a_5 a_9} Z^{a_6 a_7}_{a_2 a_7} Z^{a_8 a_9}_{a_1 a_3}\\
v_{a_6} v^{\ast a_1} Z^{a_2 a_3}_{a_4 a_8} Z^{a_4 a_5}_{a_5 a_9} Z^{a_6 a_7}_{a_2 a_7} Z^{a_8 a_9}_{a_1 a_3}\\
v_{a_6} v^{\ast a_1} Z^{a_2 a_3}_{a_4 a_8} Z^{a_4 a_5}_{a_5 a_7} Z^{a_6 a_7}_{a_2 a_9} Z^{a_8 a_9}_{a_1 a_3}\\
v_{a_6} v^{\ast a_1} Z^{a_2 a_3}_{a_4 a_7} Z^{a_4 a_5}_{a_5 a_8} Z^{a_6 a_7}_{a_2 a_9} Z^{a_8 a_9}_{a_1 a_3}\\
v_{a_6} v^{\ast a_1} Z^{a_2 a_3}_{a_4 a_5} Z^{a_4 a_5}_{a_8 a_9} Z^{a_6 a_7}_{a_2 a_7} Z^{a_8 a_9}_{a_1 a_3}\\
v_{a_4} v^{\ast a_1} Z^{a_2 a_3}_{a_8 a_9} Z^{a_4 a_5}_{a_5 a_6} Z^{a_6 a_7}_{a_2 a_7} Z^{a_8 a_9}_{a_1 a_3}\\
v_{a_4} v^{\ast a_1} Z^{a_2 a_3}_{a_6 a_8} Z^{a_4 a_5}_{a_5 a_9} Z^{a_6 a_7}_{a_2 a_7} Z^{a_8 a_9}_{a_1 a_3}\\
v_{a_4} v^{\ast a_1} Z^{a_2 a_3}_{a_6 a_8} Z^{a_4 a_5}_{a_5 a_7} Z^{a_6 a_7}_{a_2 a_9} Z^{a_8 a_9}_{a_1 a_3}\\
v_{a_4} v^{\ast a_1} Z^{a_2 a_3}_{a_6 a_7} Z^{a_4 a_5}_{a_5 a_8} Z^{a_6 a_7}_{a_2 a_9} Z^{a_8 a_9}_{a_1 a_3}\\
v_{a_6} v^{\ast a_1} Z^{a_2 a_3}_{a_4 a_8} Z^{a_4 a_5}_{a_5 a_9} Z^{a_6 a_7}_{a_2 a_3} Z^{a_8 a_9}_{a_1 a_7}\\
v_{a_6} v^{\ast a_1} Z^{a_2 a_3}_{a_4 a_8} Z^{a_4 a_5}_{a_5 a_7} Z^{a_6 a_7}_{a_2 a_3} Z^{a_8 a_9}_{a_1 a_9}\\
v_{a_6} v^{\ast a_1} Z^{a_2 a_3}_{a_2 a_8} Z^{a_4 a_5}_{a_4 a_9} Z^{a_6 a_7}_{a_5 a_7} Z^{a_8 a_9}_{a_1 a_3}\\
v_{a_6} v^{\ast a_1} Z^{a_2 a_3}_{a_2 a_4} Z^{a_4 a_5}_{a_5 a_8} Z^{a_6 a_7}_{a_7 a_9} Z^{a_8 a_9}_{a_1 a_3}
\end{align}

\subsection{\label{app:larger}Larger CP-odd invariants}

In addition to the smaller invariants discussed previously, we also found some larger CPIs with up to 9 $Z$ tensors. Using the notation established in Eq.~(\ref{eq:CPI_I_c}), we show only half of the CPI , which is sufficient to uniquely define it.

\bea
I^{(7,2)}_1 &=& 
Z_{b_1^{}b_2^{}}^{b^{}_3b^{}_4} ~
Z_{b_3^{}a_1^{}}^{b^{}_1a^{}_5} ~
Z_{b_4^{}a_2^{}}^{b^{}_2a^{}_6} ~
Z_{c_1^{}c_2^{}}^{c^{}_3c^{}_4} ~
Z_{c_3^{}a_3^{}}^{c^{}_1a^{\prime}_1} ~
Z_{c_4^{}a_4^{}}^{c^{}_2a^{}_2} ~
Z_{a^{}_5a^{}_6}^{a'_3a^{}_4} ~
Y_{a'_1}^{a^{}_1} ~
Y_{a'_3}^{a^{}_3}
\ ,\label{eq:deltaCPoddinv}\\[2mm]
I^{(8)}_1 &=&  
 Z_{b^{}_1 b^{}_2}^{b^{}_3 b^{}_4}   
~Z_{b^{}_3 b^{}_5}^{b^{}_1 b^{}_6}   
~ Z_{b^{}_4 a^{}_1}^{b^{}_2 a^{}_5}  
~Z_{b^{}_6 a^{}_2}^{b^{}_5 a^{}_6} 
~Z_{c^{}_1 c^{}_2}^{c^{}_3 c^{}_4}   
~Z_{c^{}_3 a^{}_3}^{c^{}_1 a^{}_1}
~Z_{c^{}_4 a^{}_4}^{c^{}_2 a^{}_2}
~Z_{a^{}_5 a^{}_6}^{a^{}_3 a^{}_4} 
  \ , \\[2mm]
I^{(9)}_1 &=&  
Z_{b^{}_1 b^{}_2}^{b^{}_3 b^{}_4}   
~Z_{b^{}_3 b^{}_5}^{b^{}_1 b^{}_6}   
~Z_{b^{}_4 b^{}_7}^{b^{}_2 b^{}_8}  
~Z_{b^{}_6 a^{}_1}^{b^{}_5 a^{}_5} 
~ Z_{b^{}_8 a^{}_2}^{b^{}_7 a^{}_6} 
~Z_{c^{}_1 c^{}_2}^{c^{}_3 c^{}_4}   
~Z_{c^{}_3 a^{}_3}^{c^{}_1 a^{}_1}
~Z_{c^{}_4 a^{}_4}^{c^{}_2 a^{}_2}
~Z_{a^{}_5 a^{}_6}^{a^{}_3 a^{}_4} \ .
\eea
The respective CPIs $\mathcal{I}^{(7,2)}_1$, $\mathcal{I}^{(8)}_1$, $\mathcal{I}^{(9)}_1$ can be obtained by subtracting from the $I$ above the $I^*$ obtained by swapping the upper and lower indices, as described in general in Section~\ref{CPIs}.

\section{$\bs{\Delta(6 n^2)}$ potentials as particular cases of $\bs{\Delta(3
    n^2)}$ potentials \label{6n2_from_3n2}}
\cleqn
The triplet generators of $\Delta(6n^2)$ with $n\in \mathbb N$ are~\cite{Escobar:2008vc}
\be
a = \begin{pmatrix} 0&1&0\\
0&0&1\\
1&0&0 \end{pmatrix} \ , \qquad
b = \pm \begin{pmatrix} 0&0&1\\
0&1&0\\
1&0&0 \end{pmatrix} \ , \qquad
c = \begin{pmatrix} \eta^l&0&0\\
0&\eta^{-l}&0\\
0&0&1 \end{pmatrix} \ , \label{eq:D6n2}
\ee
where $\eta=e^{2\pi i/n}$ and $l\in \mathbb N$. If the field transforms as a
faithful triplet, any $c$-invariant operator $\mathcal O$ will also be
invariant under the phase transformation\footnote{For faithful
  representations, $l$ and $n$ have to be coprime. As a consequence, there
  must be an integer $p$ such that $c^p=c_0$.} 
\be
c_0=\begin{pmatrix} \eta&0&0\\
0&\eta^{-1} &0\\
0&0&1\end{pmatrix} \ , 
\ee
and therefore also under 
\be
c_0^{-l}=\begin{pmatrix} \eta^{-l}&0&0\\
0& \eta^{l}&0\\
0&0&1 \end{pmatrix}   \qquad \text{and}\qquad
c_0^{k}=\begin{pmatrix} \eta^{k}&0&0\\
0&\eta^{-k} &0\\
0&0&1\end{pmatrix} \ .
\ee
Imposing additionally invariance  under $a$, we quickly find that
the operator $\mathcal O$ is also invariant under 
\be
a c_0^{-l} a^2 =\begin{pmatrix} \eta^{l}&0&0\\
0&1&0\\
0&0&\eta^{-l} \end{pmatrix}
 \qquad \text{and}\qquad
a^2 c_0^{k} a =\begin{pmatrix} 1&0&0\\
0&\eta^{k}&0\\
0&0&\eta^{-k} \end{pmatrix} \ .
\ee
As a result, the operator $\mathcal O$ is symmetric under the successive
application of $a c_0^{-l} a^2$ and $a^2 c_0^{k} a$, i.e.
\be
\begin{pmatrix} \eta^l&0&0\\
0&\eta^{k}&0\\
0&0&\eta^{-k-l} \end{pmatrix} \ .
\ee
Demanding invariance under $a$ and $c$ of Eq.~\eqref{eq:D6n2}
therefore leads to the set of $\Delta(3n^2)$ invariant operators where the
triplet generators are given by~\cite{Luhn:2007uq}
\be
a' = \begin{pmatrix} 0&1&0\\
0&0&1\\
1&0&0 \end{pmatrix} \ , \qquad
c' = \begin{pmatrix} \eta^l&0&0\\
0&\eta^k&0\\
0&0&\eta^{-k-l} \end{pmatrix} \ .
\ee

We thus conclude that the $\Delta(6n^2$) symmetric potential can be deduced
from the  $\Delta(3n^2)$ invariant potential by simply dropping all terms
which are {\it not} symmetric under $b$ of Eq.~\eqref{eq:D6n2}.
Therefore, in each of these cases it is sufficient to use the already obtained
expressions for the CPIs and set constraints on the coefficients to make all
the terms in the potential invariant under the $b$ generator.

\end{appendix}

\end{document}